\documentclass[12pt,aps,prd,showpacs,amsmath,amssymb,floatfix,nofootinbib,superscriptaddress,preprint]{revtex4-1}
\usepackage{graphicx}  
\usepackage{multirow,bigdelim}
\usepackage[dvipsnames]{xcolor}
\usepackage[colorlinks,linkcolor=Blue,citecolor=Green]{hyperref}

\usepackage[sicmds,freestanding]{hepunits}
\DeclareSIUnit{\fm}{\femto\metre}

\usepackage{physics}
\newcommand{\olsi}[1]{\,\overline{\!{#1}}} 

\usepackage{latexsym}
\usepackage{amsfonts}
\usepackage{amsmath,amssymb}
\usepackage{slashed}
\usepackage{color}

\usepackage{supertabular} 
\usepackage{epsfig}
\usepackage{lipsum}
\usepackage{orcidlink}
\usepackage{accents}
\usepackage[inline]{enumitem}

\usepackage{dsfont}

\usepackage{gentium}
\usepackage[cochineal,bigdelims,vvarbb]{newtxmath}

\usepackage[T1]{fontenc}

\newcommand{\vpp}{\vec{p}{}'}
\newcommand{\vp}{\vec{p}}
\newcommand{\oT}{\widehat{T}}
\newcommand{\oTC}{\widehat{T}_C}
\newcommand{\oTSC}{\widehat{T}_{SC}}
\newcommand{\oH}{\widehat{H}}
\newcommand{\oHF}{\widehat{H}_0}
\newcommand{\oHC}{\widehat{H}_C}
\newcommand{\oV}{\widehat{V}}
\newcommand{\oVS}{\widehat{V}_S}
\newcommand{\oVC}{\widehat{V}_C}
\newcommand{\sFp}{\phi_{\vp}^{(+)}}
\newcommand{\sFpmenosp}{\phi_{-\vp}^{(+)}}
\newcommand{\sFm}{\phi_{\vp}^{(-)}}
\newcommand{\sFpm}{\phi_{\vp}^{(\pm)}}
\newcommand{\sCp}{\psi_{\vp}^{(+)}}
\newcommand{\sCpmenosp}{\psi_{-\vp}^{(+)}}
\newcommand{\sCm}{\psi_{\vpp}^{(-)}}
\newcommand{\sCmbis}{\psi_{\vp}^{(-)}}
\newcommand{\sCpm}{\psi_{\vp}^{(\pm)}}

\newcommand{\GCp}{\widehat{G}_C^{(+)}}
\newcommand{\GFp}{\widehat{G}_0^{(+)}}
\newcommand{\GSCp}{\widehat{G}_{SC}^{(+)}}
\newcommand{\GCpm}{\widehat{G}_C^{(\pm)}}
\newcommand{\GFpm}{\widehat{G}_0^{(\pm)}}
\newcommand{\GSCpm}{\widehat{G}_{SC}^{(\pm)}}
\newcommand{\ie}{i\epsilon}
\newcommand{\LJ}{\mathcal{L}_0}
\newcommand{\tresmom}{\widehat{\vec{p}}}

\newcommand{\CorrName}{FRP}
\usepackage{setspace}
\usepackage[bottom]{footmisc}

\usepackage[compact]{titlesec}
\titleformat{\section}
   {\normalfont\bfseries\MakeUppercase}{\thesection}{1em}{}
\titlespacing*{\section}
{0pt}{4ex plus 1ex minus 1ex}{2ex plus 1ex}
\titlespacing*{\subsection}
{0pt}{3ex plus 1ex minus 1ex}{0.5ex plus 0.5ex}

\usepackage{titletoc}

\titlecontents{section}[0em]
  {}
  {\small\bfseries\contentslabel[\thecontentslabel]{1.5em}\MakeUppercase}
  {\hspace*{-8em}\small\bfseries\MakeUppercase}
  {\hfill\small\contentspage}

\titlecontents{subsection}[1em]
  {}
  {\small\contentslabel{1.5em}}
  {\hspace*{-8em}}
  {\hfill\small\contentspage}
   
\usepackage[normalem]{ulem}

\begin{document}
\count\footins = 1000

\onehalfspacing
\setlength{\parskip}{2pt}
\abovedisplayskip=6pt plus 3pt minus 3pt
\abovedisplayshortskip=0pt plus 3pt
\belowdisplayskip=6pt plus 3pt minus 3pt
\belowdisplayshortskip=5pt plus 3pt minus 3pt

\allowdisplaybreaks

\title{\boldmath Femtoscopy correlation functions and hadron-hadron scattering amplitudes in presence of Coulomb potential}

\newcommand{\ific}{\affiliation{\small%
Instituto de F\'isica Corpuscular (centro mixto CSIC-UV), \\
Institutos de Investigaci\'on de Paterna, Apartado 22085, 46071, Valencia, Spain}}

\author{Miguel Albaladejo\orcidlink{0000-0001-7340-9235}}\email{Miguel.Albaladejo@ific.uv.es}
\ific 

\author{Amador García-Lorenzo\orcidlink{0009-0001-3200-9996}}\email{Amador.Garcia@ific.uv.es}
\ific 

\author{Juan Nieves\orcidlink{0000-0002-2518-4606}} \email{Juan.M.Nieves@ific.uv.es}
\ific

\renewcommand{\abstractname}{\vspace{20pt}Abstract}

\begin{abstract}
\vspace{10pt}%
This work addresses the incorporation of Coulomb interactions into femtoscopy correlation functions (CFs) used to probe hadron interactions. Combining strong contact potentials with Coulomb effects, the derived scattering amplitudes and wave functions are used to compute  CFs, accounting for both interactions coherently. Next, we analyze the nature of the corrections due to the finite range of the strong interaction, closely linked to off-shell effects, and propose an approximate method to account for them. We show how these corrections turn out to be essential to accurately extract the strong hadron-hadron scattering amplitudes from CF data.  We compare the obtained expressions for the CFs with those reported in previous theoretical frameworks. Also, using proton-proton systems as a case study, we obtain a good comparison of our results with previous realistic theoretical calculations and with accurate recent data from the ALICE experiment. The framework enables precise CF calculations without numerical solution of the Schr\"odinger equation, offering a practical tool for femtoscopy analyses involving charged hadrons. In addition, the scheme allows for an easy connection with scattering amplitudes obtained from Effective Field Theories. 
\end{abstract}

\maketitle

\setcounter{tocdepth}{3}

\tableofcontents

\section{Introduction}

Despite the enormous recent progress of the last decades, our knowledge of the QCD spectrum and hadron interactions is often limited by the technical difficulties faced in performing direct scattering experiments between hadrons. Experimental access to their interactions is therefore obtained in reactions where a given set of particles is produced in the final state. Lattice QCD, a first principles approach based on the numerical simulations of the QCD action, also provides excellent insights into hadron interactions.

The Hanbury Brown-Twiss interferometry of stellar photons \cite{HanburyBrown:1954amm,HanburyBrown:1956bqd} has  been adapted and employed in particle physics as a tool to study the hadronic fireball and the quark-gluon plasma in relativistic heavy-ion collisions \cite{Goldhaber:1960sf,Kopylov:1974th,Ezell:1977mh}, a technique now known as \textit{femtoscopy}. Given the importance of final-state interactions in the correlations of the outgoing particles \cite{Koonin:1977fh,Lednicky:1981su}, femtoscopy has emerged more recently as a new technique to improve our knowledge of hadron interactions. The momentum-space correlations  of a pair of hadrons are measured through their correlation function (CF), which is the quotient of the number of particle pairs with the same relative momentum produced in the same collision event over the number of pairs originated from mixed events. The hadron production yields in high-multiplicity events of proton-proton, proton-nucleus, and nucleus-nucleus collisions are well known based on statistical models. Thus, the CFs encode information about the interactions of hadrons. Further details on the developments and references can be found \textit{e.g.} in Refs.\,\cite{Lisa:2005dd,Fabbietti:2020bfg,ALICE:2020mfd}.

Even if one wants to focus on strong interactions, Coulomb effects are of particular relevance in CFs, given that they are quite significant for low momenta and that the effect of the phase-space factors of a hadron pair cancels in the ratio that gives the CF. Therefore, one often needs to consider strong and Coulomb interactions simultaneously. For instance, in a recent work \cite{Albaladejo:2023pzq}, we have studied femtoscopy correlation functions (CFs) for $H_c \phi$ final states, where $H_c$ is a pseudoscalar charm meson and $\phi$ stands for a (light) pseudo-Goldstone boson. In these CFs clear and distinct signatures, coming from the two states around the $D^{\ast}_0(2300)$ structure \cite{Albaladejo:2016lbb,Du:2017zvv}, show up. However, these studies were carried out only for channels in which there are no Coulomb interactions, \textit{i.e.}, $D^+ \pi^0$, $D^0 \pi^+$, $D^0 K^+$, $D^+ K^0$ etc. The Barcelona group \cite{Torres-Rincon:2023qll} has studied similar CFs, but including also channels in which the Coulomb interaction is present such as $D^+ \pi^+$, $D^+ \pi^-$ etc, using the momentum-space Coulomb potential obtained by truncating its Fourier-transform at a given distance $\mathcal{R}_C \sim 60\,\fm$ (where the results are stable, according to the authors). In the present work, we aim to include the Coulomb effects in a different way, taking into account the methods of Refs.~\cite{Kong:1999sf,Nieves:2003uu}.

Kong and Ravndal presented in Ref.~\cite{Kong:1999sf} a formalism to compute scattering amplitudes in the simultaneous presence of Coulomb and strong (contact) interactions. This formalism has been used by Guo \textit{et al.} \cite{Zhang:2020mpi,Shi:2021hzm} for the case of $D \olsi{D}{}^{(\ast)}$ interactions (where Coulomb effects appear in the $D^0 \olsi{D}{}^{(\ast)0}$--$D^+ D^{(\ast)-}$ coupled channels). On the other hand, Nieves \cite{Nieves:2003uu} used Distorted Wave Theory \cite{Barford:2002je} (which follows the two-potential trick \cite{Newton:1982qc,Birse:2005um}) to derive the equations for isovector $NN$ S-wave scattering amplitudes taking into account simultaneously one-pion exchange and contact interactions, including derivative ones. One-pion exchange is a long-range interaction and, with due care, the scheme of Ref.~\cite{Nieves:2003uu} can be applied to the case of the Coulomb potential. One of the goals of this manuscript is to show the equivalence between both formalisms.

Given the importance of Coulomb interactions for CFs, there have been studies about the subject already at the early stages of the development of femtoscopy, \textit{e.g.} the seminal works of Koonin \cite{Koonin:1977fh} or Lednicky and Lyuboshitz \cite{Lednicky:1981su}. In Ref.\,\cite{Koonin:1977fh} the $p$-$p$ CF is computed from a numerically obtained wave-function taking into account Coulomb and elaborated potentials. In Ref.\,\cite{Lednicky:1981su}, the Coulomb interaction is exactly incorporated in an analytical way into the $p$-$p$ asymptotic wave function and the CF, and the strong interaction is accounted for through the effective range expansion (ERE). In the present manuscript, we will apply the same techniques employed in Refs.\,\cite{Kong:1999sf,Nieves:2003uu} to incorporate strong contact potentials and Coulomb interactions into the scattering amplitudes to obtain, in a coherent scheme, the wave functions that enter into the femtoscopy CFs. By doing so, we will obtain expressions for the latter that are equivalent to those in Ref.\,\cite{Lednicky:1981su}. Furthermore, given the equivalence that we will prove between the formalisms in Refs.\,\cite{Kong:1999sf} and \cite{Nieves:2003uu}, more elaborate Effective Field Theories (EFTs) for the strong-interaction part can be applied within our formalism. 

We also study in detail the nature  of the corrections due to the finite range of the strong interaction, closely linked to off-shell effects, and propose an approximate method to include these effects in the calculation of the CFs. We show how taking these corrections into account is essential to accurately extract the strong hadron-hadron scattering amplitudes from CF data.

The manuscript is organized as follows. In Sec.\,\ref{sec:formalism-amplitudes} we discuss the scattering amplitudes for strong interactions in the presence of a Coulomb potential, distinguishing the repulsive (\ref{subsec:repulsive}) and the attractive (\ref{subsec:attractive}) cases, and introduce  the Coulomb ERE (\ref{subsec:phaseEREuni}). The relation between the formalisms employed in Refs.\,\cite{Kong:1999sf} and \cite{Nieves:2003uu} is derived in Subsec.\,\ref{subsec:JuanFormalism}. In Sec.\,\ref{sec:CFs} we discuss CFs, incorporating the Coulomb interactions into their evaluation (\ref{subsec:CoulombintoCFs}), some possible improvements of the LL approximation (\ref{subsec:deltaC}), and the required modifications when dealing with identical particles (\ref{subsec:identical}). In Sec.\,\ref{sec:results} we present comparisons of $p$-$p$ CF calculations with other theoretical approaches and with ALICE data. Section \ref{sec:summary} is devoted to our summary and conclusions. Additionally, in  Appendices \ref{app:beta} and  \ref{app:esfera-dura}, we discuss  the corrections due to the finite range of the strong interaction for simple potentials and analytically solve the scattering problem for  a spherical square-well in presence of the Coulomb potential, respectively.

\section{Strong interactions in the presence of Coulomb potential} \label{sec:formalism-amplitudes}

\subsection{General features}
To combine strong and Coulomb interactions, we follow Ref.\,\cite{Kong:1999sf}. Let us consider a hadron pair of reduced mass $\mu$, and let us consider that their relative motion is described by a full Hamiltonian $\oH = \oHF + \oVC + \oVS$, which splits into free (kinetic), Coulomb, and strong-interaction terms, $\oHF$, $\oVC$, and $\oVS$, respectively. Also, $\oHC = \oHF + \oVC$ denotes the purely Coulomb Hamiltonian, and $\oV = \oVS + \oVC$ the total potential. We then define the eigenfunctions of the different Hamiltonian operators as follows:\footnote{We use here the same  normalization of the states as in Ref.~\cite{Kong:1999sf}, $\braket{ \vec{p}^{\,\prime} }{ \vec{p}\,}=(2\pi)^3\delta^3(\vec p - \vec p\,')$ and therefore $\int \frac{d^3\vec p}{(2\pi)^3}\, \dyad{\vec{p}} = \mathbb{1}$. On the other hand, $\braket{\vec{r}^{\,\prime}}{\vec{r}\,}=\delta^3(\vec r - \vec r\,')$ and therefore $\int d^3\vec r\,\dyad{\vec{r}} = \mathbb{1}$. These conventions lead  in coordinate space to $\braket{\vec r}{\vec p}=e^{i\vec p\cdot\vec r}$. In addition, the spherical waves are introduced by the decomposition
\begin{align*}
\ket{\vec{p} }  & = \sum_{\ell,m_\ell} \,  Y^*_{\ell m_\ell}(\hat p) \ket{ p,\ell,m_\ell}\,, \\ 
\braket{ p',\ell',m_{\ell'} }{p,\ell,m_\ell} &= (2\pi)^3\frac{\delta(p'-p)}{p^2}\delta_{\ell\ell'}\delta_{m_\ell m_{\ell'}}\,.
\end{align*}
Note that we use natural units $\hbar=1=c$.}
\begin{subequations}\begin{align}
\oHF \ket*{\vec{p}}    & = E_p \ket*{\vec{p}}\,, \label{eq:sc-freehamiltonian}\\ 
\oHC \ket*{\sCpm} & = E_p \ket*{\sCpm}\,,\\
\oH \ket*{\sFpm}  & = E_p \ket*{\sFpm}\,,
\end{align}\end{subequations}
with $2\mu\,E_p = \vec{p}^{\,2}\equiv p^2$ and where the label $(\pm)$ refers to the fact that the states $\ket*{\sCpm}$ and $\ket*{\sFpm}$ are those obtained by acting with the $\Omega_\pm$ and $\Omega_\pm^{C}$ M\"oller operators  on the  state $\ket* {\vec{p}}$ \cite{Pascual:2012,Taylor:2006}, respectively. While the ket $\ket*{\vec{p}}$ is an eigenstate of the three momentum operator $\tresmom$, the interacting ones $\ket*{\sCpm}$ and $\ket*{\sFpm}$ are not. Finally, we also define the free, Coulomb and full propagator operators, as respectively follows:
\begin{subequations}\begin{align}
\GFpm(E)  & = \frac{1}{E - \oHF \pm \ie}\,,\\
\GCpm(E)  & = \frac{1}{E - \oHC \pm \ie}\,,\\
\GSCpm(E) & = \frac{1}{E - \oH \pm \ie}\,.
\end{align}
\label{eq:propagators}
\end{subequations}
There are some relations between the different solutions and operators, that we just enumerate here. First, the propagators satisfy:
\begin{subequations}\label{eq:relation-propagators}
    \begin{align}
           \GCpm - \GFpm & = \GFpm\, \oVC\, \GCpm \,, \label{eq:rel-GC-GSC}\\
           \GSCpm - \GCpm & = \GCpm\, \oVS\, \GSCpm\,.
    \end{align} 
\end{subequations}
There are also Lippmann-Schwinger type relations that will be employed below:
\begin{subequations}\begin{align}
\oV \ket*{\sFpm} & = \oT(E_p\pm \ie) \ket*{\vec{p}}\,,\\
\ket*{\sCpm} & = \ket{\vec{p}} + \GCpm(E_p) \oVC \ket{\vec{p}}= \ket{\vec{p}} + \GFpm(E_p) \oTC(E_p\pm \ie) \ket{\vec{p}}\,,\label{eq:psiCFromFree}\\
\ket*{\sFpm} & = \ket*{\sCpm} + \GSCpm (E_p)\oVS \ket*{\sCpm} = \ket*{\sCpm} + \sum_{n=1}^{\infty} \left( \GCpm(E_p) \oVS \right)^n \ket*{\sCpm}
\,,\label{eq:phiCFrompsiC}
\end{align}\end{subequations}
where we have introduced the full $T$-matrix operator $\oT(z)$, which satisfies the Lippmann-Schwinger equation (LSE),
\begin{equation}
\widehat{T}(z)=\widehat{V} + \widehat{V} \frac{1}{z-\widehat{H}_0}\widehat{T}(z)= \widehat{V}+ \widehat{V} \frac{1}{z-\widehat{H}}\widehat{V} \,.  
\label{eq:TdeJuan}
\end{equation}
The Coulomb $T$-matrix operator $\oTC$ is readily obtained from the above equation by the obvious replacements $\oV$ by $\oVC$ and $\oH$ by $\oHC$. In addition, we  define, 
\begin{equation}
\oTSC(z)=\oVS + \oVS \frac{1}{z-\oHC} \oTSC\,.
\end{equation}

The full amplitude, for energy $E_p+\ie$, can be rearranged or decomposed into the pure Coulomb one plus a strong amplitude modified by Coulomb interactions as follows~\cite{Taylor:2006}:
\begin{align}
T(\vpp,\vp) & = \mel*{\vpp}{\oT}{\vp} = \mel*{\vpp}{\oV}{\phi_{\vp}^{(+)}} = \mel*{\vpp}{\oVS}{\sFp} + \mel*{\vpp}{\oVC}{\sFp} \nonumber \\
& = \mel*{\vpp}{\oVS}{\sFp} + \mel*{\vpp}{\oVC}{\sCp} + \mel*{\vpp}{\oVC\,\GSCp\,\oVS}{\sCp} \nonumber \\
& = \mel*{\vpp}{\oVS}{\sFp} + \mel*{\vpp}{\oVC}{\sCp} + \mel*{\vpp}{\oVC\,\GCp\,\oTSC}{\sCp} \nonumber \\
& = \mel*{\vpp}{\oVS}{\sFp} + \mel*{\vpp}{\oVC}{\sCp} + \mel*{\vpp}{\oVC\,\GCp\,\oVS}{\sFp} \nonumber \\
& = \mel*{\vpp}{\oVC}{\sCp} + \mel*{\vpp}{(\mathbb{1}+\oVC\GCp)\oVS}{\sFp} \nonumber \\
& = \mel*{\vpp}{\oVC}{\sCp} + \mel*{\sCm}{\oVS}{\sFp} = \mel*{\vpp}{\oTC}{\vp} + \mel*{\sCm}{\oTSC}{\sCp} \nonumber \\
& \equiv T_C(\vpp,\vp) + T_{SC}(\vpp,\vp)\,. \label{eq:T_TC_TSC_decomposition}
\end{align}
This arrangement, called the Gell-Mann-Goldberger decomposition \cite{Gell-Mann:1953dcn,Zorbas:1976cd}, is also obtained in Ref.~\cite{Nieves:2003uu}, as we will see below in Sec.~\ref{subsec:JuanFormalism}. 

The Coulomb wave functions for two hadrons of charges $e_1$ and $e_2$ in coordinate space are~\cite{Kong:1999sf}:
\vspace{-15pt}
\begin{subequations}\begin{align}
\psi^{(\pm)}_{\vec{p}}(\vec{r}\,) &= \braket*{\vec{r}}{\psi_{\vp}^{(\pm)}} = e^{-\frac{\pi\,\eta}{2}} \Gamma(1\pm i \eta) M(\mp i\eta,1;\pm ipr - i \vec{p}\cdot\vec{r})e^{i \vec{p}\cdot\vec{r}}\,, \label{eq:WFC}  \\
\braket*{\psi_{\vp}^{(\pm)}}{\psi_{\vpp}^{(\pm)}} & = (2\pi)^3 \delta^3(\vec{p}-\vec{p}')\,,
\end{align}\end{subequations}
and $\eta = {\rm sign}[e_1e_2]\,\alpha \mu/p$ with $\alpha\simeq 1/137.036$, the fine-structure constant in terms of which the local Coulomb potential is $\langle \vec r\,'|\oVC | \vec r\,\rangle= {\rm sign}[e_1e_2]\,\alpha\,\delta^3(\vec{r}-\vec{r}')/r$. For simplicity we have assumed $|e_1|=|e_2|=e$, with $e$ the proton charge. To extend the results below to other values of the charges one  should simply  replace  $\alpha  \to (|e_1e_2|/e^2)\,\alpha$ in all formulae. In addition,  $M(a,b;z) \equiv {}_{1}\!\,F_{1}(a;b;z)$ is the Kummer confluent hypergeometric function,
\begin{equation}
 M(a,b;z) = \sum_{n=0}^{+\infty}   \frac{(a)_n z^n}{(b)_n n!}
\end{equation}
with $(a)_n=a(a+1)\cdots (a+n-1)$ and $(a)_0=1$.  Note that $\psi^{(-)}_{\vec{p}}(\vec{r}\,)= \left[\psi^{(+)}_{-\vec{p}}(\vec{r}\,)\right]^*$ as a consequence of the transformation of the M\"oller operators under time-reversal transformations~\cite{Taylor:2006}.  

The Sommerfeld factor $C_\eta$ is given by:
\begin{equation}\label{eq:SommerfeldFactor}
C_\eta^2 = \lvert \psi^{(\pm)}_{\vec{p}}(\vec{0}) \rvert^2 = e^{-\pi\,\eta}| \Gamma(1\pm i \eta)|^2=\frac{2\pi \eta}{e^{2\pi\eta}-1}\,,
\end{equation}
whereas the $S$-wave Coulomb phase-shift $\sigma_0(E_p)$ fixes 
\begin{subequations}\begin{align}
\psi^{(+)}_{\vec{p}}(\vec{0}) \left( \psi^{(-)}_{\vec{p}}(\vec{0}) \right)^* & = \left({\psi^{(+)}_{\vec{p}}(\vec{0})}\right)^2 = e^{-\pi \eta} \Gamma(1 + i\eta)^2 = C_\eta^2 e^{2i\sigma_0}\,,\\
e^{2i\sigma_\ell} & = \frac{\Gamma(1+\ell+i\eta)}{\Gamma(1+\ell-i\eta)}\,.
\end{align}\end{subequations}

The on-shell Coulomb amplitude reads~\cite{Pascual:2012}:
\begin{align}
T^{\rm on}_C(\vpp,\vp) & = -\frac{2\pi}{\mu}\left(-\eta\frac{\Gamma(1+i \eta)}{\Gamma(1-i \eta)}\frac{e^{-2i\eta\ln[\sin\Phi/2]}}{2p\sin^2\Phi/2}\right)
\end{align}
with $\vp\cdot\vpp=p^2\cos\Phi$. Both the full and the half off-shell Coulomb $T$-matrices are discussed in \cite{Mukhamedzhanov:2012qv,Okubo:1960zz}. In these references, the needed renormalization of the amplitude, due to the long range character of the Coulomb potential which makes that the charged particles are not free even when the distance between them increases to infinity, are also considered.

The partial wave decomposition of the Coulomb wave functions is given by \cite{Pascual:2012}: 
\begin{subequations}\begin{align}
\psi^{(\pm)}_{\vec{p}}(\vec{r})\, &= \sum_{\ell=0}^{\infty}(2\ell+1)i^\ell\frac{u^{(\pm)}_{\ell,p}(r)}{pr}P_\ell(\cos\theta) \,, \label{eq:partialCW} \\
u^{(+)}_{\ell,p}(r)&=\frac{2^\ell e^{-\pi \eta/2} \Gamma(\ell+1 + i\eta)}{(2\ell+1)!}e^{ipr}(pr)^{\ell+1}M(\ell+1+ i\eta,2\ell+2;- 2ipr)\label{eq:ucoul}\\
&\sim e^{i\sigma_\ell}\sin[pr-\ell\pi/2+\sigma_\ell-\eta\ln(2pr)] + {\cal O}\left(\frac{1}{pr}\right)\,,
\end{align}\end{subequations}
with $\vec{p}\cdot\vec{r}=pr\cos\theta$ and normalization given by
\begin{equation}
  \int_0^{+\infty} dr [u^{(+)}_{\ell,p'}(r)]^*u^{(+)}_{\ell,p}(r)= \frac{\pi}{2}\delta(p-p') \,.
\end{equation}
On the other hand, the relation  $\psi^{(-)}_{\vec{p}}(\vec{r}\,)= \left[\psi^{(+)}_{-\vec{p}}(\vec{r}\,)\right]^*$  leads to  
$[u^{(-)}_{\ell,p}(r)]^*=u^{(+)}_{\ell,p}(r)$.
Coulomb partial wave-amplitudes and phase-shifts read\footnote{The $-1$ contribution to all partial waves recovers the divergence of the Coulomb amplitude in the forward direction, since  $ \sum_{\ell=0}^{\infty} (2\ell +1) P_\ell(\cos\Phi) \propto \delta(1-\cos\Phi)$ \cite{Landau:1977}.}
\begin{subequations}\begin{align}
    T^{\rm on}_C(\vpp,\vp)  & = \sum_{\ell=0}^{\infty} (2\ell +1) T_C^{\ell}(E_p) P_\ell(\cos\Phi)\,, \\
    T_C^{\ell}(E_p) & = -\frac{2\pi}{\mu}\left[\frac{e^{2i\sigma_\ell} - 1}{2ip}\right]\,.  
\end{align}\end{subequations}
Finally, using $\oTSC|\sCp\rangle=\oVS|\sFp\rangle$ and Eq.~\eqref{eq:phiCFrompsiC}, one obtains the following important formal relation:
\begin{equation}\label{eq:TSC-KR-general}
T_{SC}(\vpp,\vp) = \sum_{n=0}^{\infty}\mel*{\sCm}{\oVS\left( \GCp \oVS \right)^n} {\sCp}\,,
\end{equation}
essential to connect Coulomb and strong interactions.

\subsection[Contact potential in the presence of repulsive Coulomb potential ($\eta=\alpha\mu/p$)]{\boldmath Contact potential in the presence of repulsive Coulomb potential ($\eta=\alpha\mu/p$)} \label{subsec:repulsive}

We will consider, in the presence of the Coulomb potential, a strong contact interaction:
\begin{subequations}\label{eq:contact-potential}
\begin{equation}\label{eq:contact-potential-coordinate}
\mel*{\vec{r}'}{\oVS}{\vec{r}\,} = C_0 \delta^3(\vec{r})\delta^3(\vec{r}')\,,
\end{equation}
which in momentum space reads:
\begin{equation}\label{eq:contact-potential-momentum}
\mel*{\vec{q}'}{\oVS}{\vec{q}\,} = C_0\,.
\end{equation}
\end{subequations}
In the case of repulsive Coulomb potential, taking $\vec{p}$ and $\vec{p}'$ on the mass shell, the series of Eq.~\eqref{eq:TSC-KR-general} can be shown to have the following structure,
\begin{align}\label{eq:TSC-KR}
    T_{SC}(E_p)=\mel*{\sCm}{\oTSC}{\sCp} = C_\eta^2\, e^{2i\sigma_0} C_0 \left( 1 + C_0 \, J_0(E_p) + \cdots \right) =  \frac{C_\eta^2(E_p)\, e^{2i\sigma_0(E_p)}}{C_0^{-1} - J_0(E_p)}\,,
\end{align}
with the Coulomb-dressed loop function given by 
\begin{equation}
    J_0(E_p)   = \left. \mel*{\vec{r}}{\GCp(E_p)}{\vec{r}}\right|_{\vec{r}=\vec{0}}=\int \frac{d^3 \vec{q}}{\left(2\pi\right)^3} \frac{\lvert \psi_{\vec{q}}^{(+)}(0) \rvert^2 }{E_p - \frac{\vec{q}^2}{2\mu} + \ie}\,,\quad \lvert \psi_{\vec{q}}^{(+)}(0) \rvert^2 = \frac{2\pi \eta_q}{e^{2\pi\eta_q}-1}\,,\label{eq:J0sctat}
\end{equation}
with $\eta_q=\alpha \mu/|\vec{q}\,|$ for a repulsive Coulomb interaction. The integral is ultraviolet divergent and must be renormalized. Performing one subtraction, one finds a finite  Coulomb-dressed loop function  $\bar{J}_0(E_p)=J_0(E_p) - J_0(0)$ given by \cite{Kong:1999sf}:
\begin{subequations}\begin{align}
   \bar{J}_0(E_p) & = -\frac{\mu^2\alpha}{\pi} H(\eta)\,,\\ \quad H(\eta) & = \psi(i\eta)+\frac{1}{2i\eta}-\ln{(i\eta)}\,,
\end{align}\end{subequations}
where $\psi(x)=\Gamma'(x)/\Gamma(x)$  is the logarithmic derivative of the $\Gamma$-function,\footnote{It is defined on the entire complex plane, except for zero and negative integers,  by the sum of the series
\begin{equation}
\psi(z) = -\gamma_E +  \sum_{n=1}^\infty \frac{z-1}{n(n+z-1)}\, ,  
\end{equation}
with $\gamma_E=0.57721 \cdots$, the Euler-Mascheroni  constant. For ${\rm Re}\,z>0$, $\psi(z) $ can be numerically computed  as
\begin{equation}
\psi(z) =  \int_0^\infty dt \left(\frac{e^{-t}}{t}-\frac{e^{-zt}}{1-e^{-t}}\right)\,\quad {\rm Re}\,z>0
\end{equation}
and analytically prolonged  to its domain using the recurrence relation $\psi(1+z)=\psi(z)+1/z$. Finally, for $x$ a positive real number, ${\rm Im}[\psi(i x)]= \frac{1}{2x} + \frac{\pi}{2} \coth(\pi x)$, and ${\rm Re}[\psi(ix)] \sim \log|x|+ \frac{1}{12x^2} + {\cal O }\left(\frac{1}{x^4}\right)$.} and the principal argument of the logarithm is taken between $(-\pi,\pi]$, in such a way that $\text{Im}[H(\eta)]= C^2_\eta/2\eta$ and hence:
\begin{equation}\label{eq:imag-barJ}
\text{Im}[\bar{J}_0(E_p) ] = -C^2_\eta\, \frac{\mu p}{2\pi}\,.
\end{equation}

\subsection[Contact potential in the presence of attractive Coulomb potential ($\eta=-\alpha\mu/p $)]{\boldmath Contact potential in the presence of attractive Coulomb potential ($\eta=-\alpha\mu/p $)} \label{subsec:attractive}
The result in Eq.~\eqref{eq:TSC-KR} has been obtained inserting complete sets of Coulomb scattering states. In the case of an attractive Coulomb interaction, the resolution of the identity operator involves also bound levels in addition to the scattering states: 
\begin{equation}
\mathbb{1} = \sum_{n\,,\ell\,, m_\ell} \ketbra*{n\ell m_\ell}{n\ell m_\ell} + \int \frac{d^3 \vec{q}}{\left(2\pi\right)^3} \ketbra*{\psi_{\vec{q}}^{(\pm)}}{\psi_{\vec{q}}^{(\pm)}}\,.
\end{equation}
Then, the Coulomb-dressed loop function can be written~\cite{Kong:1999sf} as
the sum $ J_0(E_p)=  J_0^b(E_p) + J^s_0(E_p) $, where the first term
\begin{equation}
  J_0^b(E_p) = \sum_{n} \frac{|\Psi_n(\vec{r}=\vec{0})|^2}{E_p-E_n} =  \frac{\alpha\mu^2}{\pi}\sum_{n=1}^\infty\frac{2\eta^2}{n(n^2+\eta^2)} =\frac{\alpha\mu^2}{\pi}\left[\psi(i\eta)+\psi(-i\eta)+2\gamma_E\right] \label{eq:J0bound}
\end{equation}
comes from the $S$-wave bound states of energy $E_n=-\mu\alpha^2/2n^2$ with $n=1,2,\cdots$ (or equivalently $p=p_n=i\sqrt{-2\mu E_n}$ ), with $|\Psi_n(\vec{0})|^2= (\alpha\mu)^3/(\pi n^3)$ the probability to find the particles at the origin of the bound state.  The digamma function $\psi(i\eta)$ diverges for all Coulomb energies $E_p=E_n$ since for them $i\eta=-n$  with $n=1,2,\ldots$

In addition, the Coulomb attractive scattering contribution $J^s_0(E_p)$ is still given by Eq.~\eqref{eq:J0sctat}, but using now $\eta_q=-\alpha \mu/|\vec{q}\,|$. However, here there is a subtlety and it cannot be renormalized just by subtracting $J^s_0(0)$, since the latter quantity is not only ultraviolet divergent, but it is also ill defined in the infrared. Therefore in the Coulomb attractive case, $\bar J^s_0(E_p)= J^s_0(E_p)-J^s_0(0)$ has an infrared divergence, while $J^s_0(E_p)$ for $E_p \ne 0$ does not.\footnote{In the repulsive case, the factor $(e^{2\pi\mu\alpha/|\vec{q}\,|}-1)$ in the denominator of the integrand of Eq.~\eqref{eq:J0sctat} avoids any divergence in the infrared limit ($|\vec{q}\,|\to 0$). However in the attractive case, the latter factor becomes $(e^{-2\pi\mu\alpha/|\vec{q}\,|}-1)$ which tends to $-1$ in the $|\vec{q}\,|\to 0$ limit and thus, one finds, not only an ultraviolet, but also a logarithmic divergence in  $J^s_0(0)$.} To render $J^s_0(E_p)$ finite, we should follow a different strategy, which is also discussed in Ref.~\cite{Kong:1999sf}. We note that:
\begin{equation}
   \frac{2\pi \eta_q}{e^{2\pi\eta_q}-1} = \frac{-2\pi \eta_q}{e^{-2\pi\eta_q}-1}- 2\pi \eta_q\,,
\end{equation}
and this can be inserted in Eq.~\eqref{eq:J0sctat} giving rise to two contributions that we denote as $J^{s,a1}_0(E_p)$  and $J^{s,a2}_0(E_p)$, respectively.  Since $-\eta_q=\alpha\mu/|\vec{q}\,|$,  we see that  $J^{s,a1}_0(0)$ is now free of infrared divergences  and we immediately conclude that 
\begin{equation}
 J^{s,a1}_0(E_p) =   J^{s,a1}_0(0) -\frac{\mu^2\alpha}{\pi} H(-\eta)
\end{equation}
with $J^{s,a1}_0(0)$ an ultraviolet  subtraction constant. On the other hand $J^{s,a2}_0(E_p)$ has also an ultraviolet divergence, which can not be renormalized by performing a subtraction at zero energy because that would induce an infrared divergence. Thus, we take    
\begin{equation}
 J^{s,a2}_0(E_p) =\int \frac{d^3 \vec{q}}{\left(2\pi\right)^3} \frac{2\pi\alpha\mu}{|\vec{q}\,|}\frac{1}{E_p - \frac{\vec{q}^2}{2\mu} + \ie}= K'-\frac{2\mu^2\alpha}{\pi}\ln{(-i\eta)} \,,\label{eq:jsa2}
\end{equation}
with $K'$ also an ultraviolet  (real) subtraction constant.\footnote{The constant $K'$ is determined by a subtraction at a negative energy $E_p=-\mathcal{E}_0<0$, for which $p=i\sqrt{2\mu\mathcal{E}_0}$ 
\begin{equation*}
 K'= J^{s,a2}_0(-\mathcal{E}_0)  + \frac{2\mu^2\alpha}{\pi}\ln{\left[\mu\alpha/\sqrt{2\mu\mathcal{E}_0}\right]}\,.  
\end{equation*}} Gathering $ J^{s,a1}_0(0) $ and $K'$ in a single unknown subtraction constant that we denote as $K$, we find in the attractive Coulomb case:
\begin{subequations}\begin{align}
 J_0^s(E_p) & =J^{s,a1}_0(E_p) + J^{s,a2}_0(E_p) =K-\frac{\mu^2\alpha}{\pi} H^{(a)}(\eta)\,,\\
 H^{(a)}(\eta) & = \psi(-i\eta)-\frac{1}{2i\eta}+\ln{(-i\eta)}\,,
\end{align}\end{subequations}
that satisfies ${\rm Im}[\bar J_0^s(E_p) ] = -C^2_\eta\, \mu p/2\pi$, as it should. Furthermore, summing up the bound state  and the scattering contributions $J_0^b(E_p)+J_0^s(E_p)$, we find~\cite{Kong:1999sf} 
\begin{subequations}\begin{align}
 J_0(E_p)&=\mel*{\vec{0}}{\GCp(E_p)}{\vec{0}}=J^b_0(E_p) + J^s_0(E_p) = \left( K+\frac{2\alpha\mu^2\gamma_E}{\pi}\right)+\frac{\mu^2\alpha}{\pi} \olsi{H}(\eta)\,, \\
 \olsi{H}(\eta) & = \psi(i\eta)+\frac{1}{2i\eta}-\ln{(-i\eta)}\,,
\end{align}\end{subequations}
where the $2\alpha\mu^2\gamma_E/\pi$ term could be reabsorbed in the unknown constant $K$. We also have $\text{Im}[\olsi{H}(\eta)]= C^2_\eta/2 \eta$ and hence:
\begin{equation}\label{eq:imag-J}
\text{Im}[ J_0(E_p) ] = -C^2_\eta\, \frac{\mu p}{2\pi}\,,
\end{equation}
since the ultraviolet subtraction constant $K$ is real. Obviously ${\rm Im}[ J_0(E_p) ]$ coincides with the term ${\rm Im}[J_0^s(E_p) ]$ coming from the scattering states because the piece $J_0^b(E_p) $  is real. We note that the term $\ln{(-i\eta)}$ in Eq.~\eqref{eq:jsa2} diverges in the $p\to 0$ limit, as expected, and cancels the divergence that, in this limit,  also appears in the term $(\psi(i\eta)+\psi(-i\eta))$ of $J^b_0(E_p)$ [see Eq.~\eqref{eq:J0bound}]. Hence, $J_0(E_p=0)$, after the ultraviolet renormalization, is well defined and is just the undetermined constant $K+\frac{2\alpha\mu^2\gamma_E}{\pi}$ and thus we can write for the attractive case:
\begin{eqnarray}
 J_0(E_p)&=&\mel*{\vec{0}}{\GCp(E_p)}{\vec{0}}=J_0(0)+\frac{\mu^2\alpha}{\pi} \olsi{H}(\eta)
\end{eqnarray}

Finally, we point out that the total on-shell $S$-wave amplitude,
\begin{equation}
    T_{\ell=0}^{\rm on}(E_p) = -\frac{2\pi}{\mu}\left[\frac{e^{2i\sigma_0(E_p)} - 1}{2ip}\right] + \frac{C_\eta^2(E_p)\, e^{2i\sigma_0(E_p)}}{C_0^{-1} - J^s_0(E_p)- J_0^b(E_p)  } \label{eq:shifted-poles}
\end{equation}
does not have poles at the Coulomb bound energies ($E_n=-\mu\alpha^2/2n^2$) because, although the three factors $e^{2i\sigma_0(E_p)}$, $C_\eta^2(E_p)$ and $J_0^b(E_p)$ diverge, cancellations are produced such that the total amplitude $T_{\ell=0}^{\rm on}(E_p=E_n) $ remains finite.  The actual positions of the bound states are modified by the zero-range potential $C_0$, and are determined by the solutions of the equation $[C_0^{-1} - J^s_0(E_p)- J_0^b(E_p)]=0$, for $p=i\gamma$ and $\gamma>0$. Because of the factors $1/(E_p-E_n)$ in $J_0^b(E_p)$, the new poles are expected to appear in the vicinity of the purely Coulomb energies $E_n$ for small $C_0$ short-range interaction. Actually for $E_p<0$, where $J^s_0(E_p)$ is real, the piece $J_0^b(E_p)$ covers the full range between $-\infty$ and $+\infty$ when $E_n<E_p < E_{n+1}$, and hence it is always guaranteed the existence of a bound state in each of these intervals. This means that all Coulomb bound states appear but their energies are shifted due to the strong potential. In addition, if the short-distance interaction is sufficiently attractive, one might also have a new bound state of strong origin. 

\subsection{Phase-shift, ERE, and unitarity} \label{subsec:phaseEREuni}
In this subsection we make definitions and derivations regardless of whether the Coulomb potential is attractive or repulsive. One can define a phase-shift for strong interactions in the presence of Coulomb ones. Due to Eqs.\,\eqref{eq:imag-barJ} and \eqref{eq:imag-J}, one can write:
\begin{subequations}\begin{align}
T_{SC}(E_p) & = - \frac{2\pi}{\mu p} e^{2i\sigma_0} e^{i \delta_0} \sin{\delta_0}\,,\\
e^{2i\sigma_0} \left[ T_{SC}(E_p) \right]^{-1} & = -\frac{\mu p}{2\pi} \left( \cot\delta_0 - i \right)\,,
\end{align}\end{subequations}
so that the total amplitude can be written as:
\begin{subequations}\label{eq:phase-shift-total}\begin{align}
T_{\ell=0}^{\text{on}} & = T^{\ell=0}_C + T_{SC} = - \frac{2\pi}{\mu p} \left( e^{i\sigma_0}\sin\sigma_0 + e^{2i\sigma_0}e^{i\delta_0}\sin\delta_0  \right) = - \frac{2\pi}{\mu p} e^{i(\sigma_0 + \delta_0)} \sin(\sigma_0 + \delta_0)\,,\\
\left[ T_{\ell=0}^{\text{on}} \right]^{-1} & = - \frac{\mu p}{2\pi} \left( \cot(\sigma_0 + \delta_0) - i \right)\,.
\end{align}\end{subequations}
Hence, $(\sigma_0 + \delta_0)$ is the total phase-shift. The phase-shift $\delta_0$ is due to the strong contact interaction in presence of the Coulomb potential, but it is not equal to the phase-shift that would be obtained from the strong potential alone. As a byproduct, Eq.\,\eqref{eq:phase-shift-total} shows that $T_{\ell=0}^{\text{on}}$ fulfills exact unitarity.

The Coulomb ERE follows thanks to the previous definition of an $S$-wave phase-shift from the strong amplitude. Introducing $\lambda=\pm 1$ and the function $H^\lambda(\eta)$, with $H^{\lambda=+1}(\eta)=H(\eta)$ and $H^{\lambda=-1}(\eta)=\olsi{H}(\eta)$ for the repulsive and attractive cases, respectively, the generalized (Coulomb-modified) ERE reads~\cite{Bethe:1949yr,Jackson:1950zz,vanHaeringen:1981pb,Kong:1999sf,Konig:2012prq}:
\begin{equation}
 C^2_\eta  p(\cot\delta_0 -i)+ 2\lambda\alpha\mu H^\lambda(\eta) = -\frac{1}{a_0} + \frac{1}{2}r_0 p^2 + v_2 p^4 + \cdots \label{eq:ERE}
\end{equation}
with $a_0$, $r_0$, etc. the $S$-wave Coulomb-modified scattering length, effective range, etc. for the elastic
scattering process under consideration. Note that one can write:
\begin{subequations}\label{eq:Hlambda_and_hlambda}\begin{align}
H^\lambda(\eta) &= h^\lambda(\eta) + i \frac{C^2_\eta}{2\eta}\,,\\
 h^\lambda(\eta) & = \sum_{n=1}^\infty\frac{\eta^2}{n(n^2+\eta^2)}-\ln(\lambda\eta)-\gamma_E\, =  \frac{1}{12\eta^2} +   \frac{1}{120\eta^4}+ {\cal O}\left(\frac1{\eta^6}\right)
\end{align}\end{subequations}
and thus, one finds the usual ERE expansion in presence of Coulomb interaction reads:
\begin{equation}
 C^2_\eta  p\cot\delta_0 + 2p\eta h^\lambda(\eta) = -\frac{1}{a_0} + \frac{1}{2}r_0 p^2 + v_2 p^4 + \cdots \label{eq:EREreal}
\end{equation}
It is convenient for later reference to define here the amplitude $f_{SC}$ as:
\begin{equation} \label{eq:fsc}
 f_{SC}^{-1}(p) \equiv C^2_\eta  p(\cot\delta_0 - i)= \left(-\frac{1}{a_0} + \frac{1}{2}r_0 p^2 + v_2 p^4 + \cdots\right) -2\lambda\alpha\mu H^\lambda(\eta)\,,
\end{equation}
which is proportional to $T_{SC}$:
\begin{equation}\label{eq:relation-TSC-fSC}
T_{SC}(E_p) = -\frac{2\pi}{\mu} C_\eta^2(p) e^{2i\sigma_0(p)} f_{SC}(p)\,.
\end{equation}

Using the contact potential $\mel*{\vec{q}'}{\oVS}{\vec{q}\,} = C_0$ (constant), one can always fix this parameter to reproduce the scattering length. One might consider that $C_0$ is a function of the energy $E_p=p^2/2\mu$, as discussed later in Subsec.\,\ref{subsec:JuanFormalism}, which would allow to fix more terms of the ERE:
\begin{equation}
   - \frac{2\pi}{\mu}\left(C_0^{-1}(p^2)-J_0(0)\right) = -\frac{1}{a_0} + \frac{1}{2}r_0 p^2 + v_2 p^4 + \cdots \label{eq:codep}
\end{equation}
with $J_0(0)$ an undetermined subtraction constant, as mentioned above.\footnote{One could adopt a different point of view and use an ultraviolet cutoff to calculate $J_0(0)$. This might allow to discount the Coulomb effects affecting the ERE parameters in Eq.~\eqref{eq:codep}  and find, in this way, the  ERE associated uniquely to the strong interaction. To illustrate this, let us consider the repulsive Coulomb case and a sharp cutoff $\Lambda$, 
\begin{equation*}
J_0(0) = -\frac{2\mu}{\pi}   \int_0^\Lambda  \!\! dq\, \frac{ \eta_q}{e^{2\pi\eta_q}-1}\,,
\end{equation*}
Once $\Lambda$ is fixed from experiment, which obviously include Coulomb, one can use this cutoff to compute $J^{SC}_0(0)= -\mu\Lambda/ \pi^2$, the zero-energy loop function in absence of Coulomb effects ($\eta_q=0$ in the equation above). Finally, the ERE genuinely associated  to the strong interaction might be approximated by 
\begin{equation*}
- \frac{2\pi}{\mu}\left(C_0^{-1}(p^2)-J^{SC}_0(0)\right) = p \cot\delta^{SC}=-\frac{1}{a^{SC}_0} + \frac{1}{2}r^{SC}_0 p^2 + \cdots
\end{equation*}
The above ERE parameters are not observables, and depend on the renormalization treatment. However, this procedure allows to pass from proton-proton scattering lengths of the order of $a_0=-7.8$ fm \cite{Jackson:1950zz} to $a_{SC}\sim 23$ fm, in better agreement to that measured for the isotriplet neutron-proton $S$-wave.}  

A final remark here is in order. The amplitude $T_{SC}(E_p)$ is automatically projected into $S$-wave because we have only considered a contact interaction. In this case, the contribution to the full amplitude from higher orbital angular momentum $\ell>0$ is uniquely given by the Coulomb partial wave which, in the attractive case, has poles at $E_p=E_n, n=\ell+1,\cdots$, where $e^{2i\sigma_\ell(E_p)}$ diverges. We note, however, that  Eqs.~\eqref{eq:T_TC_TSC_decomposition} and \eqref{eq:TSC-KR-general} could include, if necessary, all possible waves, and that $T_C(E;\vp,\vpp)$ is the full Coulomb amplitude, including all waves. Moreover, in addition to the right-hand unitarity cut, the left-hand cut of the Coulomb $T$-matrix is inherited by the complete one.

\subsection{Relation with Distorted Wave Theory}\label{subsec:JuanFormalism}

In Ref.~\cite{Nieves:2003uu}, standard distorted wave theory techniques and dimensional regularization are used to find out solutions of the spin-singlet $S$-wave nucleon–nucleon Lippmann–Schwinger equation with a potential that includes a long-range interaction (finite one-pion exchange) and additional contact terms with derivatives. The main result of Ref.~\cite{Nieves:2003uu} is given in its Eq.~(14), which has obvious resemblances with the results presented in the previous subsection within the scheme of Kong and Ravndal \cite{Kong:1999sf}. Actually, it is also shown in Ref.~\cite{Nieves:2003uu} that the on-shell full amplitude ($T_{\ell=0}^{\rm on}$) is the sum of that ($T_{\ell=0}^{\pi;\, \rm on}$)  obtained when only the long-range potential is considered  plus another one ($T_{SC}$) driven by the short-range interactions, in the presence of one-pion exchange.

Coulomb and pion exchange interactions can be compared by introducing a finite mass for the photon. This allows us to directly use Eq. (14) from Ref.~\cite{Nieves:2003uu}, where, as we will see, the photon massless limit can be safely taken. In this way, we obtain\footnote{We have adapted the normalization of Nieves \cite{Nieves:2003uu} to that of Kong and Ravndal \cite{Kong:1999sf} to make the relation between formulae more transparent. The $T$-matrix obtained in Ref.\,\cite{Nieves:2003uu} should be multiplied by the factor $\pi^2/\mu$ to be consistent with that  used in Ref.\,\cite{Kong:1999sf}.} for the short range amplitude $T_{SC}$,
\begin{equation}\label{eq:TSC-J}
    T_{SC}(E_p)  = \frac{\left( 1 + \LJ(E_p) \right)^2}{C_0^{-1}+ i \frac{\mu p}{2\pi} - \mathcal{J}_0(E_p)}\,,
\end{equation}
where the $S$-wave half off-shell and full off-shell Coulomb $T_C^{\ell=0}(E_p)$ amplitudes appear in the definition of the $\LJ(E_p)$ and $\mathcal{J}_0(E_p)$ integrals
\begin{align}
    \LJ(E_p) & = \int \frac{d^3q}{(2\pi)^3}  \frac{T_C^{\ell=0}(E_p;p,q)}{E_p - \frac{q^2}{2\mu} + \ie}\,,\\
    \mathcal{J}_0(E_p) & = \int  \frac{d^3q}{(2\pi)^3}\frac{d^3q'}{(2\pi)^3}\,  \frac{T_C^{\ell=0}(E_p;q,q')}{\left(E_p - \frac{q^2}{2\mu} + \ie\right)\left(E_p - \frac{{q'}^2}{2\mu} + \ie\right)}\,.
\end{align}
One can see the similitude between Eqs.~\eqref{eq:TSC-KR} and \eqref{eq:TSC-J}, although the equivalence between the different factors needs to be proven. We start with the term $(1 + \LJ(E_p))$. From the second relation of Eq.~\eqref{eq:psiCFromFree}
\begin{align}
   \sCp(\vec{r}) = e^{i \vec{p} \cdot \vec{r} } + \mel*{\vec{r}}{\GFp(E_p) \oTC(E_p+\ie)}{\vec{p}} =  e^{i \vec{p} \cdot \vec{r} }+ \int \frac{d^3 \vec{q}}{\left( 2\pi \right)^3}\,e^{i \vec{q}\cdot\vec{r}}\, \frac{T_{C}(E_p;\vec{k},\vec{q})}{E - \frac{\vec{q}^2}{2\mu} + \ie }\,.
\end{align}
Then, one gets
\begin{align}
    \sCp(\vec{0}) = 1 + \int \frac{d^3 q}{(2\pi)^3} \, \frac{T^{\ell=0}_C(E_p;k,q)}{E_p - \frac{q^2}{2\mu} + \ie} = 1 + \LJ(E_p)\,,
\end{align}
since for $\vec{r}=\vec{0}$, the integration over the angles of $\vec{q}$ selects the $S$-wave contribution of the amplitude. Thus, we finally find
\begin{equation}\label{eq:RelationL0}
    \left( 1 + \LJ(E_p) \right)^2 = \left( \sCp(\vec{0}) \right)^2 = C_\eta^2(E_p) \, e^{2i\sigma_0(E_p)}\,,
\end{equation}
which is precisely the factor in Eq.~\eqref{eq:TSC-KR}. The relation between $J_0(E)$ and $\mathcal{J}_0(E)$ is more elaborate, but essentially it stems from that between the free and Coulomb propagators, Eq.~\eqref{eq:rel-GC-GSC}. We have
\begin{eqnarray}
-i \frac{\mu p}{2\pi} +\mathcal{J}_0(E) &=&-i \frac{\mu p}{2\pi}+ \int  \frac{d^3q}{(2\pi)^3}\frac{d^3q'}{(2\pi)^3}\,  \frac{T^{\ell=0}_C(E_p;q,q')}{\left(E_p - \frac{q^2}{2\mu} + \ie\right)\left(E_p - \frac{{q'}^2}{2\mu} + \ie\right)} \nonumber \\
&=& \int \frac{d^3\vec{q}}{(2\pi)^3} \frac{1}{E_p - \frac{\vec{q}^2}{2\mu} + \ie}  + \int  \frac{d^3q}{(2\pi)^3}\frac{d^3q'}{(2\pi)^3}\,  \frac{T_C(E_p;\vec{q},\vec{q}')}{\left(E_p - \frac{q^2}{2\mu} + \ie\right)\left(E_p - \frac{{q'}^2}{2\mu} + \ie\right)} \nonumber \\
 & = &    \left. \mel*{\vec{r}}{ \GFp(E_p)+\GFp(E_p)\, \oTC(E_p)\, \GFp(E_p)}{\vec{r}}\right|_{\vec{r}=\vec{0}} \nonumber \\
 &= & \mel*{\vec{0}}{ \GFp(E_p)+\GFp(E_p)\,\oVC\, \GCpm (E_p)}{\vec{0}} = \mel*{\vec{0}}{\GCp(E_p)}{\vec{0}}\nonumber  \\  
  &=&     J_0(E_p) =  J_0^b(E_p) + J^s_0(E_p) \,,
  \label{eq:J0}
\end{eqnarray}
where we have used dimensional regularization, since this is the method employed  in Ref.~\cite{Nieves:2003uu}, to compute the matrix element of the free propagator $\GFp(E_p)$ and $J_0(E_p)$ is the Coulomb-dressed loop function  introduced in the formalism of Ref.~\cite{Kong:1999sf} (Subsec.~\ref{subsec:repulsive}).

The above result for $\mathcal{J}_0(E_p)$, together with the one in Eq.~\eqref{eq:RelationL0} for $\LJ(E_p)$, show the equality between Eq.~\eqref{eq:TSC-J} and Eq.~\eqref{eq:TSC-KR} or the equivalent equation for the Coulomb attractive case. In the latter,  the Coulomb bound states appear both in $J_0(E_p)$ and $\mathcal{J}_0(E_p)$.  

Note that the Eqs.~\eqref{eq:RelationL0} and \eqref{eq:J0} allow to define the  $\LJ(E_p)$ and $\mathcal{J}_0(E_p)$ integrals, which involve the off-shell $T_C^{\ell=0}(E_p)$ amplitudes, in terms of the Coulomb wave function $ \psi_{\vec{q}}^{(+)}$ at the origin,  providing a well defined limit when the mass of the photon is set to zero.  

Furthermore, even considering  $C_0$ a function of $p^2$ as in Eq.~\eqref{eq:codep}, the equivalence between the formalisms derived in Refs.~\cite{Kong:1999sf} and \cite{Nieves:2003uu} still holds. This is because  the short-distance potential $V_s$ introduced in Eq.~(14) of the  latter reference is not renormalized by the Coulomb interaction since the photon is massless and hence $\hat k ^2 = k^2$ in the mentioned equation.    

\section{Correlation functions} \label{sec:CFs}

\subsection{Incorporating Coulomb interactions to femtoscopy CFs for strong contact potentials}\label{subsec:CoulombintoCFs}

The CF is computed from the emitting source $\widetilde{S}(\vec{r})$ of two distinguishable particles and the full outgoing wave function $\sFm(\vec{r})$ \cite{Albaladejo:2024lam},
\begin{equation}
C(\vp) = \int d^3 \vec{r}\, \widetilde{S}(\vec{r}) \left\lvert \sFm(\vec{r})\right\rvert^2\,.\label{eq:correla}
\end{equation}
Analogously to the computation of $T_{SC}(E_p)$, after summing up the  formal  series that appear in Eq.~\eqref{eq:phiCFrompsiC} we find:
\begin{equation}\label{eq:full-wf-Coulomb-strong}
    [\sFm(\vec{r})]^*= \sFpmenosp(\vec{r}) = \sCpmenosp(\vec{r}) + \frac{T_{SC}(E_p)\ G_C^{(+)}(E_p;r) }{e^{-\pi \eta/2} \Gamma(1 + i\eta)}\,.
\end{equation}
It can be seen that the complete and Coulomb wave functions\footnote{Recall that $\sCpmenosp(\vec{r})=[\sCmbis(\vec{r})]^*$.} differ only in an $S$-wave contribution, driven by the short-distance physics and the spatial matrix element of the Coulomb $\GCp(E_p)$ propagator
\begin{eqnarray}
 G_C^{(+)}(E_p;r) &=& \mel{\vec{r}}{\GCp(E_p)}{\vec{0}}= \frac{\Theta\left(-{\rm sign}[e_1e_2]\right)}{4\pi}\sum_{n=1}^\infty \frac{R_n(r)R_n(0)}{E_p-E_n} \nonumber \\
 && + \int \frac{d^3 q}{(2\pi)^3}\frac{e^{-\pi \eta_q/2} \Gamma(1 + i\eta_q)}{E_p - \frac{q^2}{2\mu} + \ie}\  \frac{[u^{(+)}_{\ell=0\,,q}(r)]^*}{qr} \label{eq:gmascr}
\end{eqnarray}
with the $S$-wave bound radial wave functions given by~\cite{Pascual:2012} 
\begin{equation}
    R_n(r)= 2\left(\frac{\mu\alpha}{n}\right)^\frac{3}{2}\ e^{-\mu\alpha r/n}\ \frac{1}{n}\, L^1_{n-1}(2\mu\alpha r/n)\,,\quad n=1,\ldots
\end{equation}
where $L^k_n(x)$ are the generalized Laguerre polynomials. $\Theta(\cdots)$ is the Heaviside function, such that the first contribution is only present for attractive Coulomb interaction. The $d^3q$ integration is finite for $r\ne 0$ and $G_C^{(+)}(E_p;r)$ can be expressed in terms of the Whittaker function $W_{k,m}(z)$~\cite{Hostler:1963zz}
\begin{equation}
G_C^{(+)}(E_p;r) =- \frac{\mu}{2\pi r} \Gamma(1+i\eta) W_{-i\eta, 1/2}(-2ipr) = \frac{ip\mu}{\pi} e^{ipr} \ \Gamma(1+i\eta) U(1+i\eta,2;-2ipr)\,, \quad r>0\,, \label{eq:whittaker}
\end{equation}
where the confluent hypergeometric function $U(a,b;z)$ can be calculated as:\footnote{Although this expression is undefined for integer $b$, it can be extended to any integer $b$ by continuity. For ${\rm Re}z>0$,  the function $U$ can be computed also from the Laplace integral and analytically continued to the complex plane,
\begin{equation*}
U(a,b;z) = \frac{1}{\Gamma(a)} \int_0^{+\infty} e^{-zt}t^{a-1}(1+t)^{b-a-1}dt\, , \quad \text{Re}\,a>0\,.
\end{equation*}
}
\begin{equation}
  U(a,b;z) = \frac{\Gamma(1-b)}{\Gamma(a+1-b)} M(a,b;z) +  \frac{\Gamma(b-1)}{\Gamma(a)}z^{1-b} M(a+1-b,2-b;z)\,.
\end{equation}
$G_C^{(+)}(E_p;r)$ diverges at $r=0$, and exhibits the usual $\sim 1/r$ Coulomb behavior. However, it would not be seen in the computation of the correlation function in Eq.~\eqref{eq:correla} because of the $r^2$ piece of the $d^3r$ measure.

Let us come back to the evaluation of the correlation function in Eq.~\eqref{eq:correla}. In what follows and to simplify the notation, we will introduce $\phi(\vec{r},\vp)=\sFm(\vec{r})$ and $\psi^*(\vec{r},\vp)=\sCpmenosp(\vec{r})$. In analogy to Eq.~\eqref{eq:partialCW},  we consider the partial-wave decomposition of $\phi(\vec{r},\vp)$ as
\begin{equation}
\phi(\vec{r},\vp) = \sum_{\ell=0}^{\infty} i^\ell (2\ell+1) P_\ell(\cos\theta) \phi_\ell(r,p)\,,
\end{equation}
which, inserted into $C(\vp)$ and assuming a spherically symmetric source, gives a function of the on-shell energy $E_p$
\begin{equation}
C(E_p) = \sum_{\ell=0} (2\ell+1) \int_{0}^{\infty}\!\!\!\!\! dr\, S(r) \left\lvert \phi_\ell(r,p) \right\rvert^2 = 1 + \sum_{\ell=0} (2\ell+1) \int_{0}^{\infty}\!\!\!\!\! dr\, S(r) \left( \left\lvert \phi_\ell(r,p) \right\rvert^2 - j_\ell(pr)^2 \right) \,,
\end{equation}
where $j_\ell(z)$ are the spherical Bessel functions and  we have defined:
\begin{equation}
S(r) = 4\pi r^2 \widetilde{S}(\vec{r})\,.
\end{equation}
We now add and subtract each of the partial wave components of the purely Coulomb wave functions, $\psi_\ell(r,p) =  [u^{(+)}_{\ell\,,p}(r)]^*/pr$ [see Eq.~\eqref{eq:ucoul}], and obtain:
\begin{equation}
C(E_p) = C_C(E_p) + \sum_{\ell=0} (2\ell+1)\int_0^{\infty} dr\,S(r) \left( \left\lvert \phi_\ell(r,p) \right\rvert^2 - \left\lvert \psi_\ell(r,p) \right\rvert^2 \right)\,,
\end{equation}
with $C_C(E_p)$ the purely Coulomb correlation function, given by: 
\begin{align}
    C_C(E_p) &= \int d^3 \vec{r}\, \widetilde{S}(r) \left\lvert \psi(\vec{r},\vp) \right\rvert^2 
    = \int d^3 \vec{r}\, \widetilde{S}(r) \left\lvert \sCpmenosp(\vec{r}) \right\rvert^2  = \int d^3 \vec{r}\, \widetilde{S}(r) \left\lvert \sCp(\vec{r}) \right\rvert^2 
    \nonumber \\
    &= 1 + \sum_{\ell=0} (2\ell+1) \int_{0}^{\infty}dr\, S(r) \left( \left\lvert \psi_\ell(r,p) \right\rvert^2 - j_\ell(pr)^2 \right) \,. \label{eq:CF-pure-Coulomb}
\end{align}
We note that the purely Coulomb CF $C_C(E_p)$ is completely dominated at low momenta by the Sommerfeld factor $C_\eta^2$ [\textit{cf.} Eq.\,\eqref{eq:SommerfeldFactor}], and hence $C_C(E_p)$ diverges or is zero for an attractive or repulsive Coulomb interaction, respectively.

When Coulomb interaction is present, all partial waves contribute. If, for the strong interactions, only $S$-wave is considered, as in this work, one finds:
\begin{equation}
C(E_p) = C_C(E_p) + \int_0^{\infty} dr\,S(r) \left( \left\lvert \phi_0(r,p) \right\rvert^2 - \left\lvert \psi_0(r,p) \right\rvert^2 \right)\,,\label{eq:correla-coul}
\end{equation}
where $ \psi^*_0(r,p) =  u^{(+)}_{\ell=0\,,p}(r)/pr$, as introduced above, and: 
\begin{eqnarray}
   \phi^*_0(r,p) =  \frac{u^{(+)}_{\ell=0\,,p}(r)}{pr}+ \frac{T_{SC}(E_p)\ G_C^{(+)}(E_p;r)}{e^{-\pi \eta/2} \Gamma(1 + i\eta)}\,.
\label{eq:phi0}
\end{eqnarray}
We note that the decomposition in Eq.\,\eqref{eq:correla-coul} is also given in Ref.\,\cite{Torres-Rincon:2023qll} (see Eq.\,(4) of that reference).

We make a remark here, closely connected with the discussion after Eq.\,\eqref{eq:shifted-poles}. The partial wave function $\phi^*_0(r,p)$ has poles of electromagnetic origin. In the purely Coulomb wave-function $u^{(+)}_{\ell=0\,,p}(r)$, they are located at the usual values of the Coulomb bound states. In the complete $S$-wave function, these poles are shifted\footnote{In the limit $E_p\to E_n=-\mu\alpha^2/2n^2$  with $n=1,2,\cdots$, or equivalently $p\to p_n=i\sqrt{-2\mu E_n}=i\mu\alpha/n$ that also leads to $\eta \to \eta_n=i\,n$, one finds $u^{(+)}_{\ell=0\,,p}(r)/pr= e^{-\pi \eta_n/2}\Gamma(1+i\eta_n) R_n(r)/R_n(0)$, where the singularities at the Coulomb energies become explicit in the $\Gamma(1+i\eta_n)$ factor,
\begin{equation}
 \lim_{p\to p_n} \Gamma(1+i\eta)\left(\frac{p^2}{2\mu}-E_n\right) = \frac{(-1)^n}{(n-1)!}\frac{\mu\alpha^2}{n^3}\, , \quad   n=1,2,\cdots 
\end{equation}
In turn, the short-distance physics correction $T_{SC}(E_p)\ G_C^{(+)}(E_p;r)e^{\pi \eta/2}/ \Gamma(1 + i\eta)$ also diverges for the Coulomb energies and cancels the pole of  $u^{(+)}_{\ell=0\,,p}(r)/pr$. As a result, the complete $S$-wave function $\phi^*_0(r,p)$ is finite in the $E_p\to E_n=-\mu\alpha^2/2n^2$ limit.} because of the strong short-range interaction $T_{SC}$, which non-purely Coulomb singularities are inherited by $\phi^*_0(r,p)$. In fact, $T_{SC}$ diverges both for the Coulomb energies ($E_n=-\mu\alpha^2/2n^2$) and for the zeros of $[C_0^{-1} - J^s_0(E_p)- J_0^b(E_p)]=0$. The former are not poles of the complete amplitude $T_{\ell=0}^{\rm on}(E_p)$, as discussed below Eq.~\eqref{eq:shifted-poles}, while the latter ones become the physical bound states of the system and they are also singularities of the complete $\phi^*_0(r,p)$ wave function.

\subsection{Effective range and improvement of the CF formula for contact potentials}
\label{subsec:deltaC}
The $S$-wave $\phi^*_0(r,p)$ given in Eq.~\eqref{eq:phi0} is not correct below the actual range of the strong interaction, because we have approximated the latter by a contact potential [\textit{cf}. Eqs.\,\eqref{eq:contact-potential}], neglecting the finite range of the strong interaction. As a consequence, the quantity  $\Delta C(E_p) =C(E_p) - C_C(E_p)$, that contains the information on the non-electromagnetic hadron-hadron interaction, is being computed using the asymptotic wave function of Eq.~\eqref{eq:phi0} for all relative distances $r$, which is however only correct for values of $r$ where the strong potential is already negligible. For definiteness, we will refer to this as the Lednicky-Lyuboshits (LL) approximation~\cite{Lednicky:1981su} (see also Refs.~\cite{ExHIC:2017smd, Albaladejo:2024lam} for recent discussions), and introduce the subindex LL to refer to it, and call:
\begin{equation}\label{eq:LLterm}
  \Delta C_{\rm LL}(E_p) =  \int_0^{\infty} dr\,S(r) \left( \left\lvert \phi_0(r,p) \right\rvert^2 - \left\lvert \psi_0(r,p) \right\rvert^2 \right)\,,
\end{equation}
in the understanding that $\phi_0(r,p)$ of Eq.~\eqref{eq:phi0} is not correct in regions where the strong interaction is not zero.

The correlation function calculated from the asymptotic wave function can be improved  by using
\begin{equation}\label{eq:r0eff}
\frac{d_0(p)}{2} \equiv \frac{d}{dp^2}\left({\rm Re}\left[f_{SC}^{-1}(p)\right]+2\lambda\alpha\mu h^\lambda(\eta)\right)=\int_0^{\infty} dr\, r^2 \frac{\left\lvert \phi_0(r,p) \right\rvert^2-\left\lvert \phi_0^{\rm exact }(r,p) \right\rvert^2}{C^2_\eta \,|f_{SC}(p)|^2} \,,\end{equation}
where $\phi_0^{\rm exact}(r,p)$ is the exact radial $S$-wave function, solution of the Schr\"odinger equation for a local, energy-independent and finite-range strong potential, in the presence of the Coulomb interaction. We note that ${\rm Re}\left[f_{SC}^{-1}(p)\right]+2\lambda\alpha\mu h^\lambda(\eta)$ can be written as $f_{SC}^{-1}(p)+2\lambda\alpha\mu H^\lambda(\eta)$ [see Eqs.\,\eqref{eq:Hlambda_and_hlambda} and \eqref{eq:fsc}], which allows to continue analytically $d_0(p)$ in the entire complex plane. We also recall that the amplitude $f_{SC}(p)$, defined in Eq.\,\eqref{eq:fsc}, is proportional to $T_{SC}$ [\textit{cf.} Eq.\,\eqref{eq:relation-TSC-fSC}]. In the $p\to 0$ limit, $d_0(p)\to r_0$ and Eq.~\eqref{eq:r0eff} reduces to the usual effective range formula (see for instance Appendix B of Ref.~\cite{Preston:1993}\footnote{To follow the proof in that text-book, it is useful to note that an asymptotic reduced radial wave function is used, normalized so that it is unity at $r=0$. This is $e^{\pi\eta/2}f_{SC}^{-1}(p)[r\,\phi^*_0(r,p)]/\Gamma(1+i\eta)$, with $\phi_0(r,p)$ given in Eq.~\eqref{eq:phi0}.}).

Expressing the exact correlation function as:
\begin{subequations}\label{eq:CF-decomp-and-deltaC}\begin{align}
C(E_p) & = C_C(E_p) + \Delta C_\text{LL}(E_p) + \delta C(E_p)\,, \label{eq:CF-general-decomposition}\\
\delta C(E_p) & = 4\pi \int_0^{\infty} dr\,r^2 \widetilde{S}(r)\left( \left\lvert \phi^{\rm exact}_0(r,p) \right\rvert^2 - \left\lvert \phi_0(r,p) \right\rvert^2 \right)\,, \label{eq:deltaC-general}
\end{align}\end{subequations}
we can use the result of Eq.~\eqref{eq:r0eff} to approximate the last contribution, for sufficiently extended sources, by:
\begin{equation}\label{eq:delC}
\delta C(E_p) \simeq  -2\pi d_0(p) C^2_\eta \,|f_{SC}(p)|^2 \widetilde{S}(0)\,. \end{equation}
For a Gaussian source of radius $R$:
\begin{equation}\label{eq:gaussian-source}
\widetilde{S}(r) = \frac{1}{(4\pi R^2)^\frac{3}{2}} \exp \left( -\frac{r^2}{4R^2} \right) \,,    
\end{equation}
we would obtain,
\begin{subequations}\label{eq:FRC:particular-gral-todas}\begin{align}
 \delta C(E_p) &\simeq - \frac{d_0(p) C^2_\eta \,|f_{SC}(p)|^2 }{4\sqrt{\pi}R^3}\,,\label{eq:FRC:particular-gral}\\
 C(E_p)&\simeq C_C(E_p) + \Delta C_\text{LL}(E_p) - \frac{d_0(p) C^2_\eta \,|f_{SC}(p)|^2 }{4\sqrt{\pi}R^3} \,.\label{eq:FRC:particular-gral0}
 \end{align}
\end{subequations}
In the literature, it is common to further approximate $d_0(p)$ at low momenta by the effective range $r_0$ leading to 
\begin{subequations}\label{eq:FRC:particular-todas}\begin{align}
 \delta C(E_p) &\simeq - \frac{r_0 C^2_\eta \,|f_{SC}(p)|^2 }{4\sqrt{\pi}R^3}\,,\label{eq:FRC:particular}\\
 C(E_p)&\simeq C_C(E_p) + \Delta C_\text{LL}(E_p) - \frac{r_0 C^2_\eta \,|f_{SC}(p)|^2 }{4\sqrt{\pi}R^3} \,.\label{eq:FRC:particular0}
 \end{align}
\end{subequations}
However, we remark that  when $d_0(p)$ is simply approximated by $r_0$, some ${\cal O}(p^2)$ corrections are being neglected~\cite{Preston:1993}, since it is further assumed  that:
\begin{eqnarray}
   \frac{\left\lvert \phi^{\rm exact}_0(r,p) \right\rvert^2 - \left\lvert \phi_0(r,p) \right\rvert^2 }{C^2_\eta \,|f_{SC}(p)|^2} \approx \lim_{p\to 0 }  \left[\frac{ \left\lvert \phi^{\rm exact}_0(r,p) \right\rvert^2 - \left\lvert \phi_0(r,p) \right\rvert^2 }{C^2_\eta \,|f_{SC}(p)|^2}\right]\,.
\label{eq:approx-zero-energy}
\end{eqnarray}
For future reference in the discussion of our results, we shall call Eq.\,\eqref{eq:delC} and, in general, its application in Eqs.\,\eqref{eq:FRC:particular-gral} or in \eqref{eq:FRC:particular} at low momenta, the finite-range potential (\CorrName) correction.\footnote{We stress here that the \CorrName{} correction in Eq.\,\eqref{eq:delC} is already present in the original work of Lednicky \textit{et al.} (\textit{cf.} Eq. (A.16) of Ref.\,\cite{Lednicky:1981su} or Eq. (153) of Ref.\,\cite{Lednicky:2005tb}). More recently, its derivation without Coulomb interactions can also be found in Refs.\,\cite{Ohnishi:2016elb,ExHIC:2017smd}.}

As discussed in the Introduction, Coulomb interactions were considered in CFs already at the early stages of femtoscopy \cite{Koonin:1977fh,Lednicky:1981su}. In Ref.\,\cite{Lednicky:1981su} the Coulomb-plus-strong wave-function of Eq.~\eqref{eq:full-wf-Coulomb-strong} is also presented, and some approximations are done in the latter in order to express the CF in simpler terms. For instance, the purely Coulomb contribution $C_C(p)$ is essentially approximated by $C_\eta^2$, which is the dominant factor as we have said in the discussion immediately after Eq.\,\eqref{eq:CF-pure-Coulomb}, although it does not completely determine $C_C(p)$. In a later work \cite{Lednicky:2005tb}, the CF is written in terms of the  wave-function of Eq.~\eqref{eq:full-wf-Coulomb-strong}, and similar approximations are performed to pick up the leading terms. Certainly, nothing prevents to use the exact form of the asymptotic full wave-function in the calculation of the CF, as done \textit{e.g.} in Ref.\,\cite{ALICE:2022wwr}. If this is done, our expressions in Eq.~\eqref{eq:FRC:particular-gral-todas} would reproduce the results of the aforementioned references. Our derivation, more detailed and pedagogical, has nonetheless several features that make it interesting. By explicitly performing the separation $T = T_C + T_{SC}$ or, equivalently, that of the complete wave function in Eq.\,\eqref{eq:full-wf-Coulomb-strong} into the purely Coulomb term and that due to the strong interactions, we directly obtain the equivalent separation in the CF in Eq.\,\eqref{eq:CF-general-decomposition}. We have obtained the complete wave function and, therefore, its $S$-wave contribution [\textit{cf}. Eqs.\,\eqref{eq:full-wf-Coulomb-strong} and \eqref{eq:phi0}], in terms of the same summation that appears when obtaining the strong scattering amplitude in the presence of Coulomb interaction, $T_{SC}(E_p)$ in Eq.\,\eqref{eq:TSC-KR}. Moreover, this derivation starts from the re-summation of a contact interaction [\textit{cf}. Eqs.\,\eqref{eq:contact-potential}], which makes it easily generalizable to any EFT, particularly considering the connection done in Subsec.\,\ref{subsec:JuanFormalism}, which allows to take into account momentum-dependent potentials.

In the derivation of Eq.\,\eqref{eq:delC}, the source $\widetilde{S}(r)$ is approximated by its value at origin, $\widetilde{S}(0)$. Since the integration in $ \delta C(E_p)$ runs from zero to the range of the strong potential, this approximation is justified only when the source does not appreciably vary within the range of the interaction. In case of dealing with source-sizes comparable to the range of the strong interaction, the variation of the source-intensity  in the integration of Eq.~\eqref{eq:deltaC-general} should be taken into account. One might supplement/improve the \CorrName{} correction given in Eq.\,\eqref{eq:delC} by adding a phenomenological parameter $\beta$ and approximate
\begin{eqnarray}
\delta C(E_p) &=& 4\pi \int_0^{\infty} dr\,r^2 \widetilde{S}(r)\left( \left\lvert \phi^{\rm exact}_0(r,p) \right\rvert^2 - \left\lvert \phi_0(r,p) \right\rvert^2 \right) \nonumber\\
&\simeq & 4\pi \beta\widetilde{S}(0) \int_0^{\infty} dr\,r^2 \left( \left\lvert \phi^{\rm exact}_0(r,p) \right\rvert^2 - \left\lvert \phi_0(r,p) \right\rvert^2 \right)=-2\pi \beta  \widetilde{S}(0)d_0(p) C^2_\eta \,|f_{SC}(p)|^2\,.
\label{eq:delC-improved}
\end{eqnarray}
This effective parameter $\beta$ should be fitted to some CF data for a particular value or a limited range of momenta, and it should approach one as the source-size increases. Actually, if $\beta$ is determined from the exact CF at some cm momentum $p$, for which the function $\left\lvert \phi^{\rm exact}_0(r,p) \right\rvert^2 - \left\lvert \phi_0(r,p) \right\rvert^2$ has a well defined sign in the entire $r$-integration interval, it will provide an average, in units of $\widetilde{S}(0)$, over the values that the source takes in the integration. Therefore, in this dynamical situation, one would find $0\leqslant \beta\leqslant 1$. However in the most general case, where  $ \left\lvert \phi^{\rm exact}_0(r,p) \right\rvert^2 - \left\lvert \phi_0(r,p) \right\rvert^2$ changes sign in the integration, the phenomenological parameter $\beta$ might be larger than one and even become negative, depending on the details of the source and of the exact and asymptotic radial wave-functions. In the rest of the main text of this paper, we will always set $\beta$ to 1 except for one comparison with a realistic theoretical CF calculation performed for a source-radius significantly smaller than the range of the strong interaction.  In the Appendix \ref{app:beta}, we will show further examples of the improvement achieved by the inclusion of this $\beta$ parameter in the theoretical calculation of the CF within the \CorrName{} approximation.
 
We stress that for a sufficiently extended source the correction $\delta C(E_p)$ in Eq.~\eqref{eq:FRC:particular-gral}, or  \eqref{eq:FRC:particular} at low momenta, should provide the correct correlation function  and that the incorporated corrections could become quite large in the case of large (in modulus) scattering amplitudes.

In Appendix \ref{app:esfera-dura}, we present in detail an alternative method to compute the  $\delta C(E_p)$ term, which has also been used in the past in the context of femtoscopy, see \textit{e.g.} Ref.\,\cite{Gmitro:1986ay}. It consists in solving the Schr\"odinger equation for a (strong) spherical square-well potential plus the Coulomb one. The depth and width of the square-well are determined such that the scattering length and effective range match with those of the scattering amplitude. Then, the exact wave function is used in the inner region to compute the correction $\delta C(E_p)$ in Eq.\,\eqref{eq:deltaC-general}. In this way, effective range corrections are incorporated, similarly as done in the \CorrName{} approximation. This method has two advantages with respect to Eq.~\eqref{eq:FRC:particular-todas}. First, one does not need to resort to the $\widetilde{S}(r) \simeq \widetilde{S}(0)$ and to the approximation of Eq.\,\eqref{eq:approx-zero-energy}, although, admittedly, at the expense of using a simple model. Second, the correlation function remains finite in the point-like source limit ($R \to 0$), while the LL approximation diverges in this limit, as noted in Ref.\,\cite{Albaladejo:2024lam}, and so does the \CorrName{} correction in Eqs.~\eqref{eq:FRC:particular-gral} and \eqref{eq:FRC:particular}. 

\subsection{Identical particles}
\label{subsec:identical}
The computation of the correlation function for a pair of identical particles needs to be modified to be consistent with quantum statistics~\cite{Koonin:1977fh, ExHIC:2017smd}. We illustrate the change for the case of $pp$. Since the strong interaction is of short range, for simplicity we will only consider $S$-wave strong interaction. Assuming that the protons emitted by the source are unpolarized, the $p$-$p$ correlation function should be averaged over the spin states as:
\begin{equation}
C(E_p) = \frac14 C_0(E_p) + \frac34 C_1(E_p)\,,
\end{equation}
where $C_0(E_p)$ and $C_1(E_p)$ are the correlation functions corresponding to $S= 0,\,1$, respectively. The weight factors $1/4$ and $3/4$ reflect the numbers of spin states in the two channels. These CFs should be computed using the singlet $^1\sFm(\vec{r})$ and triplet $^3\sFm(\vec{r})$ $p$-$p$ scattering wave-functions for protons of relative  momentum $|\vec p\,|=\sqrt{2\mu E_p}$. They are respectively symmetric and anti-symmetric under the interchange of $\vec r \to - \vec r$ and satisfy the two-body $p$-$p$ Schr\"odinger equation containing strong and Coulomb potentials. Hence, $^1\sFm(\vec{r}) = \left[\sFm(\vec{r})+ \sFm(-\vec{r})\right]/\sqrt{2}$, while $^3\sFm(\vec{r}) = \left[\sFm(\vec{r})- \sFm(-\vec{r})\right]/\sqrt{2}$, and the former (latter) contains only even (odd) orbital angular momentum waves. All together, we find that Eq.~\eqref{eq:correla-coul} should be replaced by:
\begin{subequations}\begin{align}
C(E_p)&= \frac14  C_{[S]C}(E_p) + \frac34  C_{[A]C}(E_p)+ \frac12 \int_0^{\infty} dr\,S(r) \left( \left\lvert \phi_0(r,p) \right\rvert^2 - \left\lvert \psi_0(r,p) \right\rvert^2 \right)\,,\label{eq:cfpp}\\
 C_{[S]C}(E_p) &= \int d^3 \vec{r}\, \widetilde{S}(r) \left\lvert \frac{[\psi(\vec{r},\vp)+\psi(-\vec{r},\vp)]}{\sqrt{2}} \right\rvert^2 \,,   \\
  C_{[A]C}(E_p) &= \int d^3 \vec{r}\, \widetilde{S}(r) \left\lvert \frac{[\psi(\vec{r},\vp)-\psi(-\vec{r},\vp)]}{\sqrt{2}} \right\rvert^2 \,,
\end{align}\end{subequations}
with $\psi^*(\vec{r},\vp)=\sCpmenosp(\vec{r})$ , $ \psi^*_0(r,p) =  u^{(+)}_{\ell=0\,,p}(r)/pr$ (see Eqs.~\eqref{eq:WFC} and \eqref{eq:ucoul}, where these Coulomb functions are defined) and $\phi_0(r,p)$ given in Eq.~\eqref{eq:phi0}.\footnote{Note that in the absence of Coulomb interaction, $ C_{[S,A]C}(E_p) \to  1 \pm e^{-4p^2R^2}$, when $\alpha \to 0$.} Hence, we see that the relative wave function is modified from the simple plain wave $e^{i\vec p \cdot \vec r}$ due to both the pairwise Coulomb and strong interactions and the quantum statistics. Finally, one can add to Eq.~\eqref{eq:cfpp} the term $\delta C(E_p)/2$ to partially account for the finite range of the interaction, as discussed in Subsec.~\ref{subsec:deltaC}.

\section{Results}\label{sec:results}

\begin{figure*}[t]\centering
\begin{tabular}{rr}
\includegraphics[height=6.25cm,keepaspectratio]{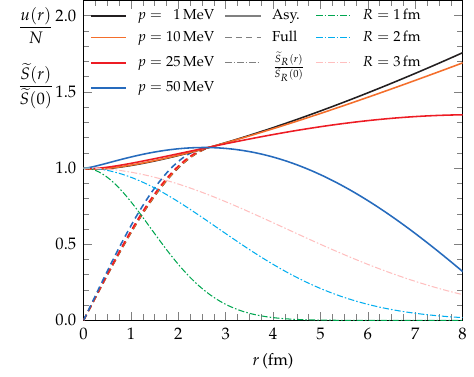} &
\includegraphics[height=6.25cm,keepaspectratio]{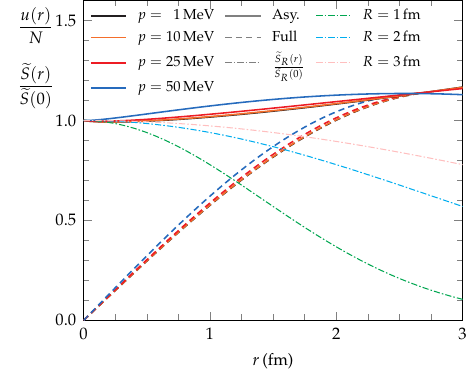} \\
\includegraphics[height=6.25cm,keepaspectratio]{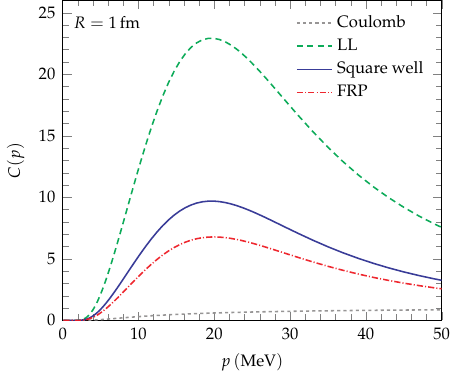} &
\includegraphics[height=6.25cm,keepaspectratio]{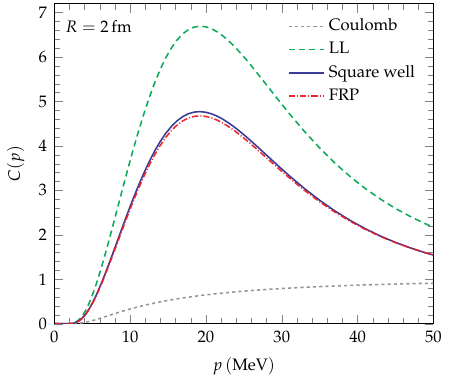}
\end{tabular}
    \caption{Left Top: Dashed and solid curves stand for the $p$-$p$ $S$-wave reduced radial functions $u^{\rm exact}_0(r)/N$ and asymptotic  $u_0(r)/N$, respectively, for different momenta, and with $N=e^{-\pi\eta/2}\Gamma(1+i\eta)f_{SC}(p)$. They have been obtained using the repulsive Coulomb interaction plus an attractive strong spherical square-well potential, as detailed in Appendix~\ref{app:esfera-dura}. We take a depth $V_0= 12.46\,\MeV$ and a radius $R_S=2.70\,\fm$. In addition, dot-dashed lines stand for the normalized Gaussian sources  $\widetilde{S}(r)/\widetilde{S}(0)$ for $R=1,\,2$ and $3\,\fm$ sizes. In the right-top panel, we zoom the $r< R_S$  region. In the bottom left and right panels, we show the CFs for source-sizes of $R=1\,\fm$  and $R = 2\,\fm$, respectively. In addition to the only Coulomb $C_C(E_p)$ (gray short dashed-lines) and exact ones (blue solid lines), we also display the CFs obtained within the LL approximation of Eq.~\eqref{eq:LLterm} $[C_C(E_p) +\Delta C_{\rm LL}(E_p)]$ (green dashed lines) and the improved-LL one of Eq.~\eqref{eq:FRC:particular-todas} that incorporates  \CorrName{} corrections  (red dot-dashed lines). All CFs are calculated using the wave-functions displayed in the top panels.}
    \label{fig:wvfandOhnshi}
\end{figure*}

Here, we  illustrate the different theoretical approximations, discussed in the previous section, to compute femtoscopy CFs. For this purpose, we will use the proton-proton system as a case study, and we will consider recent CF data from ALICE~\cite{ALICE:2019buq}, as well as the predicted CFs for different source-sizes obtained by S.\,E.\,Koonin in Ref.\,\cite{Koonin:1977fh} using the Coulomb and Reid soft-core (RSC) central, spin-orbit, and diagonal tensor potentials~\cite{Reid:1968sq}.

We start by neglecting the modifications induced by the identical particle statistics, and  we will consider Eq.\,\eqref{eq:CF-general-decomposition} in the LL approximation and the \CorrName{}-improved LL scheme of Eq.\,\eqref{eq:FRC:particular0}. In both cases, the term $\Delta C_\text{LL}$ in Eq.\,\eqref{eq:LLterm} is computed with the asymptotic wave function of Eq.\,\eqref{eq:phi0} and the exact on-shell  amplitude $T_{SC}(E_p)$. Hence, the calculations differ only by the  $\delta C(E_p)$ term. We compare these two approximate CFs to the exact one derived  from an attractive strong   spherical square-well  ($V_0=12.46\,\MeV$  and $R_S=2.70\,\fm$) interaction in presence of the repulsive Coulomb potential (Appendix \ref{app:esfera-dura}). Such interaction leads to  $S-$wave spin-singlet  ($^1S_0$) scattering length and effective range of  $a_0=-7.8\,\fm$ and $r_0=2.7\,\fm$, respectively, which are similar to those obtained using the local phenomenological  RSC potential including Coulomb effects~\cite{Reid:1968sq}.\footnote{This spherical square-well potential for the $^1S_0$ neutron-proton system results in $a_0\sim -17.3\,\fm$ and $r_0=2.87\,\fm$.} The asymptotic and exact wave functions are shown in the top-panels of Fig.\,\ref{fig:wvfandOhnshi} for several values of the center of mass (cm) momentum $p$. We note that, to obtain Eq.\,\eqref{eq:FRC:particular}, some ${\cal O}(p^2)$ corrections are neglected, as seen in Eq.~\eqref{eq:approx-zero-energy}. The top-right panel of  Fig.~\ref{fig:wvfandOhnshi} shows that they, at least for this potential, are very small. Since the CFs are computed by means of a convolution [\textit{cf.} Eqs.\,\eqref{eq:LLterm} and \eqref{eq:CF-decomp-and-deltaC}] with the source $S(r)$, we also show in the top panels of Fig.\,\ref{fig:wvfandOhnshi} the profile of the gaussian source for three different sizes $R$.

In the bottom panels of Fig.~\ref{fig:wvfandOhnshi} we compare the three calculations of the CF. We see that the LL approximation (green dashed lines) heavily overestimates the exact CF, computed from the spherical square-well (blue solid lines), for source sizes $R=1\,\fm$ and $2\,\fm$. This can be easily inferred from the large difference between the exact and asymptotic radial wave functions below $r<R_S$, where the strong potential acts. As the source-size $R$ increases, the LL approximation performs better, since the relevance of the $r<R_S$ region in the computation of the CF becomes less important, because the upper limit of the integration grows with $R$.  On the other hand, the \CorrName{} correction [$\delta C(E_p)$ of Eq.\,\eqref{eq:FRC:particular} which only uses the effective range $r_0$] notably improves the description of the CF for $R=1\,\fm$, though some discrepancies are still clearly visible, and it actually leads to a excellent reproduction of the exact CF for $R=2\,\fm$. The larger source-size is, the better this finite range effective correction performs, since the approximation $\widetilde{S}(r)\approx \widetilde{S}(0)$ for $r<R_S$ becomes more accurate. The source-size should be always referred to the finite range of the strong potential, which in this case is $R_S = 2.70\,\fm$. Hence, the improved approximation with the $\delta C(p)$ term might reasonably work well even for source sizes of the order of $1\,\fm$ in the case of a strong interaction of smaller range.

Finally, we mention that the description obtained for the exact CF by means of the Lednicky-Lyuboshits approximation $\Delta C_{\rm LL}(E_p)$ and that improved by the effective range correction $\delta C(E_p)$ of Eq.~\eqref{eq:FRC:particular-gral} for a system like neutron-proton, which is not affected by the Coulomb interaction, follows a similar pattern as that highlighted in Fig.~\ref{fig:wvfandOhnshi}, and examples can be found in Appendix \ref{app:beta}. 

\begin{figure*}[t]\centering
\begin{tabular}{cc}
\multicolumn{2}{c}{%
\includegraphics[height=5.75cm,keepaspectratio]{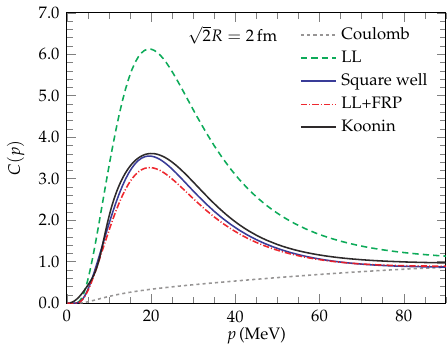}} \\
\includegraphics[height=5.75cm,keepaspectratio]{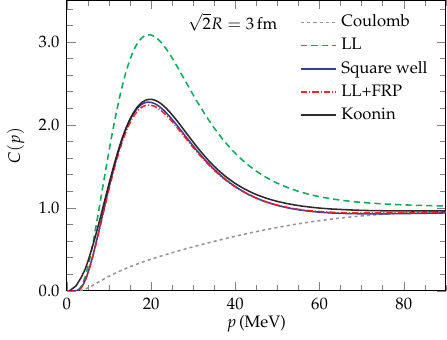} &
\includegraphics[height=5.75cm,keepaspectratio]{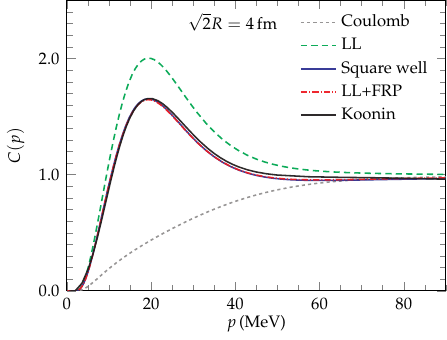} \\
\includegraphics[height=5.75cm,keepaspectratio]{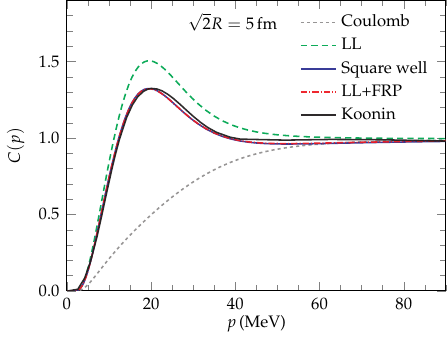} & 
\includegraphics[height=5.75cm,keepaspectratio]{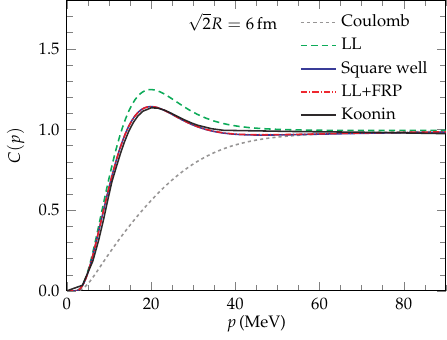}
\end{tabular}
    \caption{Comparison of the results  for $p$-$p$ CFs displayed in the upper panel of Fig.~1 of Ref.~\cite{Koonin:1977fh} (solid black lines) with different approximations (see text for details): only Coulomb $[C_{[S]C}(E_p)/4 + 3 C_{[A]C}(E_p)/4]$ (gray short dashed lines), Coulomb plus the LL term $\Delta C_{\rm LL}(E_p)/2$ (green-dashed lines),  the latter one improved by the addition of the \CorrName{} $\delta C(E_p)/2$ correction computed using the  approximation of Eq.~\eqref{eq:FRC:particular}  (red dot-dashed lines), and  finally using the spherical well potential (blue solid line). The latter CF is obtained using the exact $S$-wave reduced radial function from the Coulomb plus the strong spherical square-well potential employed in Fig.~\ref{fig:wvfandOhnshi}. This interaction provides, within 1\%, the same  ${}^1S_0$ scattering length and effective range as the RSC potential~\cite{Reid:1968sq} used in Ref.~\cite{Koonin:1977fh}.}  
    \label{fig:Koonin}
\end{figure*}

We shall now compare the LL approximation $\Delta C_{\rm LL}(E_p)$ and the \CorrName{}-improved LL one of Eq.~\eqref{eq:FRC:particular-todas} with the results for the CFs obtained by S.\,E.\,Koonin in Ref.\,\cite{Koonin:1977fh} (upper panel of Fig.~1 of that reference) with a more realistic potential. We will also show the CFs obtained  from the attractive strong   spherical square-well  ($V_0=12.46\,\MeV$  and $R_S=2.70\,\fm$) interaction in presence of the repulsive Coulomb potential, introduced above, and that was adjusted to reproduce the $^1S_0$  scattering length and effective range of the  $p$-$p$ phenomenological interaction used   in Ref.~\cite{Koonin:1977fh}. For these comparisons, we take into account  the identical-particle modifications to the evaluation of the CF discussed in Sec.\,\ref{subsec:identical}. 

As mentioned above, the CF results in Ref.\,\cite{Koonin:1977fh} were obtained using the Coulomb and RSC central, spin-orbit, and diagonal tensor potentials~\cite{Reid:1968sq} in all $p$-$p$ channels with orbital angular momenta smaller than 3. We show these results (black solid lines) in Fig.\,\ref{fig:Koonin} for several values of the source size $R$. The CFs, $C(E_p)$, are appreciably different from one only for a small range of $p$, which implies that the results are essentially unchanged if the nuclear potential is neglected in all channels except in the  $^1S_0$ wave. 

In Fig.\,\ref{fig:Koonin}, we also show the calculation within the LL approximation of Eq.~\eqref{eq:cfpp} (green-dashed lines), where the $\phi_0(r,p)$ is evaluated using Eq.~\eqref{eq:phi0} that neglects the finite range of the strong potential. The amplitude  $f_{SC}$ [proportional to $T_{SC}$, see Eq.\,\eqref{eq:relation-TSC-fSC}] is constructed from Eq.~\eqref{eq:fsc} using the RSC $^1S_0$  ERE parameters $a_0=-7.78$ fm, $r_0=2.72$ fm and $v_2=-0.028\, r_0^3$. We see that this approximation has clear limitations and it cannot properly describe the correct CF even for the most extended sources considered in the figure. On the other hand, the correction term $\delta C(E_p)/2$ computed with Eq.~\eqref{eq:FRC:particular} (depicted with red dot-dashed lines in the plots) significantly improves the LL approximation and allows an excellent reproduction of the results of Ref.~\cite{Koonin:1977fh}, except for the $\sqrt{2}R=2\,\fm$ source. Even in this latter case, it greatly improves the agreement achieved by the LL approximation. The blue solid lines in Fig.\,\ref{fig:Koonin} stand for  the $C(E_p)$ obtained using the exact $S$-wave reduced radial function from the Coulomb plus the strong spherical square-well potential employed in Fig.~\ref{fig:wvfandOhnshi}.\footnote{This is to say, we replace $\phi_0(r,p)$ in  Eq.~\eqref{eq:cfpp} by  $ [u_0^{\rm exact}(r,p)]^*/r$ from Appendix~\ref{app:esfera-dura}.} This interaction provides approximately the same scattering length and effective range as the phenomenological RSC potential and it leads to a very good description of the Koonin CFs even for the smallest source ($\sqrt{2}R=2\,\fm$), which hints to the little sensitivity of the bulk of  the proton-proton CFs reported in Ref.\,\cite{Koonin:1977fh} to higher orders in the ERE expansion.\footnote{The ERE parameter $v_2$ in the spherical square-well case is $0.033\, r_0^3$, which has opposite sign to that found ($v_2 =-0.028\, r_0^3$) using the RSC potential. Then $v_2$ is not essential to explain the disagreement between the full Koonin result of Ref.~\cite{Koonin:1977fh} and the LL approximation, supplemented by the effective range correction $\delta C(p)/2$ estimated by means of the square-well wave functions, for the $\sqrt{2}R=2\,\fm$ source.} It is also interesting to note that in the case $\sqrt{2}R =3\,\fm$, the \CorrName{} correction of Eq.~\eqref{eq:FRC:particular} and the method of Appendix~\ref{app:esfera-dura} are already very similar, being almost indistinguishable for $\sqrt{2}R \geqslant 4\,\fm$.

\begin{figure*}[t]\begin{center}%
\includegraphics[height=5.cm,keepaspectratio]{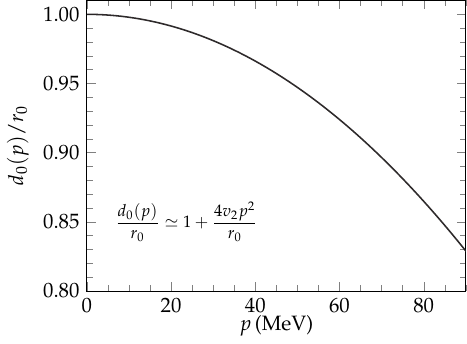}%
\hspace{15pt}%
\includegraphics[height=5.cm,keepaspectratio]{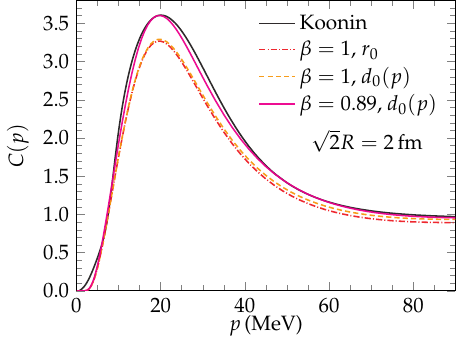}
\end{center}%
\caption{Left: $d_0(p)/r_0$ ratio  as a function of the cm momentum for the RSC phenomenological potential. We have approximated the ratio by $1+4v_2p^2/r_0$, with $v_2=-0.028\, r_0^3$~\cite{Reid:1968sq}. Right: The solid black line shows the $p$-$p$ CF calculated in Ref.~\cite{Koonin:1977fh} for the radius $\sqrt 2 R= 2$ fm. The red dot-dashed, orange dashed and magenta solid lines stand for  the \CorrName{}-improved LL CFs obtained by adding  the $\delta C(E_p)/2$ correction using  Eq.~\eqref{eq:FRC:particular} ($r_0$, $\beta=1$), Eq.~\eqref{eq:FRC:particular-gral} ($d_0(p)$, $\beta=1$), and finally  using Eq.~\eqref{eq:delC-improved} ($d_0(p)$, $\beta\ne 1$), respectively. The latter FRP-improved CF (magenta solid curve) includes the phenomenological $\beta$ parameter, which has been set to 0.89 to  reproduce the peak of the realistic CF predicted by Koonin in Ref.~\cite{Koonin:1977fh}. Note that the ($r_0$, $\beta=1$) and original Koonin CFs were already presented in the top panel of Fig.~\ref{fig:Koonin}.
\label{fig:partKR05}}
\end{figure*}

The higher momentum tail, where  the CF approaches to 1, should depend more on the higher order ERE parameters. This discussion is clearly illustrated in Fig.~\ref{fig:partKR05}. In the left plot, we show the $d_0(p)/r_0$ ratio, approximated up to the $v_2$ ERE parameter. We see that in the region of the maximum ($p$ around 25 MeV) of the Koonin realistic CF for the $\sqrt{2}R=2\,\fm$ source, the possible differences between the \CorrName{} corrections accounted for  Eq.~\eqref{eq:FRC:particular} ($r_0$, $\beta=1$) or for Eq.~\eqref{eq:FRC:particular-gral} ($d_0(p)$, $\beta=1$) are expected  to be very small.  Actually at the peak,  they cannot account for the observed disagreement  between the realistic and the \CorrName{}-improved CFs displayed in Fig.~\ref{fig:Koonin}. This can be clearly seen in the right plot of Fig.~\ref{fig:partKR05}, where we also show the  \CorrName{} CF that also includes  $d_0(p)/r_0$ corrections. However, these corrections become larger as the momentum increases, and in the latter plot, we see that the tail ($p \geqslant 40-50$ MeV) of the Koonin CF is notably better described when they are included in the \CorrName{}-improved LL CF (orange dot-dashed curve).   Nevertheless, we should note that the measurement  of the CF in this region might be affected of larger statistical and systematical uncertainties. 

To describe the peak of the realistic CF for this small source-radius ($\sqrt 2 R= 2$ fm), one should either rely  on the approximated Coulomb-plus-square-well CF, as we discussed in Fig.~\ref{fig:Koonin}, or to include the phenomenological $\beta$ parameter, introduced in Eq.~\eqref{eq:delC-improved}, to approximately account for the sizable variation of the source-strength in the region where the strong potential is not negligible. We fix $\beta=0.89$ to reproduce the peak of the Koonin's CF using the LL model supplemented by the \CorrName{} correction as in Eq.~\eqref{eq:delC-improved} ($d_0(p)$, $\beta\ne 1$)\footnote{The parameter $\beta$ cannot be determined here from the value of the CF at threshold, as it is done in Appendix~\ref{app:beta}, since the CF vanishes at zero-momentum because of the Coulomb repulsion. }.  The resulting CF is represented by the solid magenta curve in Fig.~\ref{fig:partKR05}. The agreement with the realistic CF (solid black curve) is very good for all momenta, and in the high-momentum tail, it is much better than that obtained using the square-well potential (solid blue curve) shown in the top panel of Fig.~\ref{fig:Koonin}. This is not surprising, since the latter simple interaction produces a positive $v_2$ parameter, which is of opposite sign to that found with the RSC potential~\cite{Reid:1968sq}. 

For larger source radii the effective $\beta $ parameter hardly deviates from one, and it is not really relevant in the Koonin {\it versus} LL+\CorrName{} comparisons shown in the plots of the two bottom rows of  Fig.~\ref{fig:Koonin}. However, the $d_0(p)/r_0$ corrections are still important to improve the behavior of the approximated LL+\CorrName{} CFs in the high-momentum tail.

Our final comparison is with actual $p$-$p$ CF data by the ALICE Collaboration \cite{ALICE:2019buq}. However, this comparison is more elaborate, because the experimental data contain feed-down contributions, that must be taken into account for a proper comparison. As explained in Ref.\,\cite{ALICE:2018ysd} (\textit{cf.} Eq.\,(11) of that work), we define:
\begin{equation}\label{eq:AliceCpp}
C'(p) = 1 + \lambda_{pp} (C_{pp}(p) -1) + \lambda_{pp_{\Lambda}} ( C_{pp_{\Lambda}}(p) - 1)\,,
\end{equation}
where $C'(p)$ is the CF to be compared with the data, $C_{pp}(p)$ is the genuine $p$-$p$ CF that is the one discussed/calculated in this work, and $C_{pp_{\Lambda}}$ accounts for protons stemming from the weak $\Lambda$ decay. To model $C_{pp_{\Lambda}}$, we follow the approach in Ref.\,\cite{ALICE:2018ysd}, where only the feed-down from the emission of $p$-$\Lambda$ pairs is considered, and $C_{pp_{\Lambda}}$ is obtained by fitting the experimental $C_{p\Lambda}$ and transforming it to the $p$-$p$ momentum basis, accounting for the $\Lambda \to p\pi^-$ weak decay.\footnote{More specifically, using the LL model \cite{Lednicky:1981su}, $C_{pp_{\Lambda}}(p)$ is expressed in Eq. (10) of Ref.\,\cite{ALICE:2018ysd} as:
\begin{equation*}
C_{pp_{\Lambda}}(p) = 1 + \sum_{S=0,1} \rho_S \left[ \frac{1}{2} \left|
\frac{f^S(p)}{R} \right|^2 \left( 1 - \frac{d^S_0}{2\sqrt{\pi} R}
\right) + \frac{2 \Re f^S(p)}{\sqrt{\pi} R} F_1(2 p R) -
\frac{\Im f^S(p)}{R} F_2(2 p R) \right],
\end{equation*}
where $f^S(p)$ is modeled using the scattering length $f_0^S$ and the effective range $d_0^S$ through the ERE. Assuming unpolarized emission, the fraction parameters $\rho_S$ take the values $\rho_0 = 1/4$ and $\rho_1=3/4$ for the spin singlet and triplet, respectively. The expressions for $F_{1,2}(z)$ can be found in Refs.\,\cite{STAR:2014dcy,Albaladejo:2024lam}.  The scattering lengths and effective ranges are taken from Ref.\,\cite{Shapoval:2014yha}, with values $f_0^{S=0} = 2.88\,\fm$, $d_0^{S=0}=2.92\,\fm$, $f_0^{S=1}=1.66\,\fm$, and $d_0^{S=1}=3.78\,\fm$, which are indeed compatible with some of the values explored in Ref.\,\cite{ALICE:2019buq} (\textit{cf.} Table IV of the later reference).} In addition, the parameters $\lambda_{pp}=0.67$ and $\lambda_{pp_{\Lambda}} = 0.203$ are taken from Table 1 of Ref.\,\cite{ALICE:2019buq}.

\begin{figure*}[t]\centering
\includegraphics[height=6.5cm,keepaspectratio]{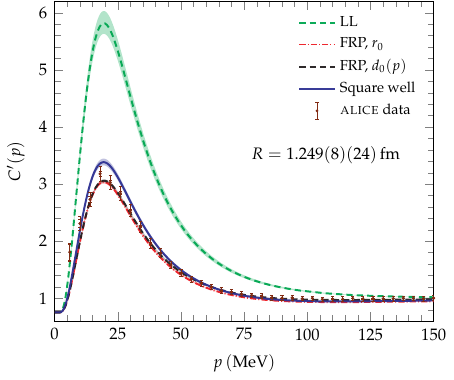}%
\includegraphics[height=6.5cm,keepaspectratio]{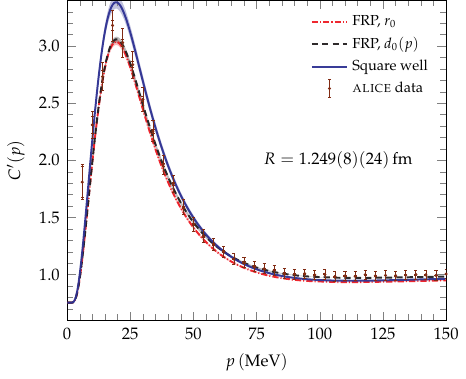}
\caption{Comparison of the $p$-$p$ CF data from ALICE \cite{ALICE:2019buq} (brown points) with the different calculations discussed in this work: LL approximation (green dashed line), the \CorrName{}-improved-LL CFs obtained by adding the $\delta C(E_p)/2$ term using  Eq.~\eqref{eq:FRC:particular} (red dot-dashed line) or Eq.~\eqref{eq:FRC:particular-gral} (black dashed line), and the square-well model (blue solid line). The latter CF is obtained using the exact $S$-wave reduced radial function from Coulomb plus the strong spherical square-well potential employed in Fig.~\ref{fig:Koonin}. The source size, as determined in Ref.\,\cite{ALICE:2019buq}, is $R=1.249(8)(24)\,\fm$.  The theoretical CFs shown are computed using Eq.\,\eqref{eq:AliceCpp}, which accounts  for the genuine $p$-$p$ CF and the feed-down contributions coming from the $\Lambda$ weak decay, as explained in the text. To evaluate the genuine $p$-$p$ LL and \CorrName{}-improved CFs, we employ the $^1S_0$ proton-proton scattering length, effective range and $v_2$ parameter deduced from the phenomenological RSC potential~\cite{Reid:1968sq}. We set  $\beta=1$ in both  \CorrName{}-improved CFs, and in the black dashed line, we approximate $d_0(p) \simeq r_0+4v_2p^2 $ as in Fig.~\ref{fig:partKR05}. The errors on the ALICE source size, added in quadrature, give rise to the 68\% confident-level uncertainty bands displayed for the predicted CFs.
\label{fig:comparisonAlice}}
\end{figure*}

In Fig.\,\ref{fig:comparisonAlice} we show the ALICE data \cite{ALICE:2019buq}, compared with the calculation of the CF in Eq.\,\eqref{eq:AliceCpp}, with the genuine $p$-$p$ CF $C_{pp}(p)$ computed with the LL approximation (green dashed line), the \CorrName{} correction of Eq.~\eqref{eq:FRC:particular} (red dot-dashed line) or the spherical square-well (blue solid line). In Ref.\,\cite{ALICE:2019buq} a Gaussian source size $R = 1.249(8)(24)\,\fm$ is obtained, where the first (second) error is statistical (systematic). As in previous comparisons, the LL approximation is quite above the rest of the results. It can be seen that the \CorrName{}-improved LL CF using Eq.~\eqref{eq:FRC:particular} and the square-well solution are close to each other, and that the experimental data lie in between both calculations. Finally in the plot, we also show the \CorrName{}-improved LL CF obtained by adding the $\delta C(E_p)/2$ term using  Eq.~\eqref{eq:FRC:particular-gral} (black dashed curve), which includes $d_0(p)/r_0$ corrections. As in the left panel of Fig.~\ref{fig:partKR05}, the changes below $p \leqslant 40\,\MeV$ are very small, but they become really significant as the momentum increases, leading to a remarkable description of the high-momentum tail of the data. This confirms that this region is very much sensitive to the off-shell corrections driven by the $v_2$ ERE parameter. For high momenta, the square-well CF performs  worse than this latter \CorrName{}-improved approach. 

We stress that no parameter has been adjusted to obtain the quite good description of the accurate ALICE data shown in Fig.~\ref{fig:comparisonAlice}, which clearly supports the validity of the CF calculation scheme, including both Coulomb and \CorrName{} off-shell effects, presented in this work.

\section{Summary and outlook}\label{sec:summary}

In this work we have discussed femtoscopy  CFs in the simultaneous presence of strong and Coulomb interactions. The method presented here allows one to take into account Coulomb interactions analytically, thus without the need of solving the Schr\"odinger equation numerically. We have discussed its connection with previous derivations by Lednicky \textit{et al.} in Refs.\,\cite{Lednicky:1981su,Lednicky:2005tb}. We also show that the formalism is amenable to account for strong scattering interactions obtained from general EFT momentum expansions in presence of Coulomb effects. This is discussed in Subsec.~\ref{subsec:JuanFormalism}. There, it is proven that the scheme of Ref.~\cite{Nieves:2003uu}, based on standard distorted wave theory techniques and dimensional regularization, can be employed to analytically find the scattering amplitude for an arbitrary contact strong potential, expanded in powers of $p^2$, in presence of the long distance Coulomb interaction. The obtained scattering amplitude satisfies exact elastic unitarity and inherits the complete Coulomb left-hand discontinuity.

We have benchmarked our results against the theoretical calculations of Ref.\,\cite{Koonin:1977fh}, where the $p$-$p$ CFs are computed using the realistic RSC potential in the presence of Coulomb interactions, which requires the numerical solution of the Schr\"odinger equation to obtain the wave functions. We have also compared our calculations with the $p$-$p$ femtoscopy CF data from the ALICE Collaboration \cite{ALICE:2019buq}. In both cases, we have found good agreement, and we have discussed the performance of the different approximations that can be used in obtaining the CFs. We have shown that the inclusion of  \CorrName{} corrections, which approximately incorporate off-shell effects that are neglected within the LL approximation, turn out be essential to obtain realistic predictions of CFs, even for extended sources with radius of the order of $4\,\fm$ or larger. The $p$-$p$ CF data provide valuable information on the two proton strong interaction, not only  for low momenta where the CF strongly deviates for one and reaches a maximum, but also in the high-momentum tail, which turns out to very much sensitive to the off-shell corrections driven by the $v_2$ ERE parameter.

All these comparisons constitute a successful test of the formalism presented in this work, which could then be applied to compute CFs of other systems involving Coulomb interactions.

\begin{acknowledgments}

We warmly thank D.\,R.\,Entem for useful discussions. This work is supported by the Spanish Ministerio de Ciencia e Innovaci\'on (MICINN) under contracts PID2020-112777GB-I00, PID2023-147458NB-C21 and CEX2023-001292-S; by Generalitat Valenciana under contracts PROMETEO 2020/023 and  CIPROM 2023/59. %
M.\,A. acknowledges financial support through GenT program by Generalitat Valencia (GVA) Grant No.\,CIDEGENT 2020/002, Ramón y Cajal program by MICINN Grant No.\,RYC2022-038524-I, and Atracción de Talento program by CSIC PIE 20245AT019. %

\end{acknowledgments}

\appendix

\section[Further examples of the performance of the \CorrName{} correction  ]{Further examples of the performance of the \CorrName{} correction}
\label{app:beta}
\begin{figure*}[t]\begin{center}%
\includegraphics[height=6cm,keepaspectratio]{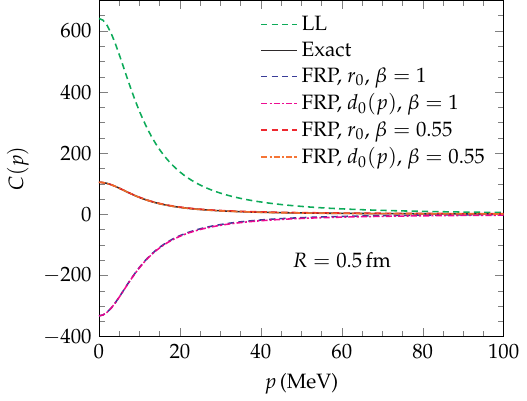}%
\includegraphics[height=6cm,keepaspectratio]{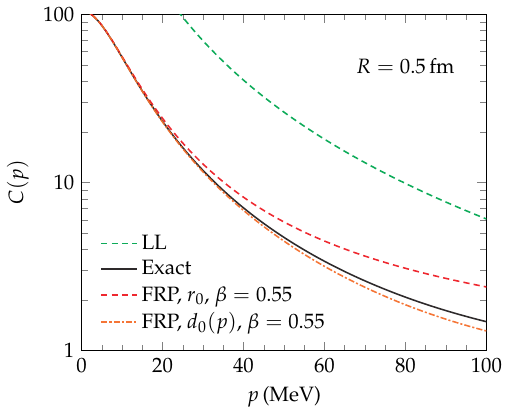}\\%
\includegraphics[height=6cm,keepaspectratio]{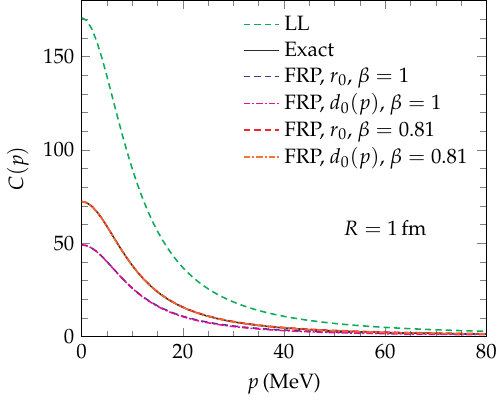}%
\includegraphics[height=6cm,keepaspectratio]{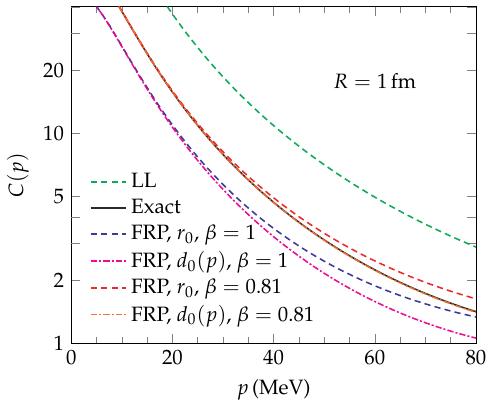}%
\end{center}%
\caption{ Left panels: Different neutron-proton CF calculations for  source-sizes of $R=0.5\,\fm$ (top) and $R=1\,\fm$  (bottom). Solid black and green dashed lines stand for the exact calculation $C(E_p)$ and the LL estimation $1+\Delta C_{\rm LL}(E_p)$, which only involves the asymptotic wave-function [see Eq.~\eqref{eq:LLterm}], respectively. The other curves add to $1+\Delta C_{\rm LL}(E_p)$, the \CorrName{} $\delta C(E_p)$ term  [Eq.~\eqref{eq:deltaC-general}] employing different approximations based on   Eq.~\eqref{eq:delC-improved}. The parameter $\beta$ is set to one in the blue dashed and magenta dot-dashed lines, which are calculated using the effective range $r_0$ and  the momentum dependent function $d_0(p)$, respectively. In the  red dashed and orange dot-dashed curves, computed with $r_0$ and  $d_0(p)$, respectively, $\beta$ is set to 0.55 (0.81) for the $R=0.5\,\fm$ ($R=1\,\fm$) source, which is the value obtained from reproducing the exact CF $C(E_p)$ at threshold [see Eq.~\eqref{eq:betadeter}]. Right panels: Zoom of the left plots. In all cases an attractive spherical square-well potential of depth $V_0= 12.46\,\MeV$ and radius $R_S=2.70\,\fm$ has been used. 
\label{fig:FRP1}}
\end{figure*}

In this section of the Appendix, we will study the validity of the improved \CorrName{} correction of Eq.~\eqref{eq:delC-improved}. To make clearer the conclusions, we simplify the discussion neglecting both Coulomb interactions and identical-particle correlations. Thus, we will present results for two different simple neutron-proton potentials in the $^1S_0$ wave,
\begin{enumerate}[leftmargin=1pt,itemindent=14pt,topsep=5pt]
    \item First, we consider the  attractive spherical square-well potential ($V_0= 12.46$ MeV and $R_S=2.70$ fm) used in the main-text of this work. In this case, the absolute value of the neutron-proton $^1S_0$ scattering length is abnormally large $\sim 17$ fm, because of the presence of a virtual state close to threshold, while the effective range is of natural order around 2.9 fm. In Fig.~\ref{fig:FRP1}, we show the femtoscopy CF calculated using various approaches and  two source-sizes [$R=0.5$ fm (top panels) and $R=1$ fm (bottom panels)] significantly smaller than the range of the strong potential. As expected, the LL estimation [$1+\Delta C_{\rm LL}(E_p)$] of Eq.~\eqref{eq:LLterm} provides a very poor description of the CF computed with the exact wave-function for these two small sources. The addition  of the \CorrName{ }$\delta C(E_p)$ term evaluated with the approximate expression of  Eq.~\eqref{eq:delC-improved} and $\beta=1$, independently of the treatment of $d_0(p)$, improves on the LL CF, but it still leads to unreliable results for  the $R=0.5$ fm source, while discrepancies persist and are clearly visible for the larger source of 1 fm. For this latter case, the pattern found here is similar to that inferred from the proton-proton CFs, calculated also for $R=1$ fm and the same strong potential, displayed in the bottom-left plot of Fig.~\ref{fig:wvfandOhnshi} above 30 MeV, where Coulomb distortion becomes less important. 

Finally, we have also examined \CorrName{} approximations with $\beta \ne 1$. We  determine this empirical parameter from the exact CF at threshold,
\begin{eqnarray}
  \beta = \frac{1+\Delta C_{\rm LL}(E_p=0)-C(E_p=0)}{2\pi r_0 \, a_0^2 \, \widetilde{S}(r=0)} = \frac{4\sqrt{\pi}R^3}{ r_0 \, a_0^2}\left[1+\Delta C_{\rm LL}(E_p=0)-C(E_p=0)\right]  \label{eq:betadeter}
\end{eqnarray}
 In Fig.~\ref{fig:FRP1},  we show different neutron-proton CF calculations for the source-sizes of $R=0.5\,\fm$ (top) and $R=1\,\fm$  (bottom). We see that both, the orange dot-dashed ($d_0(p)= 2 {\rm Re}\left[df_{SC}^{-1}(p)/dp^2\right]$, as obtained from Eq.\,\eqref{eq:r0eff} in the absence of Coulomb) and the red dashed ($d_0(p)$ approximated by $r_0$) curves provide quite good descriptions of the exact CFs, not only for $R=1\,\fm$, but also for the smaller source of radius $R=0.5\,\fm$. In particular, we find that  the predicted CFs (orange dot-dashed) supplemented with the \CorrName{} correction of  Eq.~\eqref{eq:delC-improved}, maintaining the momentum dependence of $d_0(p)$, are almost indistinguishable from the exact ones (solid black) in the whole momentum range examined in the plots. This is a remarkable result, and gives support to including the off-shell effects, neglected in the LL approximation, in the calculation of the CF by means of the \CorrName{} correction of  Eq.~\eqref{eq:delC-improved}. Note that in general $\beta$ will strongly depend on $R$, actually we find values of $\beta$ of $0.55$ and $0.81$ for $R=0.5$ fm and $R=1$ fm, respectively. For larger sources, $\beta$ approaches to one ($\beta=0.94$, $0.97$ and $0.99$ for $R=2,3$ and $6\,\fm$, respectively), and the inclusion of this new parameter becomes unnecessary. The above discussion is illustrated in the left panel of Fig.~\ref{fig:integrando1}, where we show the integrand of Eq.~\eqref{eq:deltaC-general} at threshold (\textit{i.e.}, for $E_p=0$) for various sources. The parameter $\beta$ for each of them is simply the ratio of the area under the corresponding dashed curve to  that obtained from the solid black one. In addition, in the right plot of Fig.~\ref{fig:integrando1}, we show $d_0(p)$ in units of the effective range $r_0$. We observe that the changes for momenta around 50 MeV, those relevant in  Fig.~\ref{fig:FRP1}, are moderate ($\sim$ 10\%-20\%) for the attractive  square-well potential ($V_0= 12.46$ MeV and $R_S=2.70$ fm) considered in that figure.

\begin{figure*}[t]\begin{center}%
\includegraphics[height=6cm,keepaspectratio]{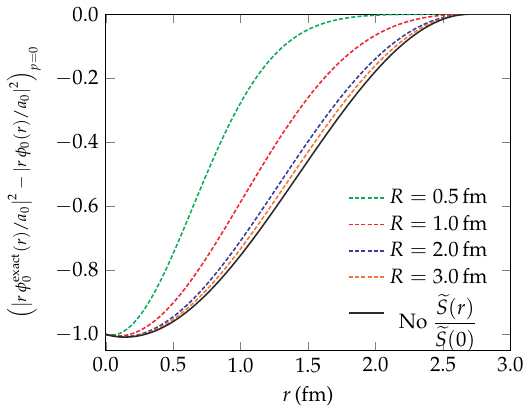}%
\includegraphics[height=6cm,keepaspectratio]{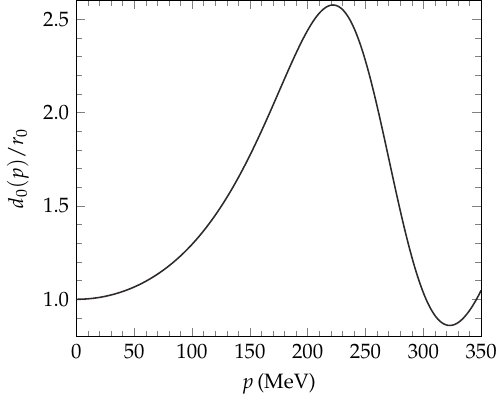}
\end{center}%
\caption{Left:   We show different curves related to the integrand of Eq.~\eqref{eq:deltaC-general} at threshold. Black solid line: difference of the squares of the exact and asymptotic reduced-radial wave functions in $a_0^2$ units $ \left( \left\lvert r\phi^{\rm exact}_0(r,p=0)/a_0 \right\rvert^2 - \left\lvert r\phi_0(r,p=0)/a_0\right\rvert^2\right)$. Dashed lines: the former function of $r$ multiplied by $\widetilde{S}(r)/\widetilde{S}(0)$,  with $R=0.5$ fm (green), 1 fm (red), 2 fm (blue) and 3 fm (orange). Right: Ratio $d_0(p)/r_0$ as a function of the  cm momentum of the neutron-proton pair. All curves in this figure have been obtained using the same neutron-proton  potential as in Fig.~\ref{fig:FRP1} (range of 2.7 fm and an attraction of 12.46 MeV).
\label{fig:integrando1}}
\end{figure*}

The determination of $\beta$ from the CF at threshold cannot be done in presence of the Coulomb interaction, which  will force  $C(E_p=0)$ to vanish or to diverge in the repulsive or attractive cases, respectively. However, $\beta$ can be fixed from the exact CF at some other finite momentum, sufficiently displaced for zero to make meaningful the determination. This is precisely the procedure that we have followed in the discussion of Fig.~\ref{fig:partKR05} in the main text.

\item Next, we supplement the previous potential with a repulsive spherical-barrier of height $V_b>0$ and width $(R_T-R_S)>0$, 
\begin{figure*}[t]\begin{center}%
\includegraphics[height=4.5cm,keepaspectratio]{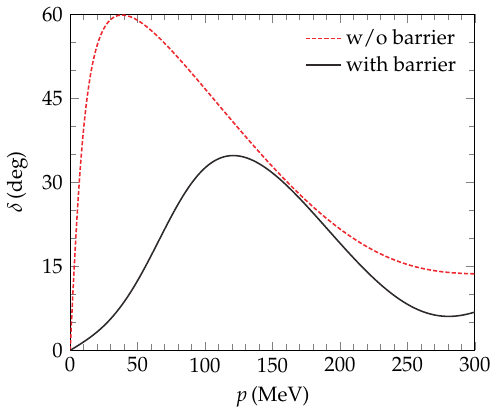}%
\includegraphics[height=4.5cm,keepaspectratio]{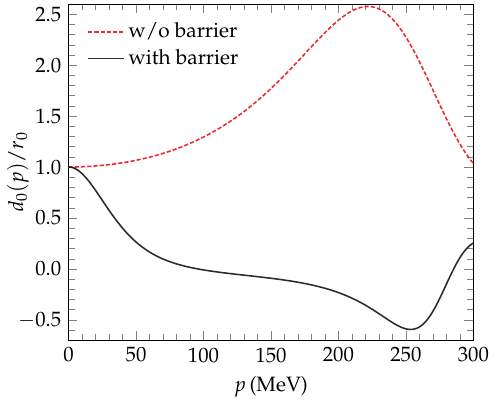}%
\includegraphics[height=4.5cm,keepaspectratio]{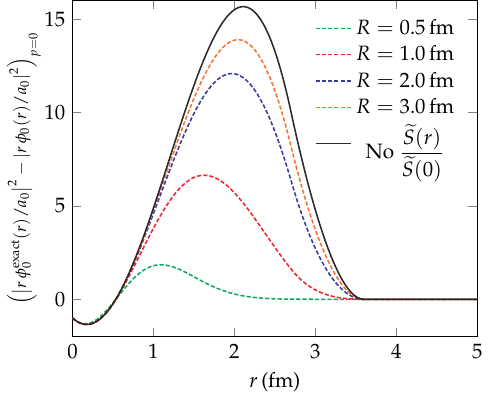}
\end{center}%
\caption{Left and middle plots: The black curves stand for the phase-shift and the ratio $d_0(p)/r_0$ as a function of the cm  momentum for the neutron-proton potential of Eq.~\eqref{eq:potbarr}, with parameters $V_0=12.46\,\MeV$, $R_S=2.7\,\fm$, $V_b= 10\,\MeV$ and $R_T=3.6\,\fm$. The dashed-red curves show the phase-shift and  the ratio $d_0(p)/r_0$ obtained removing the repulsive barrier ($V_b=0$ and $R_T=R_S)$. Right plot: Same curves as those displayed in the left plot of  Fig.~\ref{fig:integrando1}, but calculated with the spherical well-plus-barrier interaction of Eq.~\eqref{eq:potbarr}. 
\label{fig:integrando2}}
\end{figure*}
\begin{figure*}[t]\begin{center}%
\includegraphics[height=6.25cm,keepaspectratio]{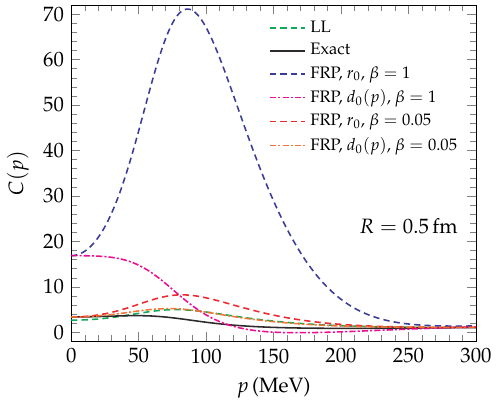}%
\includegraphics[height=6.25cm,keepaspectratio]{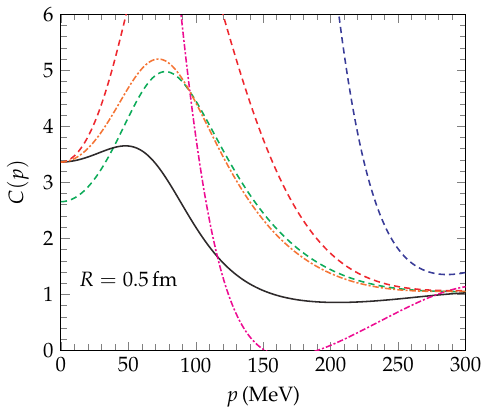}\\%
\includegraphics[height=6.25cm,keepaspectratio]{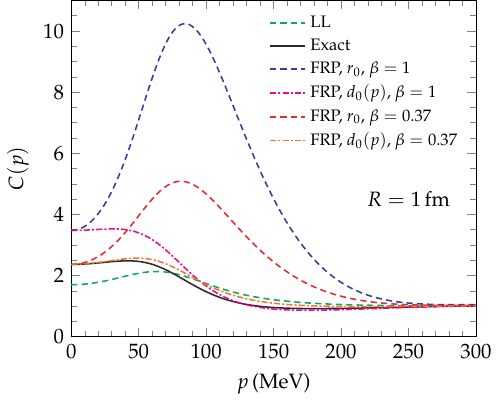}%
\includegraphics[height=6.25cm,keepaspectratio]{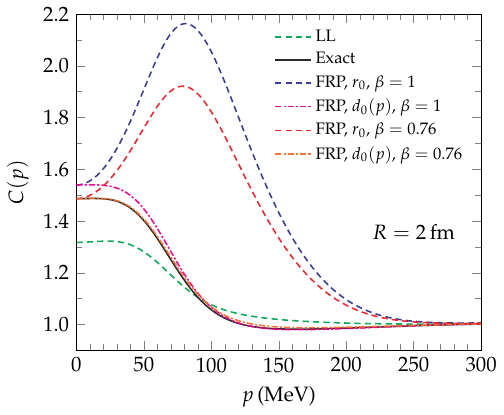}\\%
\includegraphics[height=6.25cm,keepaspectratio]{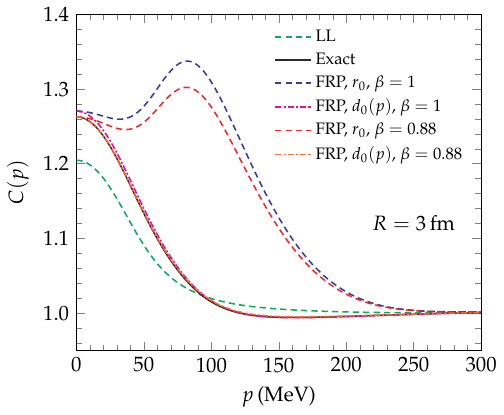}%
\includegraphics[height=6.25cm,keepaspectratio]{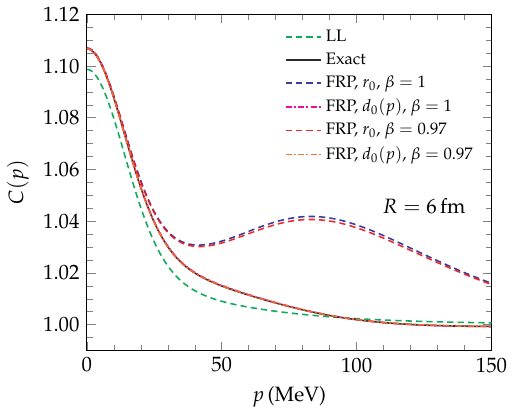}%
\end{center}%
\caption{ Different neutron-proton CF calculations, using the potential of Eq.~\eqref{eq:potbarr}, with parameters $V_0=12.46\,\MeV$, $R_S=2.7\,\fm$, $V_b= 10\,\MeV$ and $R_T=3.6\,\fm$. The source-sizes are $R=0.5\,\fm$ (top), $R=1\,\fm$  (middle-left), $R=2\,\fm$ (middle-right), $R=3\,\fm$ (bottom-left) and $R=6\,\fm$ (bottom-right). The different curves are calculated as described in Fig.~\ref{fig:FRP1}. The corresponding $\beta$ values are 0.05, 0.37, 0.76, 0.88 and 0.97 for the source-sizes $R=0.5$ fm, 1 fm, 2 fm, 3 fm and 6 fm, respectively, obtained from reproducing the exact CF $C(E_p)$ at threshold [\textit{cf.} Eq.~\eqref{eq:betadeter}]. In the right top panel ($R=0.5\,\fm$), we present a  zoom  of the left plot.
\label{fig:FRP2}}
\end{figure*}

\begin{equation}
V(r)=-V_0 \Theta[R_S-r]+V_b\Theta[r-R_S]\Theta[R_T-r]\label{eq:potbarr}\, ,
\end{equation}
where we keep $V_0=12.46$ MeV and $R_T=2.7$ fm and we additionally set $V_b= 10$ MeV and $R_T=3.6$ fm. This potential leads to a negative scattering length and to a huge negative effective range, $ a_0=-0.51$ fm and  $r_0=-49.2$ fm, respectively. It generates a resonance located at $(E_R-i\Gamma_R/2)= (1.36-i 8.43)$ MeV in the fourth quadrant of the second Riemann sheet, which provides a maximum in the phase-shift at a cm momentum around 100 MeV, as can be seen  in the left plot of Fig.~\ref{fig:integrando2}. 

As can be seen in Fig.~\ref{fig:FRP2},  the reproduction of the exact CF for this potential using the LL approximation,  complemented or not by the \CorrName{} corrections, is poorer than when the repulsive barrier was not included. The difficulty increases because of the larger range of the potential, which makes that the non-asymptotic details of the wave-function  become quite relevant even for source-sizes of $3\,\fm$. The intricate momentum dependence  of $d_0(p)$, exhibited in the middle plot of Fig.~\ref{fig:integrando2}, contributes also to the complexity of the problem, in particular for momenta around 100 MeV where it vanishes. Nevertheless,  we find that  the approximated orange dot-dashed CFs, which incorporate the \CorrName{} corrections of  Eq.~\eqref{eq:delC-improved} maintaining the momentum dependence of $d_0(p)$ and ($\beta\ne 1)$ fixed by means of Eq.~\eqref{eq:betadeter}, provide a very good approximation to the exact ones (solid black curves) in the whole momentum range examined in the plots for source-sizes of 1 fm or larger. We stress the large deviations from one that we find for the effective $\beta$ parameter (see caption of Fig.~\ref{fig:FRP2}) which are naturally expected from the inspection of the areas under the curves displayed in the right plot  of Fig.~\ref{fig:integrando2}. Thus, for instance $\beta$ is 0.37 for $R=1$ fm and it still is 0.88 for a source of radius 3 fm, since the source modifies the existing cancellation in the computation of the areas in  Fig.~\ref{fig:integrando2}.

For the smallest source of $R=0.5$ fm, discrepancies are  clearly  visible even after incorporating the full \CorrName{} corrections of  Eq.~\eqref{eq:delC-improved}, in particular in the region of $p=100\,\MeV$ where $d_0(p)$ vanishes. This means that for this small radius and this interaction, off-shell effects cannot be successfully accounted for by these approximate \CorrName{} corrections due to the large variation of the source intensity within the finite range of the strong spherical well-plus-barrier potential of Eq.~\eqref{eq:potbarr}.

\end{enumerate}

\section[\boldmath $S$-wave scattering reduced wave-function for a spherical square-well in presence of the Coulomb potential]{\boldmath $S$-wave scattering reduced wave-function for a spherical square-well in presence of the Coulomb potential}
\label{app:esfera-dura}

We solve, for positive energies $E>0$, the $S$-wave reduced radial Schr\"odinger equation for a potential $V(r)=-V_0 \Theta[R_S-r]+ \lambda \alpha/r$,  with $\lambda=\pm 1$. From the discussion in this work, one finds 
\begin{equation}
    u_0^{\rm exact}(r,p) = \left\{ \begin{array}{lr} A(p)\,u^{(+)}_{\ell=0\,, p_{\rm V_0}}(r),  & \quad r \leqslant R_S\\  \\
   \frac{u^{(+)}_{\ell=0\,,p}(r)}{p} +  B(p)C^2_{\eta}e^{2i\sigma_0} e^{\pi\eta/2}W_{-i\eta, 1/2}(-2ipr) , & \quad r \geqslant R_S \end{array} \right. \label{eq:u0exacta}
\end{equation}
where $p_{\rm V_0}= \sqrt{2\mu V_0 +p^2}$ and $p^2=2\mu E$, with  $\mu$ the reduced mass and the Whittaker function $W_{k,m}(z)$  already introduced in Eq.~\eqref{eq:whittaker}. The dependence on $\lambda$ is hidden in the factors $\eta(q)\equiv \eta_q = \lambda \mu\alpha/q$ that appear explicitly in Eq.~\eqref{eq:u0exacta}, in the computation of $u^{(+)}_{\ell=0\,, p_{\rm V_0}}(r)$ and $u^{(+)}_{\ell=0\,,p}(r)$   in the inner and outer regions, respectively, and in the Coulomb phase-shift $\sigma_0$. To avoid confusions, we note that the momentum involved in the computation of $\eta_q$ in the inner region is $p_{\rm V_0}$, while for the outer one $q=p$. The complex constants $A(p)$ and $B(p)$, both have fermi units,   are obtained by requiring continuity of $u_0^{\rm exact}(r,p)$ and $du_0^{\rm exact}(r,p)/dr$ at $r=R_S$. The normalization is such that  the scattering amplitude $f_{SC}(p)$ of Eq.~\eqref{eq:fsc} is just given by the constant $B(p)$, i.e. $f_{SC}(p)=B(p)$. Moreover, we note that $u_0^{\rm exact}(r,p)$ vanishes at $r=0$, while the outer reduced wave function ($r>R_S$)  is $r\phi^*_0(r,p)$, with the latter  given in Eq.~\eqref{eq:phi0}. 

\begin{figure*}[t]\centering
\includegraphics[height=5.5cm,keepaspectratio]{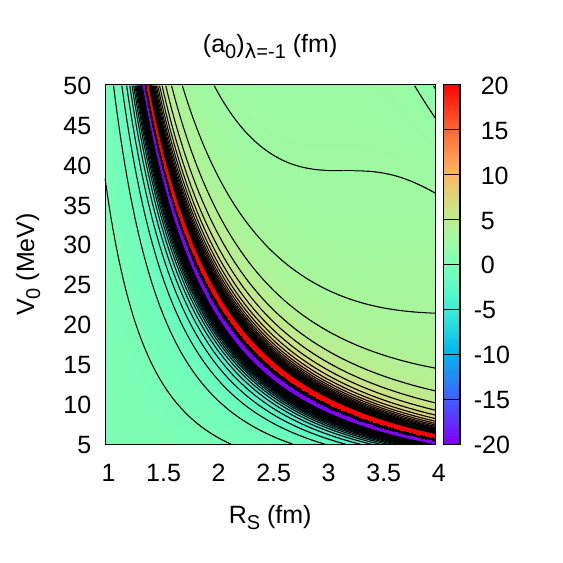}%
\includegraphics[height=5.5cm,keepaspectratio]{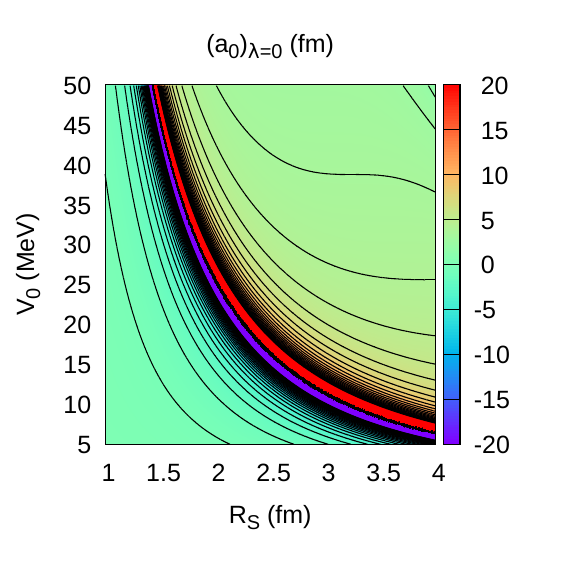}%
\includegraphics[height=5.5cm,keepaspectratio]{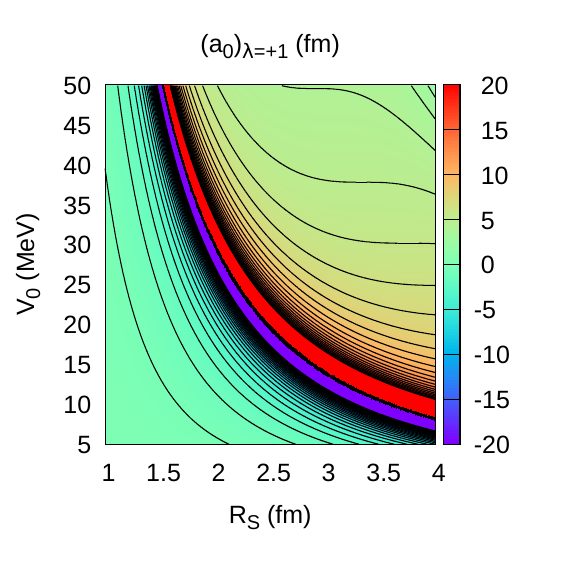}
\caption{Dependence of the scattering length (fm) on the width ($R_S$) and depth ($V_0$) of the spherical square-well strong potential for a system of reduced mass $\mu=m_p/2$, in the presence of attractive (left) and repulsive (right) Coulomb interaction. In the middle panel, we display the results in the absence of the Coulomb interaction ($a_0[\alpha= 0]$ ). The frontier-curve between the blueish and the reddish regions marks the ($R_S$-$V_0$) values that lead to a zero-energy bound state. Hence, just infinitesimally below (above) that curve, the scattering length tends to $-\infty$ ($+\infty$). We also observe the expected relative movement of this frontier-curve  from left to right panels as a consequence of the diminution of the attraction. \label{fig:a0CplusF}}
\end{figure*}

\begin{figure*}[t]\begin{center}%
\includegraphics[height=5.5cm,keepaspectratio]{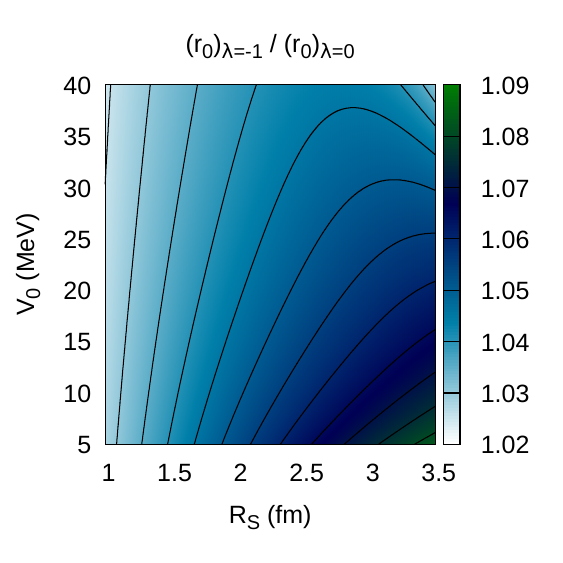}%
\includegraphics[height=5.5cm,keepaspectratio]{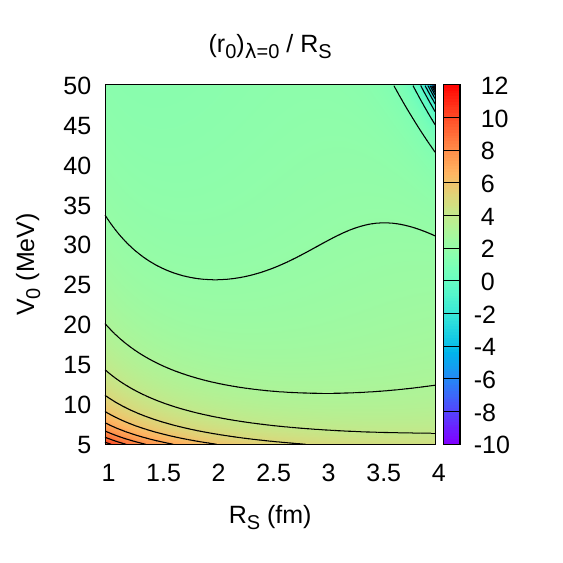}%
\includegraphics[height=5.5cm,keepaspectratio]{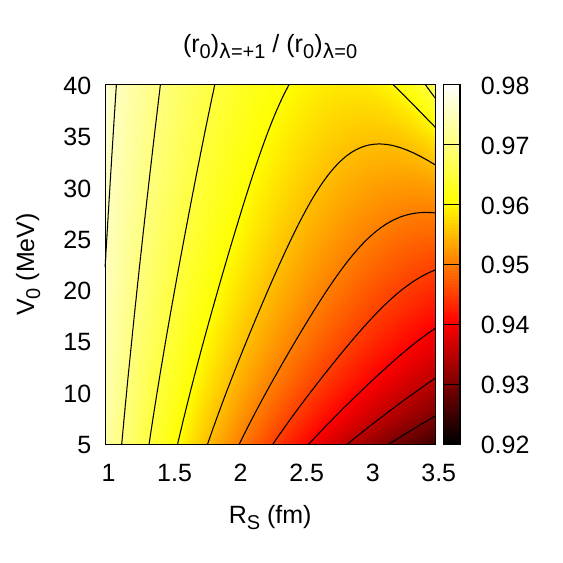}\\
\includegraphics[height=5.5cm,keepaspectratio]{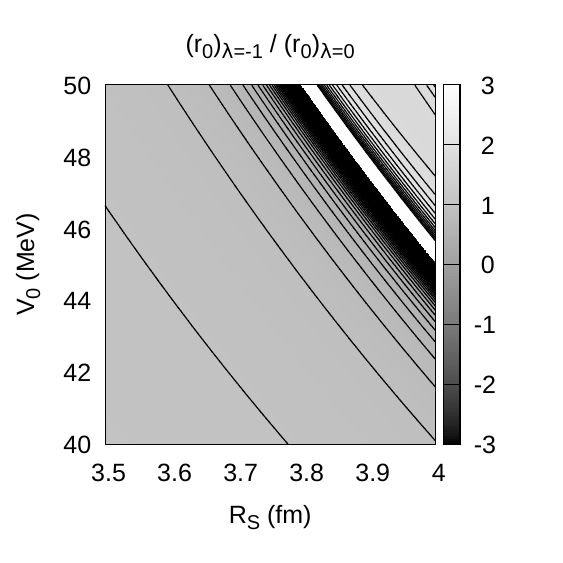}%
\includegraphics[height=5.5cm,keepaspectratio]{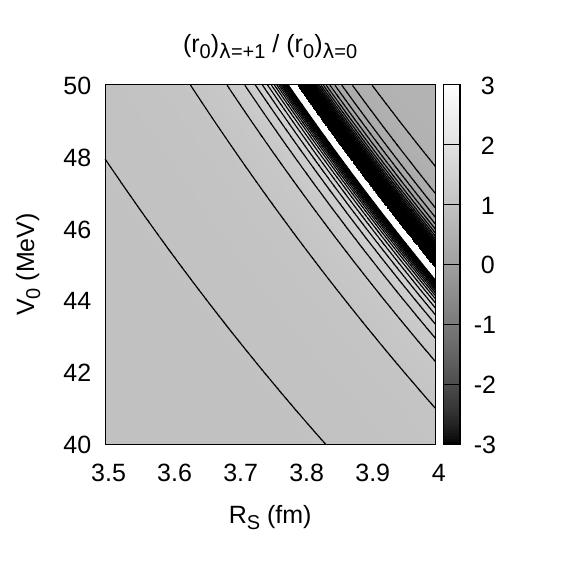}
\end{center}%
\caption{Top panels: Dependence on the width ($R_S$) and depth ($V_0$) of the spherical square-well strong potential for a system of reduced mass $\mu=m_p/2$ of the effective range ratios $r_0[\lambda=-1]/r_0[\alpha= 0]$ (attractive),  $r_0[\alpha= 0]/R_S$ [\textit{cf.} Eq.~\eqref{eq:r0neutro}] and $r_0[\lambda=+1]/r_0[\alpha= 0]$ (repulsive), with the width $R_S$ of the spherical square-well varying from $1\,\text{fm}$ to $3.5\,\text{fm}$. Bottom panels: Ratios $r_0[\lambda=-1]/r_0[\alpha= 0]$ (left) and $r_0[\lambda=+1]/r_0[\alpha= 0]$ (right)  with $R_S$ and $V_0$ taking values in the 3.5-4 fm and 40-50 MeV intervals, respectively. In this region,  the neutral effective range  $r_0[\alpha= 0]$ is very small and passes through zero, changing sign. This is reflected in the plots by the abrupt transition from black to white regions, where the  $r_0[\lambda=\pm 1]/r_0[\alpha= 0]$ ratios rapidly varies from $-\infty$ to $+\infty$.\label{fig:effrange_dep}}
\end{figure*}

In  the absence of the Coulomb interaction $u^{(+)}_{\ell=0\,, q}(r)\to \sin qr $ and $W_{-i\eta, 1/2}(-2ipr)\to e^{ipr}$, and one recovers the purely $S$-wave spherical square-well reduced wave function. In the full case, one finds
\begin{eqnarray}
f_{SC}^{-1}(p)&=& C^2_\eta\, p (\cot\delta -i) = C^2_\eta p \frac{ F'(\eta_{p_{\rm V_0}};p_{\rm V_0}R_S){\cal H}_1(\eta_p;p R_S)-F(\eta_{p_{\rm V_0}};p_{\rm V_0}R_S){\cal H}_1'(\eta_p;pR_S)}{F(\eta_{p_{\rm V_0}};p_{\rm V_0}R_S)F'(\eta_p;pR_S)-F'(\eta_{p_{\rm V_0}};p_{\rm V_0}R_S)F(\eta_p;pR_S)} \nonumber\\
&=& C^2_\eta\, p \left\{- \frac{{\cal H}_1(\eta_p;pR_S)}{F(\eta_p; pR_S)} 
+ \frac{p}{F^2(\eta_p;pR_S)\left[\frac{F'(\eta_p;\, pR_S)}{F(\eta_p; \,pR_S)}-\frac{F'(\eta_{p_{\rm V_0}};\, p_{\rm V_0}R_S)}{F(\eta_{p_{\rm V_0}};\, p_{\rm V_0}R_S)}\right]}\right\}
\end{eqnarray}
where $F'(\eta_q; qR_S)= dF(\eta_q; qr)/dr\,\Big|_{r=R_S}$ and ${\cal H}_1'(\eta_q;qR_S)= d{\cal H}_1(\eta_q; qr)/dr\,\Big|_{r=R_S}$ and we  have made use that the $z-$Wronksian of $F(\eta; z)$ and ${\cal H}_1(\eta; z)$ is one~\cite{Abramowitz:1972}, and  thus $F(\eta_q;qr){\cal H}_1'(\eta_q;qr)= -q + F'(\eta_q;qr){\cal H}_1(\eta_q;qr)$. In addition, 
\begin{eqnarray}
 F(\eta_q; qr) &=&   e^{-i\sigma_0(\eta_q)}u^{(+)}_{\ell=0\,,q}(r), \\
 \frac{dF(\eta_q;qr)}{dr}&=& q \left( \frac{1}{qr}+\eta_q\right) F(\eta_q; qr) -q\sqrt{1+\eta_q^2}\,  e^{-i\sigma_1(\eta_q)}u^{(+)}_{\ell=1\,,q}(r)  \\
 {\cal H}_1(\eta_q;qr)&=&e^{i\sigma_0(\eta_q)} e^{\pi\eta_q/2}W_{-i\eta_q, 1/2}(-2iqr)
\end{eqnarray}
On the other hand, $[{\cal H}_1(\eta_q;qr)-{\cal H}^*_1(\eta_q;qr)]= 2 i F(\eta_q; qr)$ \cite{Abramowitz:1972}, which guarantees that $F(\eta_q; qr)$ is real and that it is the imaginary part of ${\cal H}_1(\eta_q;qr)$. Thus, 
\begin{equation}
    \frac{{\cal H}_1(\eta_p;pR_S)}{F(\eta_p; pR_S)}= i + \frac{{\rm Re}[{\cal H}_1(\eta_p;pR_S)]}{F(\eta_p; pR_S)},
\end{equation}
which guarantees that $f_{SC}(p)$ fulfills unitarity and moreover we find  
\begin{eqnarray}
 C^2_\eta\, p \cot\delta   = C^2_\eta\, p \left\{ -\frac{{\rm Re}[{\cal H}_1(\eta_p;pR_S)]}{F(\eta_p; pR_S)}+ \frac{p}{F^2(\eta_p;pR_S)\left[\frac{F'(\eta_p;\, pR_S)}{F(\eta_p; \,pR_S)}-\frac{F'(\eta_{p_{\rm V_0}};\, p_{\rm V_0}R_S)}{F(\eta_{p_{\rm V_0}};\, p_{\rm V_0}R_S)}\right]} \right\}
\end{eqnarray}
Note that ${\rm Re}[{\cal H}_1(\eta_p;pr)]=G(\eta_p; \,pr)$, the irregular Coulomb $S$-wave function introduced in Ref.~\cite{Abramowitz:1972}, and hence ${\cal H}_1(\eta_p;pr)=G(\eta_p; \,pr) + iF(\eta_p; \,pr) $.

Using the asymptotic behaviors compiled in Chapter 33.10 of Ref.~\cite{DLMF} (see also \cite{Abramowitz:1972}) for $F(\eta; \,\rho), dF(\eta; \,\rho)/d\rho$ and $G(\eta; \,\rho)$ in the $\eta\to \pm \infty$ limits with $\eta\rho$ fixed, one can easily find the scattering length
\begin{subequations}\begin{align}
    a_0 & = -\frac{I_1^2(x_S)}{2\alpha\mu}\left [\frac{1}{\frac{x_S}{2}\frac{I_0(x_S)}{I_1(x_S)}-\tau_{\lambda}}-2I_1(x_S)K_1(x_S)\right]^{-1},\quad \lambda >0 \\
a_0 & = -\frac{J_1^2(x_S)}{2\alpha\mu}\left [\frac{1}{\frac{x_S}{2}\frac{J_0(x_S)}{J_1(x_S)}-\tau_{\lambda}}+\pi J_1(x_S)Y_1(x_S)\right]^{-1},\quad \lambda <0 
\end{align}
where $x_S = \sqrt{8\alpha\mu R_S}$, and there appear the [modified] Bessel functions of the first $J_n(x)$ [$I_n(x)$] and second $Y_n(x)$ [$K_n(x)$] kinds and:
\begin{equation}
\tau_{\lambda} = \frac{R_S\, F'\left(\lambda\alpha\mu/\sqrt{2\mu V_0};\,\sqrt{2\mu V_0} R_S\right)} {F\left(\lambda\alpha\mu/\sqrt{2\mu V_0};\, \sqrt{2\mu V_0}\, R_S\right)}= \frac{\sqrt{2\mu V_0}\,R_S}{\tan(\sqrt{2\mu V_0}\,R_S)}
+ {\cal O}(\alpha)
\end{equation}\end{subequations}
Finally, for both the repulsive and attractive cases, when $\alpha\to 0$ the scattering length becomes $a_0[\alpha= 0]=R_S-\tan(\sqrt{2\mu V_0}\,R_S)/\sqrt{2\mu V_0}$. The analytical expressions for the effective range $r_0$, in presence of Coulomb, are much involved. In the limit $\alpha\to 0$, one finds~\cite{Pascual:2012}
\begin{equation}
  r_0[\alpha= 0]= R_S\left( 1- \frac{1}{2\mu V_0\,R_S\,a_0[\alpha= 0]}-\frac{R_S^2}{3a^2_0[\alpha= 0]}\right)  \label{eq:r0neutro}
\end{equation}
The Coulomb interaction strongly modifies the scattering length, in particular when in its absence there exists a loosely bound state or a close-to-threshold virtual one, as can be seen in the surface plots of the top panels of Fig.~\ref{fig:a0CplusF}. However, the effective range is less  affected by Coulomb, and in general, it takes values around $R_S$ for sufficiently large depths $V_0$ of the strong potential. These behaviors can be appreciated in the top panels of Fig.~\ref{fig:effrange_dep}, where we see variations of around $\pm 10\%$, at most, due to the inclusion of the Coulomb interaction for the whole $(R_S,V_0)$ ranges of values considered in the plots. In the central panel of this first row of plots, we show the effective range for the neutral case, and observe that $r_0[\alpha= 0]$ [\textit{cf.} Eq.~\eqref{eq:r0neutro}] only significantly differs from $R_S$ in the two corner-regions of small and large depths and widths of the spherical square-well, especially in the first one where the effective range is much larger than $R_S$. 

Coulomb corrections obviously become quite large for spherical square-well strong potentials leading to small effective ranges $r_0[\alpha= 0]$ in the absence of the Coulomb interaction, as can be observed in the last row of plots displayed in Fig.~\ref{fig:effrange_dep}.

As an example for a system of two particles with mass that of the proton ($\mu=m_p/2$), the  spherical square-well potential used in Fig.~\ref{fig:wvfandOhnshi} ($V_0= 12.46$ MeV and $R_S=2.70$ fm) leads to $a_0[\alpha= 0]= -17.2$ fm  and $r_0[\alpha= 0]= 2.9$ fm in absence of Coulomb, when the latter interaction is switched on, one finds  $a_0=-7.8$ fm  and $r_0= 2.7$ fm or $a_0 = 177.1$ fm  and $r_0= 3.0$ fm, for the repulsive ($\lambda=+1$) or attractive ($\lambda=-1$) cases, respectively.

\bibliographystyle{apsrev4-1_MOD}
\bibliography{RefsCorrCoul}

\begin{thebibliography}{47}%
\makeatletter
\providecommand \@ifxundefined [1]{%
 \@ifx{#1\undefined}
}%
\providecommand \@ifnum [1]{%
 \ifnum #1\expandafter \@firstoftwo
 \else \expandafter \@secondoftwo
 \fi
}%
\providecommand \@ifx [1]{%
 \ifx #1\expandafter \@firstoftwo
 \else \expandafter \@secondoftwo
 \fi
}%
\providecommand \natexlab [1]{#1}%
\providecommand \enquote  [1]{``#1''}%
\providecommand \bibnamefont  [1]{#1}%
\providecommand \bibfnamefont [1]{#1}%
\providecommand \citenamefont [1]{#1}%
\providecommand \href@noop [0]{\@secondoftwo}%
\providecommand \href [0]{\begingroup \@sanitize@url \@href}%
\providecommand \@href[1]{\@@startlink{#1}\@@href}%
\providecommand \@@href[1]{\endgroup#1\@@endlink}%
\providecommand \@sanitize@url [0]{\catcode `\\12\catcode `\$12\catcode
  `\&12\catcode `\#12\catcode `\^12\catcode `\_12\catcode `\%12\relax}%
\providecommand \@@startlink[1]{}%
\providecommand \@@endlink[0]{}%
\providecommand \url  [0]{\begingroup\@sanitize@url \@url }%
\providecommand \@url [1]{\endgroup\@href {#1}{\urlprefix }}%
\providecommand \urlprefix  [0]{URL }%
\providecommand \Eprint [0]{\href }%
\providecommand \doibase [0]{http://dx.doi.org/}%
\providecommand \selectlanguage [0]{\@gobble}%
\providecommand \bibinfo  [0]{\@secondoftwo}%
\providecommand \bibfield  [0]{\@secondoftwo}%
\providecommand \translation [1]{[#1]}%
\providecommand \BibitemOpen [0]{}%
\providecommand \bibitemStop [0]{}%
\providecommand \bibitemNoStop [0]{.\EOS\space}%
\providecommand \EOS [0]{\spacefactor3000\relax}%
\providecommand \BibitemShut  [1]{\csname bibitem#1\endcsname}%
\let\auto@bib@innerbib\@empty
\bibitem [{\citenamefont {Hanbury~Brown}\ and\ \citenamefont
  {Twiss}(1954)}]{HanburyBrown:1954amm}%
  \BibitemOpen
  \bibfield  {author} {\bibinfo {author} {\bibfnamefont {R.}~\bibnamefont
  {Hanbury~Brown}}\ and\ \bibinfo {author} {\bibfnamefont {R.~Q.}\ \bibnamefont
  {Twiss}},\ }\href {\doibase 10.1080/14786440708520475} {\bibfield  {journal}
  {\bibinfo  {journal} {Phil. Mag. Ser. 7}\ }\textbf {\bibinfo {volume} {45}},\
  \bibinfo {pages} {663} (\bibinfo {year} {1954})}\BibitemShut {NoStop}%
\bibitem [{\citenamefont {Hanbury~Brown}\ and\ \citenamefont
  {Twiss}(1956)}]{HanburyBrown:1956bqd}%
  \BibitemOpen
  \bibfield  {author} {\bibinfo {author} {\bibfnamefont {R.}~\bibnamefont
  {Hanbury~Brown}}\ and\ \bibinfo {author} {\bibfnamefont {R.~Q.}\ \bibnamefont
  {Twiss}},\ }\href {\doibase 10.1038/1781046a0} {\bibfield  {journal}
  {\bibinfo  {journal} {Nature}\ }\textbf {\bibinfo {volume} {178}},\ \bibinfo
  {pages} {1046} (\bibinfo {year} {1956})}\BibitemShut {NoStop}%
\bibitem [{\citenamefont {Goldhaber}\ \emph {et~al.}(1960)\citenamefont
  {Goldhaber}, \citenamefont {Goldhaber}, \citenamefont {Lee},\ and\
  \citenamefont {Pais}}]{Goldhaber:1960sf}%
  \BibitemOpen
  \bibfield  {author} {\bibinfo {author} {\bibfnamefont {G.}~\bibnamefont
  {Goldhaber}}, \bibinfo {author} {\bibfnamefont {S.}~\bibnamefont
  {Goldhaber}}, \bibinfo {author} {\bibfnamefont {W.-Y.}\ \bibnamefont {Lee}},
  \ and\ \bibinfo {author} {\bibfnamefont {A.}~\bibnamefont {Pais}},\ }\href
  {\doibase 10.1103/PhysRev.120.300} {\bibfield  {journal} {\bibinfo  {journal}
  {Phys. Rev.}\ }\textbf {\bibinfo {volume} {120}},\ \bibinfo {pages} {300}
  (\bibinfo {year} {1960})}\BibitemShut {NoStop}%
\bibitem [{\citenamefont {Kopylov}(1974)}]{Kopylov:1974th}%
  \BibitemOpen
  \bibfield  {author} {\bibinfo {author} {\bibfnamefont {G.~I.}\ \bibnamefont
  {Kopylov}},\ }\href {\doibase 10.1016/0370-2693(74)90263-9} {\bibfield
  {journal} {\bibinfo  {journal} {Phys. Lett. B}\ }\textbf {\bibinfo {volume}
  {50}},\ \bibinfo {pages} {472} (\bibinfo {year} {1974})}\BibitemShut
  {NoStop}%
\bibitem [{\citenamefont {Ezell}\ \emph {et~al.}(1977)\citenamefont {Ezell},
  \citenamefont {Gutay}, \citenamefont {Laasanen}, \citenamefont {Dao},
  \citenamefont {Schubelin},\ and\ \citenamefont {Turkot}}]{Ezell:1977mh}%
  \BibitemOpen
  \bibfield  {author} {\bibinfo {author} {\bibfnamefont {C.}~\bibnamefont
  {Ezell}}, \bibinfo {author} {\bibfnamefont {L.~J.}\ \bibnamefont {Gutay}},
  \bibinfo {author} {\bibfnamefont {A.~T.}\ \bibnamefont {Laasanen}}, \bibinfo
  {author} {\bibfnamefont {F.~T.}\ \bibnamefont {Dao}}, \bibinfo {author}
  {\bibfnamefont {P.}~\bibnamefont {Schubelin}}, \ and\ \bibinfo {author}
  {\bibfnamefont {F.}~\bibnamefont {Turkot}},\ }\href {\doibase
  10.1103/PhysRevLett.38.873} {\bibfield  {journal} {\bibinfo  {journal} {Phys.
  Rev. Lett.}\ }\textbf {\bibinfo {volume} {38}},\ \bibinfo {pages} {873}
  (\bibinfo {year} {1977})}\BibitemShut {NoStop}%
\bibitem [{\citenamefont {Koonin}(1977)}]{Koonin:1977fh}%
  \BibitemOpen
  \bibfield  {author} {\bibinfo {author} {\bibfnamefont {S.~E.}\ \bibnamefont
  {Koonin}},\ }\href {\doibase 10.1016/0370-2693(77)90340-9} {\bibfield
  {journal} {\bibinfo  {journal} {Phys. Lett. B}\ }\textbf {\bibinfo {volume}
  {70}},\ \bibinfo {pages} {43} (\bibinfo {year} {1977})}\BibitemShut {NoStop}%
\bibitem [{\citenamefont {Lednicky}\ and\ \citenamefont
  {Lyuboshits}(1981)}]{Lednicky:1981su}%
  \BibitemOpen
  \bibfield  {author} {\bibinfo {author} {\bibfnamefont {R.}~\bibnamefont
  {Lednicky}}\ and\ \bibinfo {author} {\bibfnamefont {V.~L.}\ \bibnamefont
  {Lyuboshits}},\ }\href@noop {} {\bibfield  {journal} {\bibinfo  {journal}
  {Yad. Fiz.}\ }\textbf {\bibinfo {volume} {35}},\ \bibinfo {pages} {1316}
  (\bibinfo {year} {1981})}\BibitemShut {NoStop}%
\bibitem [{\citenamefont {Lisa}\ \emph {et~al.}(2005)\citenamefont {Lisa},
  \citenamefont {Pratt}, \citenamefont {Soltz},\ and\ \citenamefont
  {Wiedemann}}]{Lisa:2005dd}%
  \BibitemOpen
  \bibfield  {author} {\bibinfo {author} {\bibfnamefont {M.~A.}\ \bibnamefont
  {Lisa}}, \bibinfo {author} {\bibfnamefont {S.}~\bibnamefont {Pratt}},
  \bibinfo {author} {\bibfnamefont {R.}~\bibnamefont {Soltz}}, \ and\ \bibinfo
  {author} {\bibfnamefont {U.}~\bibnamefont {Wiedemann}},\ }\href {\doibase
  10.1146/annurev.nucl.55.090704.151533} {\bibfield  {journal} {\bibinfo
  {journal} {Ann. Rev. Nucl. Part. Sci.}\ }\textbf {\bibinfo {volume} {55}},\
  \bibinfo {pages} {357} (\bibinfo {year} {2005})},\ \Eprint
  {http://arxiv.org/abs/nucl-ex/0505014} {arXiv:nucl-ex/0505014}\BibitemShut
  {NoStop}%
\bibitem [{\citenamefont {Fabbietti}\ \emph {et~al.}(2021)\citenamefont
  {Fabbietti}, \citenamefont {Mantovani~Sarti},\ and\ \citenamefont
  {Vazquez~Doce}}]{Fabbietti:2020bfg}%
  \BibitemOpen
  \bibfield  {author} {\bibinfo {author} {\bibfnamefont {L.}~\bibnamefont
  {Fabbietti}}, \bibinfo {author} {\bibfnamefont {V.}~\bibnamefont
  {Mantovani~Sarti}}, \ and\ \bibinfo {author} {\bibfnamefont {O.}~\bibnamefont
  {Vazquez~Doce}},\ }\href {\doibase 10.1146/annurev-nucl-102419-034438}
  {\bibfield  {journal} {\bibinfo  {journal} {Ann. Rev. Nucl. Part. Sci.}\
  }\textbf {\bibinfo {volume} {71}},\ \bibinfo {pages} {377} (\bibinfo {year}
  {2021})},\ \Eprint {http://arxiv.org/abs/2012.09806} {arXiv:2012.09806
  [nucl-ex]}\BibitemShut {NoStop}%
\bibitem [{\citenamefont {Collaboration}\ \emph {et~al.}(2020)\citenamefont
  {Collaboration} \emph {et~al.}}]{ALICE:2020mfd}%
  \BibitemOpen
  \bibfield  {author} {\bibinfo {author} {\bibfnamefont {A.}~\bibnamefont
  {Collaboration}} \emph {et~al.} (\bibinfo {collaboration} {ALICE}),\ }\href
  {\doibase 10.1038/s41586-020-3001-6} {\bibfield  {journal} {\bibinfo
  {journal} {Nature}\ }\textbf {\bibinfo {volume} {588}},\ \bibinfo {pages}
  {232} (\bibinfo {year} {2020})},\ \bibinfo {note} {[Erratum: Nature 590, E13
  (2021)]},\ \Eprint {http://arxiv.org/abs/2005.11495} {arXiv:2005.11495
  [nucl-ex]}\BibitemShut {NoStop}%
\bibitem [{\citenamefont {Albaladejo}\ \emph {et~al.}(2023)\citenamefont
  {Albaladejo}, \citenamefont {Nieves},\ and\ \citenamefont
  {Ruiz-Arriola}}]{Albaladejo:2023pzq}%
  \BibitemOpen
  \bibfield  {author} {\bibinfo {author} {\bibfnamefont {M.}~\bibnamefont
  {Albaladejo}}, \bibinfo {author} {\bibfnamefont {J.}~\bibnamefont {Nieves}},
  \ and\ \bibinfo {author} {\bibfnamefont {E.}~\bibnamefont {Ruiz-Arriola}},\
  }\href {\doibase 10.1103/PhysRevD.108.014020} {\bibfield  {journal} {\bibinfo
   {journal} {Phys. Rev. D}\ }\textbf {\bibinfo {volume} {108}},\ \bibinfo
  {pages} {014020} (\bibinfo {year} {2023})},\ \Eprint
  {http://arxiv.org/abs/2304.03107} {arXiv:2304.03107 [hep-ph]}\BibitemShut
  {NoStop}%
\bibitem [{\citenamefont {Albaladejo}\ \emph {et~al.}(2017)\citenamefont
  {Albaladejo}, \citenamefont {Fernandez-Soler}, \citenamefont {Guo},\ and\
  \citenamefont {Nieves}}]{Albaladejo:2016lbb}%
  \BibitemOpen
  \bibfield  {author} {\bibinfo {author} {\bibfnamefont {M.}~\bibnamefont
  {Albaladejo}}, \bibinfo {author} {\bibfnamefont {P.}~\bibnamefont
  {Fernandez-Soler}}, \bibinfo {author} {\bibfnamefont {F.-K.}\ \bibnamefont
  {Guo}}, \ and\ \bibinfo {author} {\bibfnamefont {J.}~\bibnamefont {Nieves}},\
  }\href {\doibase 10.1016/j.physletb.2017.02.036} {\bibfield  {journal}
  {\bibinfo  {journal} {Phys. Lett. B}\ }\textbf {\bibinfo {volume} {767}},\
  \bibinfo {pages} {465} (\bibinfo {year} {2017})},\ \Eprint
  {http://arxiv.org/abs/1610.06727} {arXiv:1610.06727 [hep-ph]}\BibitemShut
  {NoStop}%
\bibitem [{\citenamefont {Du}\ \emph {et~al.}(2018)\citenamefont {Du},
  \citenamefont {Albaladejo}, \citenamefont {Fern\'andez-Soler}, \citenamefont
  {Guo}, \citenamefont {Hanhart}, \citenamefont {Mei\ss{}ner}, \citenamefont
  {Nieves},\ and\ \citenamefont {Yao}}]{Du:2017zvv}%
  \BibitemOpen
  \bibfield  {author} {\bibinfo {author} {\bibfnamefont {M.-L.}\ \bibnamefont
  {Du}}, \bibinfo {author} {\bibfnamefont {M.}~\bibnamefont {Albaladejo}},
  \bibinfo {author} {\bibfnamefont {P.}~\bibnamefont {Fern\'andez-Soler}},
  \bibinfo {author} {\bibfnamefont {F.-K.}\ \bibnamefont {Guo}}, \bibinfo
  {author} {\bibfnamefont {C.}~\bibnamefont {Hanhart}}, \bibinfo {author}
  {\bibfnamefont {U.-G.}\ \bibnamefont {Mei\ss{}ner}}, \bibinfo {author}
  {\bibfnamefont {J.}~\bibnamefont {Nieves}}, \ and\ \bibinfo {author}
  {\bibfnamefont {D.-L.}\ \bibnamefont {Yao}},\ }\href {\doibase
  10.1103/PhysRevD.98.094018} {\bibfield  {journal} {\bibinfo  {journal} {Phys.
  Rev. D}\ }\textbf {\bibinfo {volume} {98}},\ \bibinfo {pages} {094018}
  (\bibinfo {year} {2018})},\ \Eprint {http://arxiv.org/abs/1712.07957}
  {arXiv:1712.07957 [hep-ph]}\BibitemShut {NoStop}%
\bibitem [{\citenamefont {Torres-Rincon}\ \emph {et~al.}(2023)\citenamefont
  {Torres-Rincon}, \citenamefont {Ramos},\ and\ \citenamefont
  {Tolos}}]{Torres-Rincon:2023qll}%
  \BibitemOpen
  \bibfield  {author} {\bibinfo {author} {\bibfnamefont {J.~M.}\ \bibnamefont
  {Torres-Rincon}}, \bibinfo {author} {\bibfnamefont {A.}~\bibnamefont
  {Ramos}}, \ and\ \bibinfo {author} {\bibfnamefont {L.}~\bibnamefont
  {Tolos}},\ }\href {\doibase 10.1103/PhysRevD.108.096008} {\bibfield
  {journal} {\bibinfo  {journal} {Phys. Rev. D}\ }\textbf {\bibinfo {volume}
  {108}},\ \bibinfo {pages} {096008} (\bibinfo {year} {2023})},\ \Eprint
  {http://arxiv.org/abs/2307.02102} {arXiv:2307.02102 [hep-ph]}\BibitemShut
  {NoStop}%
\bibitem [{\citenamefont {Kong}\ and\ \citenamefont
  {Ravndal}(2000)}]{Kong:1999sf}%
  \BibitemOpen
  \bibfield  {author} {\bibinfo {author} {\bibfnamefont {X.}~\bibnamefont
  {Kong}}\ and\ \bibinfo {author} {\bibfnamefont {F.}~\bibnamefont {Ravndal}},\
  }\href {\doibase 10.1016/S0375-9474(99)00406-6} {\bibfield  {journal}
  {\bibinfo  {journal} {Nucl. Phys. A}\ }\textbf {\bibinfo {volume} {665}},\
  \bibinfo {pages} {137} (\bibinfo {year} {2000})},\ \Eprint
  {http://arxiv.org/abs/hep-ph/9903523} {arXiv:hep-ph/9903523}\BibitemShut
  {NoStop}%
\bibitem [{\citenamefont {Nieves}(2003)}]{Nieves:2003uu}%
  \BibitemOpen
  \bibfield  {author} {\bibinfo {author} {\bibfnamefont {J.}~\bibnamefont
  {Nieves}},\ }\href {\doibase 10.1016/j.physletb.2003.05.009} {\bibfield
  {journal} {\bibinfo  {journal} {Phys. Lett. B}\ }\textbf {\bibinfo {volume}
  {568}},\ \bibinfo {pages} {109} (\bibinfo {year} {2003})},\ \Eprint
  {http://arxiv.org/abs/nucl-th/0301080} {arXiv:nucl-th/0301080}\BibitemShut
  {NoStop}%
\bibitem [{\citenamefont {Zhang}\ and\ \citenamefont
  {Guo}(2021)}]{Zhang:2020mpi}%
  \BibitemOpen
  \bibfield  {author} {\bibinfo {author} {\bibfnamefont {Z.-H.}\ \bibnamefont
  {Zhang}}\ and\ \bibinfo {author} {\bibfnamefont {F.-K.}\ \bibnamefont
  {Guo}},\ }\href {\doibase 10.1103/PhysRevLett.127.012002} {\bibfield
  {journal} {\bibinfo  {journal} {Phys. Rev. Lett.}\ }\textbf {\bibinfo
  {volume} {127}},\ \bibinfo {pages} {012002} (\bibinfo {year} {2021})},\
  \Eprint {http://arxiv.org/abs/2012.08281} {arXiv:2012.08281
  [hep-ph]}\BibitemShut {NoStop}%
\bibitem [{\citenamefont {Shi}\ \emph {et~al.}(2022)\citenamefont {Shi},
  \citenamefont {Zhang}, \citenamefont {Guo},\ and\ \citenamefont
  {Yang}}]{Shi:2021hzm}%
  \BibitemOpen
  \bibfield  {author} {\bibinfo {author} {\bibfnamefont {P.-P.}\ \bibnamefont
  {Shi}}, \bibinfo {author} {\bibfnamefont {Z.-H.}\ \bibnamefont {Zhang}},
  \bibinfo {author} {\bibfnamefont {F.-K.}\ \bibnamefont {Guo}}, \ and\
  \bibinfo {author} {\bibfnamefont {Z.}~\bibnamefont {Yang}},\ }\href {\doibase
  10.1103/PhysRevD.105.034024} {\bibfield  {journal} {\bibinfo  {journal}
  {Phys. Rev. D}\ }\textbf {\bibinfo {volume} {105}},\ \bibinfo {pages}
  {034024} (\bibinfo {year} {2022})},\ \Eprint
  {http://arxiv.org/abs/2111.13496} {arXiv:2111.13496 [hep-ph]}\BibitemShut
  {NoStop}%
\bibitem [{\citenamefont {Barford}\ and\ \citenamefont
  {Birse}(2003)}]{Barford:2002je}%
  \BibitemOpen
  \bibfield  {author} {\bibinfo {author} {\bibfnamefont {T.}~\bibnamefont
  {Barford}}\ and\ \bibinfo {author} {\bibfnamefont {M.~C.}\ \bibnamefont
  {Birse}},\ }\href {\doibase 10.1103/PhysRevC.67.064006} {\bibfield  {journal}
  {\bibinfo  {journal} {Phys. Rev. C}\ }\textbf {\bibinfo {volume} {67}},\
  \bibinfo {pages} {064006} (\bibinfo {year} {2003})},\ \Eprint
  {http://arxiv.org/abs/hep-ph/0206146} {arXiv:hep-ph/0206146}\BibitemShut
  {NoStop}%
\bibitem [{\citenamefont {Newton}(1982)}]{Newton:1982qc}%
  \BibitemOpen
  \bibfield  {author} {\bibinfo {author} {\bibfnamefont {R.~G.}\ \bibnamefont
  {Newton}},\ }\href@noop {} {\emph {\bibinfo {title} {{Scattering theory of
  waves and particles}}}},\ edited by\ \bibinfo {editor} {\bibfnamefont
  {W.}~\bibnamefont {Beiglb{\"o}ck}}, \bibinfo {editor} {\bibfnamefont {E.~H.}\
  \bibnamefont {Lieb}}, \ and\ \bibinfo {editor} {\bibfnamefont
  {W.}~\bibnamefont {Thirring}}\ (\bibinfo  {publisher} {Springer},\ \bibinfo
  {year} {1982})\BibitemShut {NoStop}%
\bibitem [{\citenamefont {Birse}(2006)}]{Birse:2005um}%
  \BibitemOpen
  \bibfield  {author} {\bibinfo {author} {\bibfnamefont {M.~C.}\ \bibnamefont
  {Birse}},\ }\href {\doibase 10.1103/PhysRevC.74.014003} {\bibfield  {journal}
  {\bibinfo  {journal} {Phys. Rev. C}\ }\textbf {\bibinfo {volume} {74}},\
  \bibinfo {pages} {014003} (\bibinfo {year} {2006})},\ \Eprint
  {http://arxiv.org/abs/nucl-th/0507077} {arXiv:nucl-th/0507077}\BibitemShut
  {NoStop}%
\bibitem [{\citenamefont {Galindo}\ and\ \citenamefont
  {Pascual}(2012)}]{Pascual:2012}%
  \BibitemOpen
  \bibfield  {author} {\bibinfo {author} {\bibfnamefont {A.}~\bibnamefont
  {Galindo}}\ and\ \bibinfo {author} {\bibfnamefont {P.}~\bibnamefont
  {Pascual}},\ }\href@noop {} {\emph {\bibinfo {title} {{Quantum Mechanics
  II}}}}\ (\bibinfo  {publisher} {Springer-Verlag},\ \bibinfo {year}
  {2012})\BibitemShut {NoStop}%
\bibitem [{\citenamefont {Taylor}(2006)}]{Taylor:2006}%
  \BibitemOpen
  \bibfield  {author} {\bibinfo {author} {\bibfnamefont {J.}~\bibnamefont
  {Taylor}},\ }\href@noop {} {\emph {\bibinfo {title} {{Scattering Theory: The
  Quantum Theory of Nonrelativistic Collisions}}}}\ (\bibinfo  {publisher}
  {Dover Publications},\ \bibinfo {year} {2006})\BibitemShut {NoStop}%
\bibitem [{\citenamefont {Gell-Mann}\ and\ \citenamefont
  {Goldberger}(1953)}]{Gell-Mann:1953dcn}%
  \BibitemOpen
  \bibfield  {author} {\bibinfo {author} {\bibfnamefont {M.}~\bibnamefont
  {Gell-Mann}}\ and\ \bibinfo {author} {\bibfnamefont {M.~L.}\ \bibnamefont
  {Goldberger}},\ }\href {\doibase 10.1103/PhysRev.91.398} {\bibfield
  {journal} {\bibinfo  {journal} {Phys. Rev.}\ }\textbf {\bibinfo {volume}
  {91}},\ \bibinfo {pages} {398} (\bibinfo {year} {1953})}\BibitemShut
  {NoStop}%
\bibitem [{\citenamefont {Zorbas}(1976)}]{Zorbas:1976cd}%
  \BibitemOpen
  \bibfield  {author} {\bibinfo {author} {\bibfnamefont {J.}~\bibnamefont
  {Zorbas}},\ }\href {\doibase 10.1016/0034-4877(76)90063-X} {\bibfield
  {journal} {\bibinfo  {journal} {Rept. Math. Phys.}\ }\textbf {\bibinfo
  {volume} {9}},\ \bibinfo {pages} {309} (\bibinfo {year} {1976})}\BibitemShut
  {NoStop}%
\bibitem [{\citenamefont {Mukhamedzhanov}\ \emph {et~al.}(2012)\citenamefont
  {Mukhamedzhanov}, \citenamefont {Eremenko},\ and\ \citenamefont
  {Sattarov}}]{Mukhamedzhanov:2012qv}%
  \BibitemOpen
  \bibfield  {author} {\bibinfo {author} {\bibfnamefont {A.~M.}\ \bibnamefont
  {Mukhamedzhanov}}, \bibinfo {author} {\bibfnamefont {V.}~\bibnamefont
  {Eremenko}}, \ and\ \bibinfo {author} {\bibfnamefont {A.~I.}\ \bibnamefont
  {Sattarov}},\ }\href {\doibase 10.1103/PhysRevC.86.034001} {\bibfield
  {journal} {\bibinfo  {journal} {Phys. Rev. C}\ }\textbf {\bibinfo {volume}
  {86}},\ \bibinfo {pages} {034001} (\bibinfo {year} {2012})},\ \Eprint
  {http://arxiv.org/abs/1206.3791} {arXiv:1206.3791 [nucl-th]}\BibitemShut
  {NoStop}%
\bibitem [{\citenamefont {Okubo}\ and\ \citenamefont
  {Feldman}(1960)}]{Okubo:1960zz}%
  \BibitemOpen
  \bibfield  {author} {\bibinfo {author} {\bibfnamefont {S.}~\bibnamefont
  {Okubo}}\ and\ \bibinfo {author} {\bibfnamefont {D.}~\bibnamefont
  {Feldman}},\ }\href {\doibase 10.1103/PhysRev.117.279} {\bibfield  {journal}
  {\bibinfo  {journal} {Phys. Rev.}\ }\textbf {\bibinfo {volume} {117}},\
  \bibinfo {pages} {279} (\bibinfo {year} {1960})}\BibitemShut {NoStop}%
\bibitem [{\citenamefont {Landau}\ and\ \citenamefont
  {Lifshitz}(1977)}]{Landau:1977}%
  \BibitemOpen
  \bibfield  {author} {\bibinfo {author} {\bibfnamefont {L.}~\bibnamefont
  {Landau}}\ and\ \bibinfo {author} {\bibfnamefont {E.}~\bibnamefont
  {Lifshitz}},\ }\href@noop {} {\emph {\bibinfo {title} {{Quantum Mechanics:
  NonRelativistic Theory (third edition)}}}}\ (\bibinfo  {publisher} {Pergamon
  Press},\ \bibinfo {year} {1977})\BibitemShut {NoStop}%
\bibitem [{\citenamefont {Bethe}(1949)}]{Bethe:1949yr}%
  \BibitemOpen
  \bibfield  {author} {\bibinfo {author} {\bibfnamefont {H.~A.}\ \bibnamefont
  {Bethe}},\ }\href {\doibase 10.1103/PhysRev.76.38} {\bibfield  {journal}
  {\bibinfo  {journal} {Phys. Rev.}\ }\textbf {\bibinfo {volume} {76}},\
  \bibinfo {pages} {38} (\bibinfo {year} {1949})}\BibitemShut {NoStop}%
\bibitem [{\citenamefont {Jackson}\ and\ \citenamefont
  {Blatt}(1950)}]{Jackson:1950zz}%
  \BibitemOpen
  \bibfield  {author} {\bibinfo {author} {\bibfnamefont {J.~D.}\ \bibnamefont
  {Jackson}}\ and\ \bibinfo {author} {\bibfnamefont {J.~M.}\ \bibnamefont
  {Blatt}},\ }\href {\doibase 10.1103/RevModPhys.22.77} {\bibfield  {journal}
  {\bibinfo  {journal} {Rev. Mod. Phys.}\ }\textbf {\bibinfo {volume} {22}},\
  \bibinfo {pages} {77} (\bibinfo {year} {1950})}\BibitemShut {NoStop}%
\bibitem [{\citenamefont {van Haeringen}\ and\ \citenamefont
  {Kok}(1982)}]{vanHaeringen:1981pb}%
  \BibitemOpen
  \bibfield  {author} {\bibinfo {author} {\bibfnamefont {H.}~\bibnamefont {van
  Haeringen}}\ and\ \bibinfo {author} {\bibfnamefont {L.~P.}\ \bibnamefont
  {Kok}},\ }\href {\doibase 10.1103/PhysRevA.26.1218} {\bibfield  {journal}
  {\bibinfo  {journal} {Phys. Rev. A}\ }\textbf {\bibinfo {volume} {26}},\
  \bibinfo {pages} {1218} (\bibinfo {year} {1982})}\BibitemShut {NoStop}%
\bibitem [{\citenamefont {K\"onig}\ \emph {et~al.}(2013)\citenamefont
  {K\"onig}, \citenamefont {Lee},\ and\ \citenamefont
  {Hammer}}]{Konig:2012prq}%
  \BibitemOpen
  \bibfield  {author} {\bibinfo {author} {\bibfnamefont {S.}~\bibnamefont
  {K\"onig}}, \bibinfo {author} {\bibfnamefont {D.}~\bibnamefont {Lee}}, \ and\
  \bibinfo {author} {\bibfnamefont {H.~W.}\ \bibnamefont {Hammer}},\ }\href
  {\doibase 10.1088/0954-3899/40/4/045106} {\bibfield  {journal} {\bibinfo
  {journal} {J. Phys. G}\ }\textbf {\bibinfo {volume} {40}},\ \bibinfo {pages}
  {045106} (\bibinfo {year} {2013})},\ \Eprint {http://arxiv.org/abs/1210.8304}
  {arXiv:1210.8304 [nucl-th]}\BibitemShut {NoStop}%
\bibitem [{\citenamefont {Albaladejo}\ \emph {et~al.}(2024)\citenamefont
  {Albaladejo}, \citenamefont {Feijoo}, \citenamefont {Nieves}, \citenamefont
  {Oset},\ and\ \citenamefont {Vida\~na}}]{Albaladejo:2024lam}%
  \BibitemOpen
  \bibfield  {author} {\bibinfo {author} {\bibfnamefont {M.}~\bibnamefont
  {Albaladejo}}, \bibinfo {author} {\bibfnamefont {A.}~\bibnamefont {Feijoo}},
  \bibinfo {author} {\bibfnamefont {J.}~\bibnamefont {Nieves}}, \bibinfo
  {author} {\bibfnamefont {E.}~\bibnamefont {Oset}}, \ and\ \bibinfo {author}
  {\bibfnamefont {I.}~\bibnamefont {Vida\~na}},\ }\href {\doibase
  10.1103/PhysRevD.110.114052} {\bibfield  {journal} {\bibinfo  {journal}
  {Phys. Rev. D}\ }\textbf {\bibinfo {volume} {110}},\ \bibinfo {pages}
  {114052} (\bibinfo {year} {2024})},\ \Eprint
  {http://arxiv.org/abs/2410.08880} {arXiv:2410.08880 [hep-ph]}\BibitemShut
  {NoStop}%
\bibitem [{\citenamefont {Hostler}\ and\ \citenamefont
  {Pratt}(1963)}]{Hostler:1963zz}%
  \BibitemOpen
  \bibfield  {author} {\bibinfo {author} {\bibfnamefont {L.}~\bibnamefont
  {Hostler}}\ and\ \bibinfo {author} {\bibfnamefont {R.~H.}\ \bibnamefont
  {Pratt}},\ }\href {\doibase 10.1103/PhysRevLett.10.469} {\bibfield  {journal}
  {\bibinfo  {journal} {Phys. Rev. Lett.}\ }\textbf {\bibinfo {volume} {10}},\
  \bibinfo {pages} {469} (\bibinfo {year} {1963})}\BibitemShut {NoStop}%
\bibitem [{\citenamefont {Cho}\ \emph {et~al.}(2017)\citenamefont {Cho} \emph
  {et~al.}}]{ExHIC:2017smd}%
  \BibitemOpen
  \bibfield  {author} {\bibinfo {author} {\bibfnamefont {S.}~\bibnamefont
  {Cho}} \emph {et~al.} (\bibinfo {collaboration} {ExHIC}),\ }\href {\doibase
  10.1016/j.ppnp.2017.02.002} {\bibfield  {journal} {\bibinfo  {journal} {Prog.
  Part. Nucl. Phys.}\ }\textbf {\bibinfo {volume} {95}},\ \bibinfo {pages}
  {279} (\bibinfo {year} {2017})},\ \Eprint {http://arxiv.org/abs/1702.00486}
  {arXiv:1702.00486 [nucl-th]}\BibitemShut {NoStop}%
\bibitem [{\citenamefont {Preston}\ and\ \citenamefont
  {Bhaduri}(1993)}]{Preston:1993}%
  \BibitemOpen
  \bibfield  {author} {\bibinfo {author} {\bibfnamefont {M.}~\bibnamefont
  {Preston}}\ and\ \bibinfo {author} {\bibfnamefont {R.~K.}\ \bibnamefont
  {Bhaduri}},\ }\href@noop {} {\emph {\bibinfo {title} {{Structure of the
  Nucleus}}}}\ (\bibinfo  {publisher} {Westview Press},\ \bibinfo {year}
  {1993})\BibitemShut {NoStop}%
\bibitem [{\citenamefont {Lednicky}(2009)}]{Lednicky:2005tb}%
  \BibitemOpen
  \bibfield  {author} {\bibinfo {author} {\bibfnamefont {R.}~\bibnamefont
  {Lednicky}},\ }\href {\doibase 10.1134/S1063779609030034} {\bibfield
  {journal} {\bibinfo  {journal} {Phys. Part. Nucl.}\ }\textbf {\bibinfo
  {volume} {40}},\ \bibinfo {pages} {307} (\bibinfo {year} {2009})},\ \Eprint
  {http://arxiv.org/abs/nucl-th/0501065} {arXiv:nucl-th/0501065}\BibitemShut
  {NoStop}%
\bibitem [{\citenamefont {Ohnishi}\ \emph {et~al.}(2016)\citenamefont
  {Ohnishi}, \citenamefont {Morita}, \citenamefont {Miyahara},\ and\
  \citenamefont {Hyodo}}]{Ohnishi:2016elb}%
  \BibitemOpen
  \bibfield  {author} {\bibinfo {author} {\bibfnamefont {A.}~\bibnamefont
  {Ohnishi}}, \bibinfo {author} {\bibfnamefont {K.}~\bibnamefont {Morita}},
  \bibinfo {author} {\bibfnamefont {K.}~\bibnamefont {Miyahara}}, \ and\
  \bibinfo {author} {\bibfnamefont {T.}~\bibnamefont {Hyodo}},\ }\href
  {\doibase 10.1016/j.nuclphysa.2016.05.010} {\bibfield  {journal} {\bibinfo
  {journal} {Nucl. Phys. A}\ }\textbf {\bibinfo {volume} {954}},\ \bibinfo
  {pages} {294} (\bibinfo {year} {2016})},\ \Eprint
  {http://arxiv.org/abs/1603.05761} {arXiv:1603.05761 [nucl-th]}\BibitemShut
  {NoStop}%
\bibitem [{ALI(2022)}]{ALICE:2022wwr}%
  \BibitemOpen
  \href@noop {} {\emph {\bibinfo {title} {{Letter of intent for ALICE 3: A
  next-generation heavy-ion experiment at the LHC}}}}\ (\bibinfo {year}
  {2022})\ \Eprint {http://arxiv.org/abs/2211.02491} {2211.02491}\BibitemShut
  {NoStop}%
\bibitem [{\citenamefont {Gmitro}\ \emph {et~al.}(1986)\citenamefont {Gmitro},
  \citenamefont {Kvasil}, \citenamefont {Lednicky},\ and\ \citenamefont
  {Lyuboshits}}]{Gmitro:1986ay}%
  \BibitemOpen
  \bibfield  {author} {\bibinfo {author} {\bibfnamefont {M.}~\bibnamefont
  {Gmitro}}, \bibinfo {author} {\bibfnamefont {J.}~\bibnamefont {Kvasil}},
  \bibinfo {author} {\bibfnamefont {R.}~\bibnamefont {Lednicky}}, \ and\
  \bibinfo {author} {\bibfnamefont {V.~L.}\ \bibnamefont {Lyuboshits}},\ }\href
  {\doibase 10.1007/BF01598029} {\bibfield  {journal} {\bibinfo  {journal}
  {Czech. J. Phys. B}\ }\textbf {\bibinfo {volume} {36}},\ \bibinfo {pages}
  {1281} (\bibinfo {year} {1986})}\BibitemShut {NoStop}%
\bibitem [{\citenamefont {Acharya}\ \emph {et~al.}(2020)\citenamefont {Acharya}
  \emph {et~al.}}]{ALICE:2019buq}%
  \BibitemOpen
  \bibfield  {author} {\bibinfo {author} {\bibfnamefont {S.}~\bibnamefont
  {Acharya}} \emph {et~al.} (\bibinfo {collaboration} {ALICE}),\ }\href
  {\doibase 10.1016/j.physletb.2020.135419} {\bibfield  {journal} {\bibinfo
  {journal} {Phys. Lett. B}\ }\textbf {\bibinfo {volume} {805}},\ \bibinfo
  {pages} {135419} (\bibinfo {year} {2020})},\ \Eprint
  {http://arxiv.org/abs/1910.14407} {arXiv:1910.14407 [nucl-ex]}\BibitemShut
  {NoStop}%
\bibitem [{\citenamefont {Reid}(1968)}]{Reid:1968sq}%
  \BibitemOpen
  \bibfield  {author} {\bibinfo {author} {\bibfnamefont {R.~V.}\ \bibnamefont
  {Reid}, \bibfnamefont {Jr.}},\ }\href {\doibase 10.1016/0003-4916(68)90126-7}
  {\bibfield  {journal} {\bibinfo  {journal} {Annals Phys.}\ }\textbf {\bibinfo
  {volume} {50}},\ \bibinfo {pages} {411} (\bibinfo {year} {1968})}\BibitemShut
  {NoStop}%
\bibitem [{\citenamefont {Acharya}\ \emph {et~al.}(2019)\citenamefont {Acharya}
  \emph {et~al.}}]{ALICE:2018ysd}%
  \BibitemOpen
  \bibfield  {author} {\bibinfo {author} {\bibfnamefont {S.}~\bibnamefont
  {Acharya}} \emph {et~al.} (\bibinfo {collaboration} {ALICE}),\ }\href
  {\doibase 10.1103/PhysRevC.99.024001} {\bibfield  {journal} {\bibinfo
  {journal} {Phys. Rev. C}\ }\textbf {\bibinfo {volume} {99}},\ \bibinfo
  {pages} {024001} (\bibinfo {year} {2019})},\ \Eprint
  {http://arxiv.org/abs/1805.12455} {arXiv:1805.12455 [nucl-ex]}\BibitemShut
  {NoStop}%
\bibitem [{\citenamefont {Adamczyk}\ \emph {et~al.}(2015)\citenamefont
  {Adamczyk} \emph {et~al.}}]{STAR:2014dcy}%
  \BibitemOpen
  \bibfield  {author} {\bibinfo {author} {\bibfnamefont {L.}~\bibnamefont
  {Adamczyk}} \emph {et~al.} (\bibinfo {collaboration} {STAR}),\ }\href
  {\doibase 10.1103/PhysRevLett.114.022301} {\bibfield  {journal} {\bibinfo
  {journal} {Phys. Rev. Lett.}\ }\textbf {\bibinfo {volume} {114}},\ \bibinfo
  {pages} {022301} (\bibinfo {year} {2015})},\ \Eprint
  {http://arxiv.org/abs/1408.4360} {arXiv:1408.4360 [nucl-ex]}\BibitemShut
  {NoStop}%
\bibitem [{\citenamefont {Shapoval}\ \emph {et~al.}(2015)\citenamefont
  {Shapoval}, \citenamefont {Erazmus}, \citenamefont {Lednicky},\ and\
  \citenamefont {Sinyukov}}]{Shapoval:2014yha}%
  \BibitemOpen
  \bibfield  {author} {\bibinfo {author} {\bibfnamefont {V.~M.}\ \bibnamefont
  {Shapoval}}, \bibinfo {author} {\bibfnamefont {B.}~\bibnamefont {Erazmus}},
  \bibinfo {author} {\bibfnamefont {R.}~\bibnamefont {Lednicky}}, \ and\
  \bibinfo {author} {\bibfnamefont {Y.~M.}\ \bibnamefont {Sinyukov}},\ }\href
  {\doibase 10.1103/PhysRevC.92.034910} {\bibfield  {journal} {\bibinfo
  {journal} {Phys. Rev. C}\ }\textbf {\bibinfo {volume} {92}},\ \bibinfo
  {pages} {034910} (\bibinfo {year} {2015})},\ \Eprint
  {http://arxiv.org/abs/1405.3594} {arXiv:1405.3594 [nucl-th]}\BibitemShut
  {NoStop}%
\bibitem [{\citenamefont {Abramowitz}\ and\ \citenamefont
  {Stegun}(1972)}]{Abramowitz:1972}%
  \BibitemOpen
  \bibfield  {author} {\bibinfo {author} {\bibfnamefont {M.}~\bibnamefont
  {Abramowitz}}\ and\ \bibinfo {author} {\bibfnamefont {I.~A.~E.}\ \bibnamefont
  {Stegun}},\ }\href@noop {} {\emph {\bibinfo {title} {{Handbook of
  Mathematical Functions with Formulas, Graphs, and Mathematical Tables, 9th
  printing}}}}\ (\bibinfo  {publisher} {Dover Publications},\ \bibinfo {year}
  {1972})\BibitemShut {NoStop}%
\bibitem [{\citenamefont {Olver}\ \emph {et~al.}()\citenamefont {Olver},
  \citenamefont {Olde~Daalhuis}, \citenamefont {Lozier}, \citenamefont
  {Schneider}, \citenamefont {Boisvert}, \citenamefont {Clark}, \citenamefont
  {Miller}, \citenamefont {Saunders}, \citenamefont {Cohl},\ and\ \citenamefont
  {McClain}}]{DLMF}%
  \BibitemOpen
  \bibfield  {author} {\bibinfo {author} {\bibfnamefont {F.}~\bibnamefont
  {Olver}}, \bibinfo {author} {\bibfnamefont {A.}~\bibnamefont
  {Olde~Daalhuis}}, \bibinfo {author} {\bibfnamefont {D.}~\bibnamefont
  {Lozier}}, \bibinfo {author} {\bibfnamefont {B.}~\bibnamefont {Schneider}},
  \bibinfo {author} {\bibfnamefont {R.}~\bibnamefont {Boisvert}}, \bibinfo
  {author} {\bibfnamefont {C.}~\bibnamefont {Clark}}, \bibinfo {author}
  {\bibfnamefont {B.}~\bibnamefont {Miller}}, \bibinfo {author} {\bibfnamefont
  {B.~V.}\ \bibnamefont {Saunders}}, \bibinfo {author} {\bibfnamefont
  {H.}~\bibnamefont {Cohl}}, \ and\ \bibinfo {author} {\bibfnamefont {M.~E.}\
  \bibnamefont {McClain}},\ }\href {https://dlmf.nist.gov/} {\emph {\bibinfo
  {title} {{NIST Digital Library of Mathematical Functions. Release 1.2.2 of
  2024-09-15}}}}\BibitemShut {NoStop}%
\end{thebibliography}%
\end{document}